\documentstyle[12pt,twoside]{article}
\makeindex
\renewcommand{\baselinestretch}{1.4}

\def\be{\begin{equation}}
\def\ee{\end{equation}}
\def\bea{\begin{eqnarray}}
\def\eea{\end{eqnarray}}
\def\a{\alpha}
\def\b{\beta}
\def\l{\lambda}
\def\t{\tau}
\def\s{\sigma}
\def\pa{\partial}
\def\e{\varepsilon}
\def\d{\delta}
 
\def\D{\Delta}
\def\L{\Lambda}

\author{Hans-J\"urgen Schmidt}

\title{Lectures on Mathematical Cosmology }

\date{}
\begin{document}
\maketitle

\centerline{Universit\"at Potsdam, Institut f\"ur Mathematik, Am
Neuen Palais 10} 
 \centerline{D-14469~Potsdam, Germany,  E-mail:
 hjschmi@rz.uni-potsdam.de}
\centerline{http://www.physik.fu-berlin.de/\~{}hjschmi }

\bigskip

\copyright  2004, H.-J. Schmidt  

\thispagestyle{empty}

\newpage
\setcounter{page}{3}
\tableofcontents 

\newpage

\section{Preface}\label{Kap0}

We present mathematical details of several 
cosmological models, whereby the topological 
 and  the  geometrical background  will be emphasized. 

\bigskip

 This book arose from lectures I read as advanced courses in 
 the following universities: 1990 at TU Berlin, 1991 in M\"unster, 
1993 - 1996 and 2002, 2003 in  Potsdam, 
1999 and 2000 in Salerno, and 2001 at FU Berlin. 

\bigskip

 The reader is assumed to be acquainted with basic knowledge 
on general relativity, e.g. by knowing   Stephani's book \cite{stephani}. 
As a rule, the deduction is made so explicit, that one should be able 
to follow the details without further background knowledge. 
The intention is to present a piece of mathematical physics related
 to cosmology to fill the gap between standard textbooks like 
MTW\footnote{C. Misner, K. Thorne, J. Wheeler {\it Gravitation},
 Freeman, San Francisco 1973}, LL\footnote{L. Landau, E. Lifschitz
 {\it Field Theory}, Nauka, Moscow 1973},  
or Weinberg\footnote{S. Weinberg
 {\it Gravitation  and Cosmology}, J. Wiley, New York 1972}, on the
 one hand and to the current research literature on the other. 
 I hope that this will become  useful 
for both students and researchers who are interested  in  cosmology,
mathematical physics or differential geometry. 

\bigskip

\noindent 
To the reference list: If more than one source is given in one
 item, then a dot is put between them. If, however, two versions 
of the same source are  given, then a semicolon is put between them.  
The gr-qc number in the reference list refers to the number in the
preprint archive 

\centerline{http://xxx.lanl.gov/find/}

\noindent 
and the numbers in italics given at the end of an item 
 show the pages where this item is mentioned. This way,  the
Bibliography simultaneously  serves as author index. 

\bigskip

Parts of this book\footnote{The details are mentioned at the corresponding
 places.}  are composed from revised versions of papers of mine
published earlier in research journals. They are reprinted with 
the kind permissions of the copyright owners: Akademie-Verlag Berlin for 
{\it Astron. Nachr.} and Plenum Press  New York for {\it Gen. Rel. Grav.}
 The details about the originals  can also be seen in

\centerline{http://www.physik.fu-berlin.de/\~{}hjschmi }

\bigskip

\noindent 
I  thank the following colleagues:  
Luca Amendola, Michael Bachmann, H. Baumg\"artel, W. Benz, 
Claudia Bernutat, 
Ulrich  Bleyer, A. Borde,  H.-H. v. Borzesz\-kowski, Salvatore
 Capozziello, G. Dautcourt, Vladimir  Dzhu\-nu\-sha\-liev, Mauro Francaviglia, 
 T.  Friedrich, V. Frolov, H. Goenner, Stefan Gottl\"ober, V. Gurovich, 
Graham Hall, F. Hehl, Olaf Heinrich,
Alan  Held, L. Herrera, Rainer John, U. Kasper, Jerzy Kijowski, 
 Hagen Kleinert, B. Klotzek, Sabine Kluske, O. Kowalski, Andrzej Krasinski, 
W. K\"uhnel, W. Kundt, K.  Lake, K.  Leichtweiss, D.-E.  Liebscher, 
  Malcolm MacCallum, W. Mai, R. Mansouri, V. Melnikov, 
P. Michor, Salvatore Mignemi, 
Volker M\"uller, S.  Nikcevic, I. Novikov, 
F. Paiva, Axel  Pelster, V. Perlick, 
K. Peters, Ines Quandt,  L. Querella, M. Rainer, J. Reichert,  Stefan Reuter, 
W. Rinow, M. Sanchez,  R. Schimming, M. Schulz, E. Schmutzer, 
U. Semmelmann, I. Shapiro,  A. Starobinsky, P. Teyssandier, 
H.-J. Treder, H. Tuschik, L. Vanhecke  and A. Zhuk for valuable comments,
 and my wife Renate for patience and support. 

\bigskip

\bigskip

\hspace{4.5cm} {\it Hans-J\"urgen Schmidt, Potsdam, July  2004}

\vfill

\eject 

\newpage

\section[Changes of the  Bianchi type]{Cosmological models with 
changes of the  Bianchi type}\label{Kap1}
\setcounter{equation}{0}

We  investigate  such cosmological models which instead of the usual spatial
 homogeneity property only fulfil
 the condition that in a certain synchronized system of reference all spacelike 
sections    $t$ = const. are homogeneous manifolds.

\bigskip

This allows time-dependent changes of the Bianchi  type. 
Discussing differential-geometrical theorems it is shown which of them are 
permitted.  Besides the trivial case of changing into type I
 there exist some possible changes between other types. However, physical reasons
 like energy inequalities partially exclude them.

\subsection{Introduction to generalized Bianchi models}

Recently,   besides the   known Bianchi  models,
 there are investigated certain classes of inhomogeneous cosmological
 models.   This is done to
get a better representation of the really existent inhomogeneities, 
cf. e.g. Bergmann \cite{bergmann}, Carmeli  \cite{carmeli},
 Collins  \cite{collins}, Spero  \cite{spero}, 
Szekeres  \cite{szekeres} and Wainwright  \cite{wain}. 

\bigskip

We consider,   similar as in   Collins  \cite{collins}, 
such inhomogeneous  models $V_4$ which in a certain synchronized system of 
 reference possess homogeneous  sections $t$ = const., called 
$V_3(t)$. This is   analogous  to  the generalization
 of the concept of spherical symmetry in Krasinski  \cite{krasinski},
 cf. also \cite{krasinskib}.  

\bigskip

In this chapter, which is based on \cite{sch82a},
  we especially investigate which
 time-dependent changes   of the Bianchi  type are possible. Thereby 
we impose, besides the twice continuous differentiability, 
 a physically reasonable condition: the energy inequality,   
\be\label{1.1a}
T_{00} \ge \vert T_{\alpha \beta}\vert \, , 
\ee
holds in each Lorentz  frame.

\bigskip

We begin with  some globally topological properties of spacetime:  
Under the physical condition (\ref{1.1a})   the topology  of the sections $V_3(t)$ is, 
according to Lee  \cite{lee}, independent of $t$.  Hence, the 
Kantowski-Sachs models,  with underlying topology 
${\rm S}^2 \times {\rm R}$ or ${\rm S}^2 \times {\rm S}^1$  and the 
models of Bianchi  type IX with  underlying topology ${\rm S}^3$  or continuous 
images of it as  SO(3) may not change, because all other types are represented by
the R$^3$-topology, factorized with reference to a discrete subgroup of
 the group of motions. But  the remaining types can all be 
represented in R$^3$-topology itself; therefore we do not get any
further global restrictions. All homogeneous  models, the  above mentioned
Kantowski-Sachs model being excluded, possess simply transitive 
groups  of motion.  If the isometry group $G$ has dimension $\ge 4$, then
this statement means: $G$ possesses a 3-dimensional simply-transitive 
 subgroup acting transitively on the $V_3(t)$. 
 Hence we do not specialize if we deal only   with locally 
simply-transitive groups of motions. We consequently do not 
consider here trivial changes of the  Bianchi  type, e. g.  from
type III to type VIII by means of an intermediate on which a group of motions
 possessing  transitive subgroups  of both types acts.

\bigskip

Next, we consider the easily tractable case of a change to type I:
  For each  type M there  one can find a manifold $V_4$
 such that for each $t \le  0$  the section $V_3(t)$
 is flat and for each $t > 0$ it belongs to type M. Indeed, one has simply to use for 
 every  $V_3(t)$  such a  representative  of type M
that their curvature vanishes 
 as $t \to 0_+$. Choosing exponentially decreasing curvature 
one can obtain  an arbitrarily high differentiable
class for the metric, e. g.
\be\label{1.1b}
     ds^2 =  -dt^2 + dr^2 +  h^2 \cdot (d\psi^2 + \sin^2 \psi d\phi^2) \, ,
\ee 
where
\be\label{1.1c}
h=  \left\{
 \begin{array}{c}
 r \qquad {\rm   for} \qquad  t \le 0   \\
 \exp (t^{-2}) \cdot \sinh^2 \left( r \cdot \exp(-t^{-2}) \right) \quad {\rm else}\, .
\end{array}
\right.  
\ee
This is a $C^\infty$-metric whose slices  $V_3(t)$
belong to type I for $t \le 0$  and to type
V for $t > 0$.
In $t =  0$, of course, it cannot be an analytical  one.  The limiting slice belongs 
 necessarily to type I by continuity reasons.
 Applying this fact twice it becomes obvious that by the
help of a flat intermediate of finite extension or only by a single
 flat slice,  all Bianchi types
can be matched together. However, if  one does not want to use 
such a flat intermediate  the
transitions of one Bianchi type  to another become a non-trivial 
problem. It is shown from the purely differential-geometrical
 as well  as from the  physical points of view,  sections \ref{s13}  and \ref{s14} 
respectively,  which  types can be matched  together immediately 
 without a  flat  intermediate. To this end
we collect the following preliminaries.

\subsection{Spaces possessing homogeneous slices}

As one knows, in cosmology the homogeneity principle is expressed by the 
fact that to a spacetime $V_4$
 there exists a group of  motions acting transitively on the spacelike 
hypersurfaces $V_3(t)$  of a slicing of $V_4$. Then the metric
is given by
\be\label{1.1}
ds^2 = - dt^2 + g_{ab}(t) \omega^a  \omega^b
\ee
where the $g_{ab}(t)$  are  positively definite and $\omega^a$ 
are the basic 1-forms corresponding  to a certain
Bianchi  type; for details see e.g. \cite{kramer}.  If  the   $\omega^a$    are  related to a 
holonomic basis $x^i$, using type-dependent functions
$ A^a_i(x^j)   $, one 
can write:
\be\label{1.2}
\omega^a =  A^a_i \,  dx^i    \, .
\ee

\bigskip

For the spaces considered here we have however: 
there is a synchronized system  of reference
such that the slices $t= $  const. are homogeneous
 spaces $V_3(t)$. In this system of reference the
metric is given by
\be\label{1.3}
     ds^2 = - dt^2 + g_{ij} dx^i dx^j 
\ee
where   $g_{ij}(x^i,t)$   are  twice continuously differentiable and 
homogeneous  for constant $t$. Hence
it is a generalization of eqs. (\ref{1.1}) and (\ref{1.2}). 
This is a genuine generalization 
because there is only
posed the condition $g_{0\alpha} = - \delta_{0\alpha}$
 on the composition of  the homogeneous slices 
$V_3(t)$  to a $V_4$.
 This means in each  slice only the first fundamental form
 $g_{ij}$  is homogeneous;  but in the contrary 
to homogeneous models, the second fundamental form 
$\Gamma_{0ab}$  need not have this property.
Therefore, also the curvature scalar   $^{(4)}R$,    
and with it the distribution of matter,  need not be
constant within a $V_3(t)$.  

\bigskip

Now let $t$ be fixed. Then 
 one can find  coordinates     $ x^i_t (x^j,t)  $  in $V_3(t)$  
such that according to eqs.  (\ref{1.1}) and (\ref{1.2})
 the inner metric gets the form
\be\label{1.4}
 g_{ab}(t) \,  \omega^a_t \,   \omega^b_t   \qquad {\rm where} \qquad
\omega^a_t =  A^a_i (x^i_t) \,  dx^i_t \, .
\ee
Transforming this into the original coordinates $x^ i$
 one obtains for the metric of the full $V_4$
\be\label{1.5}
g_{0\alpha} = - \delta_{0\alpha} \, , \qquad 
g_{ij} = g_{ab}(t) \, A^a_k (x^i_t) \ A^b_l (x^i_t) \,  
x^k_{t,i} \,   x^l_{t,j} \, .
\ee
If  the Bianchi  type changes with  time one has to take 
such an ansatz\footnote{The word ``ansatz" (setting) stems from the German noun 
 ``der Ansatz"; that word stems from the German verb ``setzen" (to set).} 
like eqs. (\ref{1.1b}), (\ref{1.1c})  for each interval of
constant type separately; between them  one has to secure
 a $C^2$-joining.  An example of such an inhomogeneous model 
 is, cf. Ellis 1967  \cite{ellis}
\be\label{1.5a}
 ds^2 =   -dt^2  + t^{-2/3} \left[ t + C(x) \right]^2 
dx^2 +   t^{4/3} (dy^2 + dz^2) \,  .
\ee
The  slices $t =$ const. are  flat, 
but only  for constant $C$ it belongs to Bianchi type I, cf. 
also \cite{griff}  for such a model. Later, see eq. (\ref{2.19}), we will 
 present  a more   general model.

\subsection{Continuous changes of the Bianchi type}\label{s13}

Using the usual homogeneity property the
 same group of motions acts on each slice $V_3(t)$.
Hence the Bianchi  type is independent of time
 by definition. But this fails to be the case for
the spaces considered here. However, we can deduce the following:
 completing $\partial/\partial t$ to an
anholonomic basis which is connected with Killing  vectors 
in each $V_3(t)$  one obtains for the
structure constants  associated to the commutators of  the basis:
\be\label{1.7}
C^0_{\alpha \beta} =0 \, ; \quad
 C^i_{jk}(t)  \quad  {\rm depends \  continuously \  on \   time} \, . 
\ee
Note:  These structure constants  are  calculated 
 as follows: Without loss of generality  let $x^i_t(0,0,0,t)=0$.
At this point 
$\partial/\partial x^i_t$
 and 
$\partial/\partial x^i$ 
 are  taken as initial values for Killing  vectors within $V_3(t)$. 
The  structure constants obtained by these Killing  vectors are
 denoted by   
$  \bar  C^i_{jk} (t) $
 and $ C^i_{jk} (t) $
respectively,  where
$  \bar  C^i_{jk} (t) $ 
 are the canonical ones. Between them it holds at $(0,0,0,t)$:
\be\label{1.6x}%
x^i_{t,j} C^j_{kl} =  \bar  C^i_{mn}
 x^m_{t,k} x^n_{t,l} \, . 
\ee

\bigskip

Using only the usual canonical structure constants then of
 course no type is changeable continuously.
 To answer the question which Bianchi types may change 
continuously we consider all sets of  structure constants 
 $ C^i_{jk}  $   being antisymmetric in $jk$ and fulfilling
 the Jacobi   identity. Let   $C_{\rm S}$  be
such a set belonging to type S. Then we have
 according to eq. (\ref{1.7}): Type R is changeable
continuously into type S, symbolically expressed by the 
validity of    R $\to$  S, if  and only if to each
   $C_{\rm S}$   there exists a sequence 
$ C_{\rm R}^{(n)}  $   such  that one has
\be\label{1.8}
\lim_{n \to \infty}  C_{\rm R}^{(n)}   =  C_{\rm S} \, .
\ee
The limit has to be understood componentwise.

\bigskip

It holds that R $\to$   S  if and only if there are
 a     $C_{\rm S}$  and a sequence 
 $ C_{\rm R}^{(n)}  $  such that  eq. (\ref{1.8})  is fulfilled.
The equivalence of both statements is shown by 
means of simultaneous rotations of the basis.
In practice one takes as    $C_{\rm S}$ 
 the canonical structure constants and investigates for which types
R there can be found corresponding sequences 
$ C_{\rm R}^{(n)}  $:  the components of    $C_{\rm S}$  
 are subjected to a perturbation not exceeding $\varepsilon$
 and their Bianchi types  are  calculated.   Finally one  looks which types appear 
for all   $\varepsilon > 0$.   Thereby one profits, e.g., from 
 the   statement that the dimension of 
 the image space of the Lie algebra, which equals 0  for type I,
\dots, and equals 3 for types VIII and IX,   cannot increase during such changes;
 see \cite{rainer} and \cite{sch87b} for more details about these Lie algebras.

\bigskip

One obtains the  following diagram. 
The validity of R $\to$   S and S $\to$  T  implies the validity 
of R  $\to$ T, hence  the   diagram must be  continued transitively. 
The statement   VI$_\infty \  \to $ IV expresses the
 facts that there exists a sequence 
\be
C^{(n)}_{{\rm VI}_h} \to C_{\rm IV}
\ee
and that in each
such a sequence the parameter $h$  must necessarily  tend to  infinity.

\bigskip

Further VI$_h \  \to$  II holds for every  $h$.
 Both statements  hold analogously for type  VII$_h$. In
MacCallum   \cite{maccallum}   a similar diagram    is shown,   but
 there it is only investigated 
which  changes appear if some of the canonical structure
 constants are  vanishing.  So e.g., the different
transitions from  type VI$_h$  to types II and IV are not 
contained in it.


\setlength{\unitlength}{2cm}
\begin{picture}(4.5,6.4) 
\thicklines
\put(4,1.5){\vector(-1,0){1.8}} 
\put(4,0){\vector(-1,0){1.8}}
\put(4.2,1.3){\vector(0,-1){1.1}} 
\put(4.2,2.8){\vector(0,-1){1.1}} 
\put(3.45,2.96){\vector(-1,-1){1.2}} 
\put(2.85,4.46){\vector(-1,-4){0.65}} 
\put(4.85,4.32){\vector(-1,-4){0.61}} 
\put(2,1.3){\vector(0,-1){1.1}} 
\put(0.55,2.96){\vector(1,0){1.25}}   
\put(0.55,3.05){\vector(1,1){1.25}}   
\put(0.52,4.46){\vector(1,0){1.28}}   

\thinlines 
\put(1.96,-0.07){\Large I}
\put(4.16,-0.07){\Large V}
\put(4.13,1.46){\Large IV}
\put(3.9,2.89){\Large VI$_\infty$}
\put(1.93,2.89){\Large VI$_0$}
\put(0.0,2.89){\Large VIII}
\put(4.4,4.39){\Large VII$_\infty$}
\put(1.91,4.39){\Large VII$_0$}
\put(0.2,4.39){\Large IX}
\put(1.96,1.46){\Large II}

\linethickness{0.7mm} 

\put(4.3,4.46){\line(-1,0){1.83}}      
\put(3.8,2.96){\line(-1,0){1.45}}      

\end{picture}

\bigskip


\centerline{Fig. 1, taken from \cite{sch82a}}

\bigskip

\bigskip

To complete the answer to the question 
posed above it must be added: To each transition shown in the diagram, 
 there one can indeed find a spacetime $V_4$  in  which it is realized. Let
 $f(t)$  be a $C^\infty$-function
 with $f(t) = 0$  for $t \le 0$  and $f(t) > 0$ else. Then, e.g., 
\be\label{1.8b}
ds^2 = -dt^2 + dx^2
+ e^{2x} \,  dy^2 + e^{2x} \, \left( dz \,  + \,  x \,  f(t) \,   dy \right)^2 
\ee
 is a  $C^\infty$-metric  whose slices $V_3(t)$ belong to type V  
and IV for $ t \le  0$ and $t > 0$ respectively. 
Presumably it is typical that in  a neighbourhood of $t =0$  
the curvature becomes singular at spatial infinity; this is at least 
the case for metric (\ref{1.8b}).  
 And,  for  the transition R $\to$  S  the limiting slice
 be1ongs  necessarily to type S.

\subsection{Physical conditions}\label{s14}

To obtain physically reasonable spacetimes
 one has at least to secure the validity of an energy  inequality
 like (\ref{1.1a}),   $T_{00} \ge \vert T_{\alpha \beta} \vert $
 in each Lorentz  frame. Without this requirement
 the transition II $\to$ I   is possible. We prove that the requirement $T_{00} \ge  0$ 
alone is sufficient to forbid this transition.
To this end let $V_4$  be a manifold which in a certain
 synchronized system of   reference possesses
a flat slice $V_3(0)$,  and for all $t > 0$ has Bianchi  type II-slices $V_3(t)$.
 Using the notations of    eq. (\ref{1.6x}) we  have: 
$ \bar  C^1_{23} = -   \bar  C^1_{32} =1  $
 are the only non-vanishing canonical 
structure constants of Bianchi  type II. With  the
exception $A_2^1= - x^3$
    we have $A^a_i = \delta^a_i$.

\bigskip

 By the help of     eq. (\ref{1.6x})  one can calculate the structure 
constants $ C^i_{jk}(t)$. The
flat-slice condition is  equivalent to
\be\label{1.9}
\lim_{t \to 0}   \  C^i_{jk} (t) =0 \, .
\ee
First we consider the special case   $x^i_t = a^i(t) \cdot x^i$
  (no sum), i. e. extensions of  the coordinate axes. It holds 
 $x^i_{t,j} = \delta ^i_j \,   a^i $   (no sum). The only
 non-vanishing     $C^i_{jk} (t)$   are 
$ C^1_{23} = -  C^1_{32} =     a(t) $, 
where $ a = a^2 \cdot  a^3/a^1$.
 Then    eq. (\ref{1.9}) reads $\lim_{\, t \to 0} \,  a(t) = 0$. 
The coefficients $g_{ab}(t)$ have to be chosen such 
that  the $g_{ij}$, according  to  eq. (\ref{1.5}), remain positive definite
and twice continuously differentiable. Let $^{(3)}R$
 be the scalar curvature within  the slices.    $^{(3)}R > 0$
appears only in Bianchi  type IX and in the Kantowski-Sachs  models. 
But changes from these
types are just the cases already excluded by global considerations.
 Inserting      $g_{ij}$   into the
Einstein equation by means of the  Gauss-Codazzi  theorem 
one obtains
\be\label{1.10}
\kappa T_{00} = R_{00} - \frac{1}{2} g_{00} \,  {}^{(4)}R = \frac{1}{2} 
 \,  {}^{(3)}R + \frac{1}{4} \,  g^{-1} \cdot H \, , \quad 
 g = \det g_{ij} \, , 
\ee
$^{(3)}R$  being the scalar curvature within the slices, hence
\bea\label{1.11} 
^{(3)}R \le 0 \quad  {\rm and } \quad  H =
g_{11}[g_{22,0} g_{33,0} - (g_{23,0})^2] \nonumber \\
- \, 2 \cdot g_{12} [ g_{33,0} g_{12,0}
- g_{13,0} g_{23,0} ]  + {\rm cyclic \  perm}\,  .
\eea

\bigskip

In our case $H$  is a quadratic  polynomial in $x^3$ whose
 quadratic coefficient  reads
\be\label{1.11a}
- g_{11} [g_{11}g_{33} - (g_{13})^2] \cdot
 \left[ \frac{\partial a(t)}{\partial t} 
\right]^2 \, . 
\ee 
Hence, for sufficiently  large  values $x^3$  and  values $t$  with 
$\partial a /\partial t \,  \ne \,  0$  we have $H <0$,  and
therefore $T_{00}<0$.

\bigskip

Secondly we  hint at another special case, namely rotations of the
 coordinate axes  against each other, e.g.
$$
x^1_t = x^1 \cos \omega + x^2 \sin \omega \, , \quad
x^2_t = x^2 \cos \omega - x^1  \sin \omega \, , \quad
x^3_t = x^3 ,   \quad \omega = \omega (t) \, .
$$
There one obtains in analogy to eq. (\ref{1.11a}) a  negative $T_{00}$. 
Concerning the general case  we  have: loosely
speaking, each diffeomorphism  is  a composite 
of such extensions and rotations. 
Hence, for
each II $\to$ I-transition one would obtain points with
 negative  $T_{00}$. 

\bigskip

Hence, the energy condition is 
a genuine restriction to the possible transitions of  the  Bianchi
types.  Concerning the other 
transitions we remark: equations   (\ref{1.10}),  (\ref{1.11})
 keep valid, and the remaining work is to 
examine the signs of the corresponding expressions $ H$.
 Presumably one  always obtains points 
with negative $H$,  i.e., also under this weakened homogeneity
presumptions the Bianchi types 
of the spacelike hypersurfaces at different distances
 from the singularity must coincide.

\newpage

\section[Inhomogeneous  models with flat slices]{Inhomogeneous 
cosmological models with flat slices}\label{Kap2}
\setcounter{equation}{0}

A family of cosmological models
 is considered which in a certain synchronized system of
reference possess flat slices
$t=$  const. They are generated from the Einstein-De Sitter  universe 
by a suitable transformation. Under physically 
reasonable presumptions  these transformed models fulfil certain energy conditions. 
In Wainwright  \cite{wain}   a  class of inhomogeneous cosmological models
 is considered which have the following property: there
exists a synchronized system of reference of such a 
kind that the slices $t=$ const. are homogeneous manifolds. 

\bigskip

Here
 we consider a special family of such models which possess flat
 slices. Following \cite{sch82b},  we use the transformation formalism developed
 in chapter  \ref{Kap1}.  In  Krasinski \cite{krasinskib}, e.g. in Fig. 2.1., 
 it is shown   how this approach  
fits into other classes of cosmological solutions of the Einstein field 
 equation.  Additionally, we require that these transformations
 leave two coordinates unchanged; this
implies the existence of a 2-dimensional Abelian group of motions. 
A similar requirement is   posed in \cite{wain}, too.  

\bigskip
 
Starting from a Friedmann universe,  we investigate
whether the energy inequalities are fulfilled in the transformed model, too. 
In general this fails  to be the case, 
but starting from the Einstein-De Sitter  universe 
 \cite{einsteindesitter}, cf. also Tolman  \cite{tolman},  and
 requiring perfect fluid for the transformed model, the energy inequalities
in the initial model imply their validity in the transformed model. 
 A new review on exact solutions can be found in [26], and a new
 proof of Birkhoff's theorem is given in [226]. 

\subsection{Models with flat slices}

The transformation formalism of chapter \ref{Kap1}  restricted to Bianchi 
type I  reads as follows: using the same notations eq. (\ref{1.2}), 
we have $A_i^a = \delta_i^a$,  $\omega^a = dx^a$ and
\be\label{2.1}
     ds^2 = - dt^2 + g_{ab}(t) dx^a dx^b       
\ee
as the initial hypersurface-homogeneous model instead of eq. (\ref{1.1}). 
Now let us consider the time-dependent
transformation $x^a_t (x^i, t)$, where for 
each $t$ it has to be a diffeomorphism of $R^3$. 
Then one obtains in place of eq. (\ref{1.5})
\be\label{2.2}
 g_{0\alpha}    =  - \delta_{0 \alpha} , \qquad g_{ij} = g_{ab}(t)
 x^a_{t,i} x^b_{t,j}  
\ee
as the transformed model. It  is no restriction to insert 
$g_{ab}(t) = \delta_{ab}$  in eq. (\ref{2.1}), i.e. to start from
Minkowski  spacetime.  In the following we consider
 only transformations which leave two  coordinates unchanged, i.e., 
now writing $t$, $x$, $y$, $z$ instead of $x^0, \dots x^3$  resp., 
 we restrict to  transformations which read as follows 
\be\label{2.3}
     x_t(x, t) \,  , \qquad   y_t = y \,  , \qquad z_t = z \, .   
\ee
These transformations we shall   call $x$-transformations. 

\bigskip

 Using $x$-transformations,   the Killing vectors   $\partial/\partial y$  
and     $\partial/\partial z$    of the initial model remain Killing
vectors.  They form a 2-dimensional Abelian
group of motions. All others,  
including the rotation $ z(\partial/\partial y) - y (  \partial/\partial z) $, 
may fail  to remain Killing  vectors. On the contrary to the general case, the 
$x$-transformed models depend
genuinely on the initial ones. In the following,  the spatially flat Friedmann  
universe with power-law behaviour of the cosmic scale factor 
 shall  be used  as initial model, i.e.
\be\label{2.4}
 g_{ab} (t) = \delta_{ab} K^2(t) \qquad {\rm with} \qquad K(t) = t^\tau \, .
\ee
Together with  eq. (\ref{2.1}) one obtains from the Einstein field equation  
\be\label{2.5}
\kappa \mu = \kappa T_{00} = 3\tau^2 / t^2 \, , \quad
p=T^2_2 = \alpha \mu      \quad {\rm with} \quad
 \alpha = \frac{2}{3\tau} -1 \, ;
\ee
where $\mu$ is the energy density and $p$ is the pressure.

\bigskip

Now, inserting  eq. (\ref{2.4}) in  eq. (\ref{2.1}) and transforming 
to  eq. (\ref{2.2}) with restriction  eq. (\ref{2.3}) one obtains for the metric
of the $x$-transformed  model
\be\label{2.6}
     g_{11} = t^{2\tau} \cdot 
    (x_{t,1})^2 \equiv   t^{2\tau} \cdot  h(x, t) \,  , \quad 
 g_{22} = g_{33} =     t^{2\tau} \, , \quad 
 g_{\alpha \beta} = \eta_{\alpha \beta}
 \   {\rm    else} \, . 
\ee
Here, $ \eta_{\alpha \beta} = {\rm diag} (-1, \, 1, \, 1, \, 1)$ is the metric 
 of the flat Minkowski spacetime. 
This metric (\ref{2.6}) belongs   to the so-called
 Szekeres class,  cf.   \cite{szekeres}.   Defining $a(x, t)$  by
\be\label{2.7}
     g_{11} =   e^{2     a(x, t) } \,   ,
\ee
a coordinate  transformation $\tilde x(x)$ yields
 $ a(x, 1) = 0$. If $v = a_0 \, t$,  then we have
\be\label{2.8}
     a(x, t) = \int_1^t    v(x, \tilde t) \cdot \tilde t^{-1} d \tilde t \, ,    
\ee
$v(x, t)$
 being an arbitrary  twice continuously differentiable function
 which may be singular at $t = 0$ and $t = \infty$.
The initial model is included by setting $x_t = x$, hence 
$a = \tau \ln t$, $v = \tau$.

\bigskip

For  $\tau \ne 0 $
 different functions $v$ correspond to the same model  only if
 they are connected by a translation into $x$-direction. 
Inserting  eq. (\ref{2.6})  with  eq. (\ref{2.7}) and  eq. (\ref{2.8})
 into the Einstein field  equation one
obtains the energy-momentum tensor
\bea \label{2.9}
\kappa T_{00} = ( \tau^2 + 2 v \tau   )/t^2 \, ,
    \nonumber \\
\kappa T^1_1 =  (2  \tau - 3 \tau^2   )/t^2 \, ,
    \nonumber \\
\kappa T^2_2 = \kappa T^3_3 = -v_{,0} t^{-1} +
 (  \tau -  \tau^2   - v\tau - v^2 + v  )/t^2 \, ,
    \nonumber \\
\kappa T_{\alpha \beta} = 0
 \qquad {\rm else\, ,    \qquad  and}     \nonumber \\
\kappa T = \kappa T^\alpha_\alpha = - 2 v_{,0} t^{-1} -
 2 ( 3 \tau^2 - 2 \tau +2 v\tau + v^2 - v  )/t^2 \, .
\eea
The question, in which cases  equation  (\ref{2.6}) gives  a usual
  hypersurface-homogeneous model, can 
be answered as follows: metric   (\ref{2.6})  is a Friedmann  universe, 
if and only if $h_{,0} = 0$. 
For this case  it is isometric to the initial model. Metric   (\ref{2.6})  is a 
hypersurface-homogeneous model, if
 and only  if  functions $A$ and $B$
 exist for which holds $h(x, t) =  A(x) \cdot  B(t)$.
 Because of $h > 0$  this is
equivalent to   $ h_{,01} \cdot  h = h_{,0} \cdot h_{,1}$.
 In this case it is a Bianchi type I  model.

\subsection{Energy inequalities and perfect fluid models}\label{s23}

In this section it shall be discussed, 
in which manner the geometrically defined models
described by eqs.   (\ref{2.6}),   (\ref{2.7}),   (\ref{2.8}) and  (\ref{2.9})
 fulfil some energy conditions. Here we impose the following
conditions: each observer measures non-negative energy density, 
time- or lightlike energy flow and spacelike tensions which are not greater 
than the energy density. In our coordinate   system
these conditions are expressed by the following inequalities
\be\label{2.10}
     T_{00} \ge \vert T^1_1 \vert \, ,  
\ee
\be\label{2.11}   
     T_{00} \ge \vert T^2_2 \vert \, ,  
\ee
\be\label{2.12}  
 {\rm   and } \qquad   T \le 0 \, . 
\ee
For the initial model this means $\tau =   0$  or $ \tau \ge 1/2$, i.e. Minkowski  
spacetime or $-1 < \alpha \le 1/3$.  For the limiting case $\a = -1$ one gets the de Sitter 
spacetime in the original model (\ref{2.5}) which does not have the assumed
 power-law behaviour for the scale factor. Using the energy-momentum tensor 
  eq. (\ref{2.9}),  eqs. (\ref{2.11}) and    (\ref{2.12}) read
\be\label{2.13}
 v_{,0} \, t \ge -v^2 - (3\tau - 1)  v + \tau - 2 \tau^2       \, ,
\ee
\be\label{2.14}
 v_{,0} \, t \le -v^2 + (\tau +1) v + \tau   \qquad   {\rm   and }
\ee
\be\label{2.15}
 v_{,0} \, t \ge -v^2  - (2 \tau -1) v + 2\tau - 3 \tau^2     \, .
\ee

\bigskip

Now, if $\tau  < 0$, i.e. $\alpha  < -1$, then  eq. (\ref{2.10})
reads $v \le \tau -1$;  together with   eq. (\ref{2.14}) one obtains 
$v_{,0} \, t \le \tau - 2$. This
 implies the existence of  a $\tilde t > 0$ 
with   $v(\tilde t) \ge -1$ in contradiction to condition  (\ref{2.10}). 
 Therefore, an initial model with $\tau  < 0 $, which itself
 contradicts the energy inequalities,  cannot produce 
transformed  models which always fulfil them.

\bigskip

If $\tau = 0$, then  eqs. (\ref{2.13}) and  
 (\ref{2.14}) imply  $v_{,0} \, t  = v - v^2$. 
This   equation has the solutions $v = 0$  and
$ v = t(t + C)^{-1} $  with 
arbitrary $C(x) \ge 0$. This yields $a = 0$ and $a = \ln (t + C) - 
\ln (1+  C) $  with eq.  (\ref{2.7}),   and $g_{11} = 1$ and 
$ g_{11} = (t + C)^2 \cdot  (1 + C)^{-2}$   resp. with  eq. (\ref{2.8}). Then 
 eq. (\ref{2.6}) shows that this is the 
Minkowski  spacetime  itself. Therefore, the Minkowski spacetime  does
 not produce  any new  models.

\bigskip

Finally, if $\tau > 0$, hence $\alpha  > -1$,  eq. (\ref{2.10})  then reads 
 $v \ge {\rm  max} (\tau -   1, 1 - 2\tau)$. A lengthy   calculation
shows which transformed models fulfil the energy inequalities. 
For each $\tau > 0$ models exist
 which do and models which  do not fulfil them.

\bigskip

The situation described above changes if
 one requires that the transformed model consists of
 perfect fluid with an equation of state. The velocity vector must be 
 $(1, 0, 0, 0)$  and
\be\label{2.16}
\kappa(T^1_1 - T^2_2 ) \equiv 
\frac{h_{,0}}{(1+\alpha) t \cdot h}
+ \frac{1}{2 \sqrt h}\left(\frac{  h_{,0}  }{\sqrt h}
\right)_{,0}=0
\ee
must be fulfilled. Eq. (\ref{2.16}) expresses the condition that the pressure has 
to be isotropic. $T^2_2 = T^3_3$ is fulfilled anyhow due to our assumption
 eq. (\ref{2.3}).

 If $ f = h_{,0}h^{-1/2}$, then eq. (\ref{2.16})  reads
$$
\frac{f}{(1+\alpha) t}
+ \frac{f_{,0}}{2} = 0 \, ,
$$
hence $f_{,0} f^{-1}$    does not depend on $x$,
 therefore $f_{,01}\cdot f = f_{,0} \cdot  f _{,1}$, 
and we can use the ansatz   $f =   a(t) \cdot b(x)$.
Inserting this    into  eq. (\ref{2.16}), one obtains
\be\label{2.17}
h= \left[ b(x) \cdot  t^{(\alpha -1)(\alpha +1)}
 + c(x) \right]^2 \, , \quad {\rm    where } \quad 
  c(x) = \sqrt{ h(x,0)}
\ee
with arbitrary non-negative functions $b$ and $c$
 fulfilling $b(x) + c(x) > 0$ for all $x$. For energy
density and pressure we then obtain
\bea\label{2.18}
\kappa \mu = \frac{4}{3(1+\alpha)^2 t^2} -
\frac{4(1-\alpha) \cdot b
}{3(1+\alpha)^2 t (bt + ct^{2/(1+\alpha)}) } \  ,  \nonumber \\
\kappa p = \frac{4\alpha}{3(1+\alpha)^2 t^2} \ .
\eea

\bigskip

An equation of state means that $p$ uniquely depends on $\mu$.
 This takes  place if and only if $\alpha =  0$ or
 $b/c =$  const. The latter is equivalent to the
 hypersurface-homogeneity of the model and is of lower interest here.
For $\alpha =   0$      the initial
model is the dust-filled Einstein-De Sitter model \cite{einsteindesitter}.

\bigskip

 With  eq. (\ref{2.18})  we obtain
 $p =  0$  and $\mu \ge   0$  and may  formulate:   If the Einstein-De Sitter 
 universe is $x$-transformed into a perfect fluid model, then
 this model also contains dust and fulfils the energy conditions.

\bigskip

These models have the following form: 
inserting  eq. (\ref{2.17}) with $\alpha =  0$  into  eq. (\ref{2.6})
 we get a dust-filled model
\be\label{2.19}
ds^2 \,  = \,  -dt^2 \,  + \,  t^{4/3} \, 
\{ \, 
 [ \, b(x)/t + c(x) \, ]^2 \, dx^2 \,  + \,  dy^2 \,  + \,  dz^2 \,   \} 
\ee
with arbitrary $b$, $c$ as before. A subcase of
 eq.  (\ref{2.19}) is contained in Szekeres  \cite{szekeres}
 as case (iii), but the parameter $\varepsilon$ used there may now  be
non-constant. For $b=1$,  eq. (\ref{2.19}) reduces to  eq. (\ref{1.5a}). 
 Of course, locally one can achieve a constant $b$ by a transformation 
 $\tilde x = \tilde x(x)$, but, e.g. changes in the sign of $b$ then are not
 covered.  
As an illustration we give two examples of  this model  eq. (\ref{2.19}):

\bigskip
 
1.   If $b =1$, $c > 0$, then
\be
\kappa \mu = \frac{4c}{3t(1+ct)} \, ,
\ee
hence the density contrast at two different values $x_1$, $x_2$
 reads
\be
 \frac{\mu_1}{\mu_2} =
  \frac{c_2}{c_1}\cdot  \frac{1 + c_1 t}{1 + c_2 t} 
\ee
and tends to 1 as $t \to \infty$. This shows that one 
needs additional presumptions if one wants to
prove an amplification of initial density fluctuations.

\bigskip

2. If $b = 1$, $c = 0$, then  eq. (\ref{2.19})  is the axially symmetric  Kasner
vacuum solution. If now $c$ differs from zero in the
neighbourhoods of two values     $x_1$, $x_2$,  then 
the model is built up from two thin dust slices and
Kasner-like vacuum outside them. 
The invariant distance of  the slices is
\be\label{2.19c}
A t^{-1/3} + B t^{2/3}
\ee 
with certain positive constants $A$  and $B$. In
 \cite{sch96c} it is argued that the invariant distance along
 spacelike geodesics need not coincide with what 
 we measure as spatial distance. The distinction comes from the 
fact, that a geodesic within a prescribed spacelike hypersurface need 
not coincide with a spacelike geodesic in spacetime. 
 This eq. (\ref{2.19c}) looks like gravitational repulsion, because the
distance has a minimum  at a positive value. But the $t^{-1/3}$-term is due to
 the participation of the slices in cosmological expansion and 
the remaining $t^{2/3}$-term is due to an attractive
gravitational force in parabolic  motion. A more  detailed discussion 
of this behaviour eq. (\ref{2.19c})
 is given in Griffiths \cite{griff}: he argues, ``however
 inhomogeneous the mass distribution, matter on each plane of symmetry 
 has no net attraction to matter on other planes." 

\bigskip

The transformation of a hypersurface-homogeneous cosmological
 model  considered here  preserves all inner properties, expressed by 
the first  fundamental form,  of the slices $t = $ const., 
and also preserves  the property that $t$  is a synchronized
time, but may change all other outer
 properties, which are  expressed by the second fundamental
form.  The investigations of section  \ref{s23}  show that energy 
conditions are preserved under  very special presumptions only.

\bigskip

One may consider these transformations as a guide in the search
 for new exact solutions of
Einstein's  field equation. The new models  are
  close   to the initial hypersurface-homogeneous
ones, if the transformation is close
 to the identical one. Thus, one can perturb a Bianchi model
 with exact solutions without use of  any   approximations.

\subsection{A simple singularity theorem}

What do we exactly know about solutions when no exact 
solution,  in the sense of  ``solution in closed
form'',  is available? In which sense these solutions have a singularity? 
 Here we make some remarks  to these questions, see \cite{schy}.

\bigskip

In \cite{schv},  the following simple type of singularity theorems was 
discussed:  The coordinates $t, \ x, \ y, \ z$ shall cover all 
the reals, and $a(t)$ shall be an arbitrary strictly 
positive monotonously increasing smooth function defined 
for all  real values $t$, where ``smooth'' denotes 
``$C^{\infty}$-differentiable''. Then it holds: 

\bigskip

The Riemannian space defined by 
\be
ds^2= dt^2+a^2(t)(dx^2+dy^2+dz^2)
\ee
 is geodesically complete. This fact is well-known and 
easy to prove;  however, on the other hand, 
 for the same class of functions $a(t)$ it holds:

\bigskip

The  spacetime defined by 
\begin{equation}
ds^2= dt^2-a^2(t)(dx^2+dy^2+dz^2)
\end{equation}
 is lightlike geodesically complete if and only if 
\begin{equation}
\int_{- \infty}^0 \ a(t) \ dt \ \ = \ \ \infty
\end{equation}
is fulfilled. The proof is straightforwardly done by 
considering lightlike  geodesics in the $x-t-$plane. So, one directly 
concludes that the statements do not depend on
the number $n$ of spatial dimensions, only $n>0$ is
used, and the formulation for $n=3$ was
chosen as the most interesting case. Moreover, this
 statement  remains valid if we replace 
``lightlike geodesically complete'' by 
``lightlike and timelike geodesically complete''.

\bigskip

Further recent results on singularity theorems are
reviewed in Senovilla [233]. This review and most of
the research concentrated mainly on the 4--dimensional
Einstein theory. Probably, the majority of arguments 
can be taken over to the higher--dimensional Einstein
equation without change, but this has not yet worked
out up to now. And  singularity theorems for
$F(R)$-gravity are known up to now only for very
special cases only. 

\bigskip

The most frequently used $F(R)$-Lagrangian is 
\be 
L = \left( \frac{R}{2} \ - \  \frac{l^2}{12} \, R^2 \right) 
\ \sqrt{-g}  \qquad {\rm where} \  l > 0 \, , 
\ee
here we discuss the non--tachyonic case only. 
From this  Lagrangian  one gets a fourth-order field equation; 
only very few closed-form solutions,  ``exact solutions'',  
are known. However, for the class of spatially flat
Friedmann models,  the set of solutions is
qualitatively  completely described, but not in closed form, 
in  \cite{muell1985}. We call them ``rigorous solutions''. 

\bigskip

One of them, often called ``Starobinsky inflation'', 
can be approximated  with 
\begin{equation} 
a(t) \ = \ \exp( - \frac{t^2}{12 l^2} )
\end{equation}
This  approximation is valid in the region
$t \ll - l$.
However, this solution does not fulfil the above integral condition. 
Therefore,   Starobinsky inflation does not 
represent a lightlike geodesically complete cosmological 
model as has been frequently stated in the literature. 

\bigskip

To prevent a further misinterpretation let me reformulate this result as 
follows: Inspite of the fact that the Starobinsky 
 model is regular, in the sense that $a(t) > 0$ for arbitrary
values of synchronized time $t$, every past--directed 
lightlike geodesic terminates in a curvature
singularity, i.e.,  $\vert R \vert \longrightarrow \infty$, 
 at a finite value of its affine parameter. Therefore, the model is not only 
geodesically  incomplete in the coordinates chosen, moreover, it also
fails to be a subspace of a complete one. 

\bigskip
In contrast to this one can say: the spatially flat  Friedmann model  
 with $a(t) = \exp(H t)$, $H$ being a positive constant, is the inflationary 
de Sitter spacetime.  According to the above integral condition, it 
is also incomplete.  However,  contrary to the Starobinsky model, 
it is a subspace of a complete spacetime, namely of the 
the de Sitter spacetime represented as a closed Friedmann model.

\bigskip

The application of conformal transformations represents one of 
 the most powerful methods to transform 
different theories into each other.
 In many cases this related theories to each other 
 which have been originally considered to be independent ones. 
     The most often discussed question  in this context 
is ``which of these metrics is the physical one'', but 
one can, of course, use such conformal relations  
also as a simple mathematical tool to find exact solutions
without the necessity of answering this question.
     However, concerning a conformal transformation of 
 singularity theorems one must
be cautious, because typically, near a singularity in one of the
theories, the conformal factor diverges, and then the
conformally transformed metric need not be singular there. 

\bigskip
 
Finally, let me again mention the distinct properties
of the Euclidean and the Lorentzian signature: 
 Smooth connected complete Riemannian spaces are 
geodesically connected. However, 
smooth connected complete spacetimes,  
e.g. the de Sitter space--time, need not be 
geodesically connected, see  for instance 
Sanchez [196]. The topological origin of this distinction 
 is the same as the distinction  between the two 
above statements: for the positive definite  signature case one 
uses the compactness of the rotation group, whereas in 
the indefinite case the  noncompactness of the Lorentz group 
has to be observed [227].

\newpage
\setcounter{page}{31}

\section{Properties of curvature invariants}\label{Kap3}
\setcounter{equation}{0}

In section \ref{s31}, which is an extended version of
 \cite{sch96a} we answer the following question: 
Let $\l$, $\mu$, $\nu$ be arbitrary real numbers. 
Does there exist a  3-dimensional  homogeneous 
 Riemannian manifold whose eigenvalues  of the
 Ricci tensor are just  $\l$, $\mu$ and $\nu$? 

\bigskip

In section \ref{s31a} we discuss, following \cite{sch96b}, why
 all the curvature invariants of a gravitational wave vanish. 

\bigskip

In section \ref{s32} we present the  example \cite{sch95} that it may
 be possible that 
 non--isometric spacetimes with non--vanishing curvature scalar
 cannot be distinguished by curvature invariants.

\subsection[The  space of  3--dimensional 
Riemannian manifolds]{The space of  3--dimensional 
homogeneous Riemannian manifolds}\label{s31}

The curvature of a 3--manifold is completely  determined by the Ricci tensor, 
 and the Ricci tensor of a homogeneous manifold has constant eigenvalues.  
 However, it is essential to observe, that,
nevertheless, the constancy of all three eigenvalues of the Ricci 
tensor of a 3-manifold does not imply the manifold 
to be locally homogeneous,  see \cite{kowalski}  for examples. 

\bigskip

Assume  $\l$, $\mu$, $\nu$ to be arbitrary real numbers. First question: 
Does there exist a  3-dimensional  homogeneous 
 Riemannian manifold whose eigenvalues  of the
 Ricci tensor are just  $\l$, $\mu$ and $\nu$?  Second question: If it exists, is 
it uniquely determined, at least locally up to isometries? 
  The answer to these  questions   does
 not alter if we replace  ``homogeneous" by  ``locally homogeneous". 
In principle,  they  could have been answered by the year 1905
already. But in fact, it was not given  until 1995: the first of  these 
 problems have been solved 
 independently by Kowalski and Nikcevic \cite{kowalskin} 
 on the one hand, and by  Rainer and Schmidt \cite{rainer}, on the other hand; 
the second one is solved several times, e.g. in \cite{rainer}. Curiously enough,
the expected answer ``Yes{}" turns out to be wrong. 

\bigskip

Of course, both  answers \cite{kowalskin} and  \cite{rainer} are equivalent. 
But the formulations are so different, that it is useful to compare them here. 
In \cite{kowalskin} one defines 
\be\label{3.1}
\rho_1 = {\rm max} \{ \l, \mu, \nu      \} \, , \qquad 
\rho_3 = {\rm min} \{ \l, \mu, \nu      \}
\ee
and
\be\label{3.2}
  \rho_2 =  \l + \mu + \nu - \rho_1 - \rho_3 
\ee
 and  formulates the conditions as inequalities between the $\rho_i$.

\bigskip

In  \cite{rainer} the formulation uses curvature invariants.
 First step: one observes that the answer does
 not alter, if one multiplies $\l$, $\mu$, $\nu$ 
 with the same positive constant. Note:  this fails 
to be the case if we take a negative constant. Second step: we take curvature 
invariants which do not alter by this multiplication, i.e.  we look  
for suitable homothetic invariants. Third step: we formulate the answer by 
use of the homothetic invariants.

\bigskip

First step: this can be achieved by a homothetic transformation of the 
metric, i.e., by a conformal transformation with constant conformal 
factor. Then one excludes the trivial case 
$\l = \mu = \nu =0$. Note: It is due to the positive definiteness of the metric
that $\l = \mu = \nu =0$ implies the flatness of the space. 

\bigskip

Second step: 
\be\label{3.3}
R= \l + \mu + \nu \, , \qquad  N=\l^2 + \mu^2 + \nu^2
\ee
 are two curvature invariants  with $N > 0$, and 
\be\label{3.4}
\hat R = R/ \sqrt N
\ee
 is a homothetic invariant. For defining a second
 homothetic invariant we introduce the trace-free part of the Ricci  tensor
\be\label{3.5}
S_{ij}=  R_{ij}- \frac{R}{3} g_{ij} 
\ee
and its invariant 
\be\label{3.6}
S = S^j_i \,  S^k_j \, S^i_k \, .
\ee 
The desired second homothetic  invariant reads 
\be\label{3.7}
\hat S= S/ \sqrt{N^3} \, .
\ee
   Of course, one can also directly define $\hat  R$ and $\hat S$ as functions of
$\l$, $\mu$ and $\nu$, then  $S$ is the cubic polynomial, 
\be\label{3.8}
S = \left( (2\l - \mu - \nu)^3 + (2\mu - \nu - \l)^3 + (2\nu - \l - \nu)^3 \right) / 27 \, ,
\ee
see \cite{rainer} for 
details. Remark: the inequalities $S > 0$  and 
 $\rho_2 < (\rho_1+\rho_3)/2$   are equivalent.

\bigskip

Third step: The answer is ``yes'' if and only 
if one of the following conditions (\ref{3.9}) till (\ref{3.14}) are fulfilled:
\bea\label{3.9}
 N =0 \, , \quad   \mbox{flat space} \   \\
\hat R^2 =3 \quad {\rm  and} \quad  \hat S =0 \, ,  \ 
 \mbox{non-flat spaces of constant curvature} \label{3.10} \  \\
\hat R^2 =2   \quad {\rm  and} \quad  9 \hat R \hat S  = -1 \, , \ 
 \mbox{real  line  times  constant curvature space}  \label{3.11} \  \\
\hat R^2 =1    \qquad {\rm  and} \qquad   9 \hat R \hat S = 2 \label{3.12} \  \\
 18 \hat S > \hat R (9  - 5 \hat R^2) \, ,  \   
 \mbox{equivalent to} \quad  \l \mu \nu >0   \label{3.13} \  \\
  - \sqrt 3< \hat R <- \sqrt 2   \quad {\rm  and }   \quad 
0 < \hat S \le \left( 3 - \hat R^2 \right)^2 \, 
 \frac{6 - 5 \hat  R^2 }{ 18 \hat R^3} \  .  \label{3.14} \ 
\eea
Remarks: 1. In \cite{kowalskin}  the last of the four 
inequalities  (\ref{3.14}) is  more elegantly written as 
\be\label{3.15}
\rho_2 \ge 
\frac{\rho_1^2 + \rho_3^2}{\rho_1 + \rho_3} \, .
\ee
2. There exists a one--parameter set  of Bianchi-type VI manifolds,
 with  parameter $a$ where  $0 <a <1$,
having  $\rho_1 =-a-a^2$, $\rho_2 = -1-a^2$, 
 $\rho_3 = -1-a$; they represent the case where 
in (\ref{3.14})  the fourth  inequality is  fulfilled with ``$=$". 

\bigskip

Now we turn to the second question: Let the 3 numbers $\l, \, \mu, \nu$
 be given such that a homogeneous 3-space exists having  these 3 numbers 
 as its eigenvalues of the Ricci tensor; is this 3-space locally 
 uniquely determined?  The first part of the answer is trivial: 

\bigskip

If $\l = \mu = \nu$, then ``yes{}", it is the space of constant curvature. 
 The second part of the answer is only slightly more involved: If 
$\l, \, \mu$   and  $ \nu$ represent 3 different numbers, then the 3 eigendirections 
of the Ricci tensor are uniquely determined, and then we can also
 prove:  ``yes{}". The third part is really surprising: If 
\be\label{3.16}
\hat R = 1 \qquad {\rm and} \qquad \hat S = \frac{2}{9}\, 
\ee
then the answer is ``no{}", otherwise ``yes{}".

\bigskip

Clearly, eq.   (\ref{3.16}) represents a subcase of   (\ref{3.12}). 
What is the peculiarity with the values (\ref{3.16}) of the
 homothetic invariants? They belong to Bianchi type IX, have one single and one 
double eigenvalue, and with these values, a one-parameter set of 
 examples  with 3-dimensional isometry 
group, and another example with a 4-dimensional isometry group  exist.

\subsection[Why do all the invariants of a gravitational
wave  vanish?]{Why do all the curvature invariants of a gravitational
wave  vanish?}\label{s31a}

We  prove  the  theorem valid  for  Pseudo-Riemannian 
manifolds  $V_n$: ``Let $x \in V_n$ be a fixed point of a  homothetic  
motion  which  is not an isometry then all  curvature  invariants
vanish at $x$." and get the Corollary: ``All curvature invariants
of the plane wave metric 
\be\label{31x}
ds \sp 2 \quad = \quad 2 \, du \, dv \, + \, a\sp 2 (u) \, dw \sp 2 \,
 + \, b\sp 2 (u) \, dz \sp 2 
\ee
 identically vanish."
 
\bigskip

Analysing the proof we see:  The fact that for definite signature
 flatness  can  be characterized by the vanishing of  a  curvature
 invariant,  essentially  rests on the compactness of the rotation
 group $SO(n)$.  For Lorentz signature,  however, one has the
 non-compact Lorentz group $SO(3,1)$ instead of it. 

\bigskip

A  further and independent proof of the corollary uses the  fact,
 that the Geroch limit does not lead to a Hausdorff topology,  so 
a sequence  of gravitational waves can converge to the flat 
 spacetime, even if each element of the sequence is the same
pp-wave.

\bigskip

The   energy   of   the  gravitational   field, especially   of
gravitational waves, 
within  General Relativity was subject of controversies from  the
 very  beginning,   see Einstein \cite{einstein1}.  Global
considerations  - e.g.    by   considering   the   far-field   of
asymptotically  flat  spacetimes - soon  led  to  satisfactory 
answers.  Local  considerations  became fruitful if a  system  of
 reference  is  prescribed  e.g.  by choosing a  timelike 
vector field.  If, however, no system of reference is preferred then it
 is not a priori clear whether one can constructively  distinguish
 flat spacetime from a gravitational wave. This is connected with
 the  generally  known  fact,  that for a pp-wave,  see  e.g. Stephani 
 \cite{stephani} especially section 15.3. 
and \cite{ehlers} all  curvature  invariants  vanish, 
cf.  Hawking and Ellis \cite{hawkingellis} and Jordan et al. \cite{jordan},
  but on the other  hand:  in  the 
absence   of  matter  or  reference  systems   - only   curvature
 invariants are locally constructively  measurable. See also the sentence 
  ``R. Penrose has pointed out, in plane-wave solutions the
scalar polynomials are all zero but the Riemann tensor does not
vanish." taken from  page 260 of   \cite{hawkingellis}.
 At page 97 of \cite{jordan} it is  mentioned that for a pp-wave all
curvature invariants constructed from 
\be
R_{ijkl;i_1 \dots i_r} 
\ee
by products and traces  do vanish.

\bigskip

It  is the aim of this section  to explain the topological origin
of this strange property.

\bigskip

\subsubsection{Preliminaries}\label{tx2}

 Let $V_n$ be a  $C \sp{\infty}$-Pseudo-Riemannian manifold 
of arbitrary signature with dimension  $n>1$.
The metric and the Riemann tensor have components $g_{ij}$ and 
$R_{ijlm}$  resp.  The  covariant derivative  with 
respect  to  the coordinate $x \sp m $ is denoted by  
 ``;$m$" and is performed with the Christoffel affinity 
$ \Gamma \sp i _{lm} $. We define

\bigskip

  $I$  is  called  a  generalized  curvature
invariant of order $k$ if it is a scalar with dependence 
\be
 I = I(g_{ij},  \,  R_{ijlm}, \dots , R_{ijlm;i_1 \dots \, i_k} ) \, .
\ee
By specialization we get the usual 
  Definition:    $I$  is  called  a  curvature 
invariant of order $k$ if it is a  generalized  curvature 
invariant  of  order  $k$ which depends continuously on  all  its
 arguments.  The domain of dependence is requested to contain 
the  flat space, and 
\be
 I(g_{ij}, \, 0, \dots \, 0) \equiv 0 \, . 
\ee
Examples: Let 
\be
 I_0 = {\rm  sign} (\sum \sp n _{i,j,l,m=1} \vert R_{ijlm} \vert ) \, ,
\ee
where the sign function is defined by ${\rm sign}(0)=0 $ and 
${\rm sign}(x)=1 $ for $x>0$. Then  $I_0$  is  a generalized curvature 
invariant of order 0,  but  it  fails to be a curvature invariant.  It 
holds:  $V_n$ is flat    if and only if  $I_0 \equiv 0$. Let further
\be
 I_1 = R_{ijlm} R \sp{ijlm} 
\ee
which  is  a curvature invariant of order 0.  If the  metric  has
 definite signature or if $n=2$ then it holds:   $V_n$ is flat 
if and only if  $I_1 \equiv 0$.
  Proof: For definite signature $I_0 = {\rm  sign} (I_1)$; for $n=2$,
 $I_1 \equiv 0$ implies $R \equiv 0 $, hence flatness. \ q.e.d.

\bigskip

For  all other cases,  however,  the vanishing of $I_1 $ does not
imply flatness. Moreover, there does not exist another curvature 
invariant serving for this purpose, it holds:

\bigskip

 For dimension $n \ge 3$,  arbitrary order   $k$ 
and indefinite metric it holds:  To each curvature invariant  $I$
 of order $k$ there exists a non-flat $V_n$ with $I \equiv 0$.

\bigskip

 Proof: Let $n=3$. We use
\be\label{36x}
ds \sp 2 \quad = \quad 2 \, du \, dv \quad \pm
\quad a\sp 2 (u) \, dw \sp 2
\end{equation}
with a positive non-linear function $a(u)$. The ``$\pm$" covers
the two possible indefinite signatures for $n=3$. The Ricci tensor 
is $R_{ij} = R \sp m _{\quad imj}$ and has with $u = x \sp 1$ 
\be
R_{11} \quad = \quad - \, \frac{1}{a} \cdot
 \frac{d\sp 2 a}{du\sp 2}
\end{equation}
and therefore,  eq.  (\ref{36x}) represents a non-flat metric.  Now let
 $n>3$. We use the Cartesian product of  eq.  (\ref{36x})  with a flat space of
 dimension  $n-3$  and arbitrary signature.  So we have  for  each
 $n\ge 3$ and each indefinite signature an example of a non-flat 
$V_n$.  It  remains  to  show that for all  these  examples,  all
 curvature  invariants of order $k$ vanish.  It suffices to  prove
 that  at the origin of the coordinate system,  because at all 
other  points  it  can  be  shown by  translations  of  all  coordinates
 accompanied by a redefinition of $a(u)$ to $a(u-u_0)$. Let $I$ be
 a curvature invariant of order $k$.  Independent of the dimension, 
 i.e.,  how many flat spaces are multiplied to metric   eq.  (\ref{36x}),   one
 gets for the case considered here that
\be 
I   \,    =   \,   I(a\sp{(0)}(u),   a\sp{(1)}(u),   \dots , \, 
a\sp{(k+2)}(u)) 
\ee
where $a\sp{(0)}(u) = a(u)$, 
$ a\sp{(m+1)}(u)   = \frac{d}{du} a\sp{(m)}(u)$, and
\be
I( a\sp{(0)}(u),  0,  \dots , \,  0) \,  = \, 0 \, .
\ee
This is valid  because each 
\be
R_{ijlm;i_1 \dots \, i_p}
\ee
 continuously depends on 
$a\sp{(0)}(u)$, $ a\sp{(1)}(u)$, $   \dots ,  \, a\sp{(p+2)}(u)$ 
 and  on nothing else;  and for $a = $ const., metric  (\ref{36x})  represents a 
flat space.

\bigskip

Now  we apply a coordinate transformation:  Let $\e  > 0$ be
 fixed, we replace $u$ by $u \cdot \e $ and $v$ by 
$v/  \e $.  This  represents a Lorentz boost  in  the  $u-v-
$plane.  Metric    eq.  (\ref{36x}) 
 remains form-invariant by this  rotation,
 only  $a(u)$  has to be replaced by $a(u \cdot  \e   )$.  At
 $u=0$ we have    
\be 
I   \,    =   \,   I(a\sp{(0)}(0),   a\sp{(1)}(0),   \dots , \,
a\sp{(k+2)}(0)) 
\ee
which must be equal to
\be 
I_{\e } \, = \, I(a\sp{(0)}(0), \e  \cdot
 a\sp{(1)}(0),  \dots , \, \e  \sp{k+2} \cdot
a\sp{(k+2)}(0)) 
\ee
because $I$ is a scalar.  By continuity and by the fact that flat
 space belongs to the domain of dependence of $I$, we have 
\be
  \lim  _{\e  \rightarrow 0 } \,  I_{\e }  =  0 \, .
\ee
  All values  $I_{\e }  $ with $\e  >  0$  coincide,  and  so
$I=0$. \ q.e.d.

\subsubsection{Gravitational waves}\label{tx3}

New results on gravitational waves can be found in [72], and [84]. 
A  pp-wave is a plane-fronted gravitational wave with parallel  rays,
see [77]. It  is a non-flat solution of Einstein's vacuum equation
$R_{ij} = 0$ possessing a non-vanishing covariantly constant vector; this
vector is then automatically a null vector.  The 
simplest  type  of pp-waves can be represented similar as  metric
  eq.  (\ref{36x}) 
\begin{equation}
ds \sp 2 \quad = \quad 2 \, du \, dv \, +
\, a\sp 2 (u) \, dw \sp 2 \, + \, b\sp 2 (u) \, dz \sp 2 \, , 
\end{equation}
 where
\begin{equation}
b \, \cdot \, \frac{ d \sp 2 a}{d u \sp 2} +
a \, \cdot \, \frac{ d \sp 2 b}{d u \sp 2}  \quad = \quad 0 \, . 
\end{equation}
This metric  represents flat spacetime if and only if both $a$ and $b$  are
 linear  functions.   Using the arguments of  subsection \ref{tx2} one sees that
 all curvature invariants this metric  identically vanish. Here
 we present a second proof of that statement which has 
the  advantage to put the problem into a more general  framework
and to increase the class of spacetimes covered, e.g. to the
waves
\be 
ds\sp 2 = dx\sp 2 + dy\sp 2 + 2 du\, dv + H(x,y,u)du\sp 2
\ee
Hall \cite{hall1}  considers fixed points of homothetic motions, which
cannot exist in compact spacetimes,  and shows that any plane
wave, not only vacuum plane waves, admits, for each $x$, a
homothetic vector field which vanishes at $x$.  
 It holds

\bigskip

 Let $x \in V_n$ be a fixed point of a  homothetic 
motion  which  is not an isometry then all  curvature  invariants
 vanish at $x$. 

\bigskip

Proof:   The  existence  of  a homethetic motion which is 
not  an isometry  means  that $V_n$ is selfsimilar.  Let  the  underlying
 differentiable  manifold  be equipped with two  metrics  $g_{ij}$
 and 
\be
 \tilde g_{ij} = e \sp{2C}  g_{ij}
\ee
 where $C$ is  a non-vanishing  constant.  The corresponding Riemannian 
manifolds are denoted  by $V_n$ and  $ \tilde V_n$ resp.  By assumption,  there
 exists  an  isometry from  $ V_n$ to  $ \tilde V_n$  leaving  $x$
 fixed. Let $I$ be a curvature invariant. $I$ can be represented
as continuous  function which vanishes if all the arguments do   of
 finitely  many  of  the  elementary  invariants.  The  elementary
 invariants are such products of factors $g \sp{ij}$ with  factors of  type  
\be 
R_{ijlm;i_1 \dots \,  i_p}
\ee
 which lead to  a  scalar,
i.e.,  all indices are traced out.  Let $J$ be such an elementary
 invariant. By construction we have $J(x) = e \sp{qC} J(x)$ with a
 non-vanishing  natural  $q$,  which depends on the type  of $J$. 
Therefore, $J(x) = 0$. \ q.e.d.

\bigskip

From this it follows:   All  curvature  invariants  of  metric (\ref{31x})
 identically vanish.
  This statement  refers  not only to the 14 independent  elementary
invariants  of  order 0,  see Harvey \cite{harvey} and 
Lake 1993  \cite{lake1979}
 for a list  of  them,  but  for  arbitrary order.

\bigskip
  
 Proof:  We have to show that for each point $x$,  there
exists  a 
homothetic  motion with fixed point $x$ which is not an isometry.
 But  this  is  trivially  done  by  suitable  linear   coordinate
 transformations of $v$, $w$, and $z$. \ q.e.d.

\subsubsection{Topological properties}

Sometimes it is discussed that the properties of spacetime which
can  be  locally and constructively, i.e.,  by rods  and  clocks,
 measured  are not only the curvature invariants but primarily the
 projections of the curvature tensor and its covariant derivatives
 to  an  orthonormal tetrad, called 4-bein.\footnote{The German word 
``das Bein" denotes both ``bone" and ``leg", here it is used in the second sense.}
  The  continuity  presumption 
expresses the fact that a small deformation of spacetime  should
also  lead to a correspondingly small change of the result of the
measurement.  To prevent a preferred system of reference one  can
construct curvature invariants like
\begin{equation}
I_2 \quad = \quad {\rm  inf } \quad \sum _{i,j,l,m} \quad \vert R_{ijlm}
\vert
\end{equation}
where the infimum, here the same as the minimum,  is taken over all orthonormal
tetrads. From the first glance one could believe that $I_2 \, \equiv \, 0$
 if and only if  the space is flat.  But for indefinite signature this  would
contradict the result of subsection \ref{tx2}.  What is the reason?  For
 definite signature the
infimum is to be taken about the rotation group $ SO(4) $, 
or $ O(4) $ if one allows orientation-reversing  systems;  this
 group is compact.  One knows: A positive continuous function over
 a compactum possesses a positive infimum.  So, if one of the 
$ R_{ijlm}$ differs from zero, then  $I_2 > 0$ at that point.
For Lorentz signature,  however, the infimum is to be taken about
 the non-compact Lorentz group $SO(3,1)$ and so $ R_{ijlm} \ne 0$
does not imply $I_2 \ne 0$. 

\bigskip

Another  topological  argument, which underlies our 
 subsection \ref{tx2},   is
 connected  with the Geroch limit of spacetimes \cite{geroch}, 
 we use  the
version of \cite{sch87b}. Theorem 3.1 of  reference  
  \cite{sch87b}  reads:

\bigskip

1:  For  local  Riemannian manifolds  with  definite  signature,
Geroch's  limit defines a Hausdorff topology.  

\bigskip

2:  For indefinite
signature  this  topology  is not  even  $T_1$.  

\bigskip

Explanation: A  topology  is
Hausdorff if each generalized Moore - Smith sequence 
possesses at most one limit, and it is 
$T_1$ if  each constant sequence possesses at most one limit. The
 main example is a sequence, where each element of the sequence is
the  same pp-wave,  and the sequence possesses two  limits:  flat
spacetime  and that pp-wave.
  Here  the reason is:  Only for  definite  signature, 
geodetic  $\e $-balls  form a neighbourhood  basis  for  the
 topology.

\bigskip

 Final remarks:  The change from Euclidean to Lorentzian
signature of a Pseudo-Riemannian space is much more than a
purely algebraic duality - an impression  which is sometimes 
given by writing an imaginary time coordinate: One looses all the
nice properties which follow from the compactness of the rotation
group.   Lake 1993  \cite{lake1979} pointed out that also for
the Robinson-Trautman vacuum solutions of Petrov type III all 14
curvature invariants vanish.

\subsection[Spacetimes which cannot be distinguished by 
invariants]{On spacetimes which cannot be distinguished by curvature 
invariants}\label{s32}

For a positive $C^{\infty}$--function $a(u)$ let 
\be\label{3.17}
ds^2 \ = \ \frac{1}{z^2} \left[ 2 \, du \, dv 
\ - \  a^2(u) \, dy^2 \ - \ dz^2 \right] \, . 
\ee
In the region $z>0$, $ds^2$ represents the anti-de Sitter
spacetime if and only  if $a(u)$ is linear in $u$. Now, let
$d^2a/du^2 \, < \, 0$ and
\be\label{3.18}
\phi \ := \ \frac{1}{\sqrt{\kappa}} \int 
\left( - \frac{1}{a} \, \frac{d^2a}{du^2} \right) ^{1/2}
 \, du \, .
\ee
Then 
\be\label{3.19}
\Box \phi \, = \,  \phi_{,i} \, \phi^{,i} \, = \, 0 
\ee
and 
\be\label{3.20}
R_{ij} \ - \ \frac{R}{2} \, g_{ij} \ = \ \Lambda \, 
g_{ij} \ + \ \kappa \, T_{ij}
\ee
 with  $\Lambda \, = \, - 3$ and 
$ T_{ij} \, = \, \phi_{,i} \, \phi_{,j}$.
So $(ds^2, \, \phi)$ represents a solution of Einstein's 
equation with negative cosmological constant  $\Lambda $ and 
a minimally coupled massless scalar field $\phi $.

\bigskip
 
Let $I$ be a curvature invariant of order $k$, i.e., $I$ is 
a scalar 
\be\label{3.21}
 I = I \left( g_{ij},  \,  R_{ijlm}, \dots , R_{ijlm;i_1
 \dots \, i_k} \right)
\ee
depending continuously on all its arguments. Then for
 the metric $ds^2$, $I$ does not depend on the
 function $a(u)$. 

\bigskip

This seems to be the first example that
 non--isometric spacetimes with non--vanishing curvature scalar
 cannot be distinguished by curvature invariants. The
 proof essentially uses the non--compactness of the
 Lorentz group $SO(3,1)$, here of the 
boosts $u \rightarrow u \, \lambda $, $v \rightarrow
  v/\lambda$. One can see this also in the
 representation theory in comparison with the
 representations of the compact rotation group $SO(4)$. 

\bigskip
 
Let $v \in R^4$ be a vector and $g \in SO(4)$ such that 
$g(v) \uparrow \uparrow v$, then $g(v)=v$. In Minkowski 
spacetime $M^4$, however, there exist vectors $v \in M^4$
 and a $h \in SO(3,1)$ with $h(v)  \uparrow \uparrow v$ and
$h(v) \ne v$. This is the reason why the null frame 
components of the curvature tensor called Cartan ``scalars"
 are not always curvature scalars, see \cite{sch96b}.
 Further new results on curvature invariants are given in [28], [184] and [188]. 
 Limits of spacetimes are considered in [176] and [230].

\newpage

\section{Surface layers  and relativistic  surface tensions}\label{Kap4}
\setcounter{equation}{0}

For a thin shell, the intrinsic 3-pressure will be shown to be 
analogous  to $ -A$, where $A$ is the classical 
 surface tension: First, interior and exterior Schwarzschild
 solutions will be matched together such that the surface 
layer generated at the common boundary has no gravitational 
mass; then its intrinsic 3-pressure represents a surface tension 
fulfilling Kelvin's relation between mean curvature and pressure 
difference in the Newtonian limit. Second, after a suitable definition 
of mean curvature, the general relativistic analogue  to Kelvin's relation 
will be proven to be contained in the equation of motion of the surface layer.

\subsection{Introduction to surface layers} 

In general relativity, an energy-momentum tensor concentrated
 on a timelike hypersurface is called a surface layer. Via Einstein's 
equations it is related to non-spurious jumps of the 
Christoffel affinities or equivalently to jumps of the second 
fundamental tensor. In \cite{lanczos},  \cite{papatreder}, 
  \cite{dautcourt1963},  \cite{israel},  \cite{papa} and   \cite{kuchar} 
there have been given algorithms 
for their calculating, and in \cite{lanczos},   \cite{israel}, and 
\cite{boulware}, \cite{denisov},    \cite{sch83},
\cite{urbandtke},  \cite{frehland} and  \cite{lake1979}
 the spherically   symmetric case was of a special interest. The discussion of 
disklike layers as models for accretion disks was initiated in 
 \cite{morgan} and \cite{voorhees}, and \cite{lopes}
 contains surface layers  as sources of the Kerr geometry.

\bigskip
 
A surface layer has no component in the normal direction, 
otherwise the delta-like character would be destroyed. Hence, an 
ideal fluid with non-vanishing pressure cannot be concentrated on an 
arbitrarily thin region and therefore in most cases, the layer is 
considered to be composed of dust. Additionally, an intrinsic 
3-pressure is taken into account in \cite{einstein1939},   where for
 a spherically symmetric configuration each shell 
$r =$ const.  is thought to be composed of  identical particles
 moving on circular orbits without a preferred direction, and the 
tangential pressure, which is analogous to the  intrinsic 3-pressure, is due to particle
 collisions, whereas a radial pressure
does not appear. See also  \cite{sch83}, \cite{lake1979},
 and \cite{maeda},  where the equation of
 motion for a spherically symmetric layer has been discussed. 

\bigskip
 
Such an intrinsic 3-pressure as  well as surface 
tensions are both of the physical dimension
``force per unit length." There 
the question arises whether an intrinsic 3-pressure of a surface
layer may be related to a surface tension, 
and this chapter will deal  with just this question. Then
a general relativistic formulation of  thermodynamics can be 
completed by equations for surface tensions to answer,
e.g., the question how long 
a drop, say, a liquid comet, remains connected while
 falling  towards a compact object. The only paper 
concerned with such questions seems to be  \cite{krasilnikov}.
 There  the influence of surface tensions on the propagation 
of gravitational waves has been calculated
by perturbation methods, yielding a possibly measurable effect.
For the background knowledge  cf. \cite{israelstewart} or \cite{neugebauer}
 for relativistic thermodynamics 
 and \cite{ono} for non-relativistic surface tensions.  

\bigskip
 
This chapter, which is based on \cite{sch84a},   proceeds as follows: 
Section \ref{s42} contains Kelvin's relation for non-relativistic surface
tensions and  discusses the 
matching of the interior to the exterior Schwarzschild solution
such that a surface tension appears,  and compares with
Kelvin's relation in the Newtonian limit,  calculations are 
found in the appendices, sections
  \ref{s46},  \ref{s47} and  \ref{s48}. 
Sections  \ref{s44} and  \ref{s45}  are devoted to the non-spherically
symmetric case. Section  \ref{s44}  contains 
a suitable definition of mean curvature and section  \ref{s45}
 deduces Kelvin's relation  from the 
equation of motion of the surface layer without any weak  field assumptions.

\subsection{Non-relativistic surface tensions}\label{s42}

Imagine a drop of some liquid moving in vacuo. 
Its equilibrium configuration is a spherical
one, and Kelvin's relation \cite{thomson} between surface tension $A$, 
pressure difference $\Delta P$, i.e. outer minus inner
pressure, and 
mean curvature $H$, where $H = 1/R$  for a sphere of radius $R$, 
 reads
\be\label{4.1}
     \Delta P = -2H A \, .
\ee
$A$  is a material-dependent constant. 
Equation  (\ref{4.1}) means, an energy $A \cdot \Delta F$
 is needed to increase the surface area by $\Delta F$.
This supports our description of $-A$  as a kind of intrinsic pressure. 
But of course, it is a quite 
different physical process: Pressure, say, of an ideal gas, can be explained
 by collisions of freely 
moving particles, whereas the microphysical explanation of surface
tensions requires the determination 
of the intermolecular potential, which looks like
\be
\Phi(r) = - \mu r^{-6}  +N e^{-r/\rho}
\ee 
with certain constants $\mu$, $N$, 
and $\rho$; see Ono \cite{ono}. 
In this context the surface has a thickness of about
$10^{-7}$ cm but in most cases this thickness may be neglected. 

\bigskip

In addition, for a non-spherically symmetric 
surface, the mean curvature $H$  in equation eq. (\ref{4.1})  may
be obtained from the principal curvature 
radii   $R_1$, $R_2$  by means of the relation
\be\label{4.2}
H = \frac{1}{2R_1}  + \frac{1}{2R_2}   \, .
\ee

\bigskip

Now the drop shall be composed of 
an incompressible liquid. For a general relativistic
description we have to take the interior Schwarzschild  solution
\bea\label{4.3}
ds^2 = -   \left[ \frac{3}{2}(1-r_g/r_0)^{1/2}
-  \frac{1}{2}  ( 1-r_gr^2/r_0^3) ^{1/2}
\right]^2 
dt^2    \nonumber \\
+ \,   \frac{dr^2}{1-r_gr^2/r_0^3}
+  r^2   d\Omega^2    \qquad {\rm where } \qquad 
d\Omega^2 = d\psi^2 + \sin^2 \psi d \varphi^2 \, ,  
\eea
whose energy-momentum tensor 
represents ideal fluid with energy density $\mu = 
 3 r_g / \kappa  r_0^3$  and  pressure
\be\label{4.4}
p(r) = \mu \cdot 
\frac{   ( 1-r_gr^2/r_0^3) ^{1/2} - 
(1-r_g/r_0)^{1/2} }{3 (1-r_g/r_0)^{1/2} -
   ( 1-r_gr^2/r_0^3) ^{1/2}} \, .
\ee
One has $p(r_0)= 0$, and therefore 
usually $0 \le r \le r_0$  is considered. But now we require only a
non-vanishing pressure at the inner 
surface and take  eq. (\ref{4.3}) for values $r$ with $0 \le r \le R$ 
 and a fixed $R <r_0$ only. 

\bigskip
 
The gravitational mass  of the inner region equals
\be\label{4.5}
     M= \mu \cdot 4 \pi  R^3/3 = r_g(R/r_0)^3/2 
\ee
where we assumed unit such that  $G= c = 1$. 
The outer region shall be empty and
therefore we have to  insert the Schwarzschild
solution for $r \ge R$. Neglecting the gravitational mass 
 of the boundary $ \Sigma \subset  V_4$  which is defined
by $r - R =  0$,  just $M$ of equation   (\ref{4.5}) has 
to be  used as the mass parameter of the exterior
Schwarzschild metric. Then only delta-like tensions 
appear at $\Sigma $,  and $\Delta P = - p(R) <0$. 
The  vanishing of the  gravitational mass 
is required to single out the properties of an intrinsic
3-pressure. In general, a 
surface layer is  composed of  both parts, not at least to ensure the
 validity of the energy  condition $T_{00} \ge \vert T_{ik} \vert $
 which holds for all known types of matter.

\bigskip
 
On  $\Sigma $, the energy-momentum tensor is
\be\label{4.6}
T_i^k = \tau_i^k \cdot \delta_\Sigma 
\ee
the  non-vanishing components of which are
\be\label{4.7}
\tau_2^2 = \tau_3^3 =-p(R)R/2(1 - 2M/R)^{1/2}
\ee
cf. section \ref{s46}.

\bigskip
 
The $ \delta_\Sigma $ distribution is defined such 
that for all smooth scalar functions $f$  the invariant integrals
 are related by 
\be
\int_\Sigma \ 
f \ d^{(3)} x \qquad =  \qquad 
\int_{V_4} \ 
f \cdot     \delta_\Sigma \,    d^{(4)} x \, .
\ee
 For a more detailed discussion of distribution-valued tensors
 in curved spacetime, cf. \cite{taub}. The fact that all components $\tau_0^\alpha$
 in  eq. (\ref{4.6}) vanish reflects the  non-existence of a delta-like
gravitational mass on $ \Sigma $. 

\bigskip
 
Now consider the Newtonian  limit $M/R \ll 1$; then 
the mean curvature becomes again $H= 1/R$,
and together with  eq. (\ref{4.7}) Kelvin's relation  eq. (\ref{4.1}) is just 
equivalent to
\be\label{4.8}
\tau_2^2 = \tau_3^3 = - A [ 1 + O(M/R) ] \, .
\ee
Therefore: At least for static spherically symmetric 
configurations and weak fields a delta-like
negative tangential pressure coincides 
with the classical surface tension.

\bigskip

In the next two sections we investigate 
to what extent these presumptions are necessary.

\subsection{Mean curvature in curved spacetime}\label{s44}

To obtain a general relativistic analogue   to Kelvin's
 relation  eq. (\ref{4.1}) we have to define the mean
curvature $H$ of the timelike hypersurface 
$\Sigma$ contained in a spacetime $V_4$ such that for weak
fields just the usual mean curvature arises.

\bigskip
 
To get the configuration  we have in mind, we make the ansatz
\be\label{4.9}
 \tau_{\alpha \beta} = - A ( g_{\alpha \beta}
 + u_\alpha u_\beta ) \, \qquad u_\alpha u^\alpha = -1
\ee
with $A >0$.  Thereby again the gravitational 
mass of $ \Sigma$  will be neglected. Now the mean curvature
shall be defined. But there is a problem: 
In general, there does not exist a surface $ S \subset 
\Sigma$  which can
serve as ``boundary at a fixed moment" for which 
we are  to determine the mean curvature. To
circumvent this problem we start considering the special case
\be\label{4.10}
    u_{\alpha \| \beta}  =       u_{\beta \|  \alpha }   \,     .
\ee
Then there exists a scalar $t$ on $\Sigma$  such that 
$u_\alpha  = t_{\| \alpha}$, 
and the surface $S \subset  \Sigma$  defined by $t =0$
 may be called ``boundary at a fixed moment." 
Now $S$ has to be embedded into a ``space at a fixed
moment'': We take intervals of geodesics starting 
from  points of $S$  in the normal direction $n_i$  and
the opposite one. 

\bigskip
 
The union $V_3$ of these geodetic 
segments will be called ``space at a fixed moment," and $S \subset V_3$
 is simply a two-surface in a three-dimensional positive 
definite Riemannian manifold, for which mean curvature has a definite sense: 
Let $v^\alpha$, $w^\alpha$  be the principal curvature directions inside $S$  and
 $R_1$, $R_2$ the corresponding principal 
curvature radii, then equation   (\ref{4.2}) applies to obtain $H$. 

\bigskip
 
Of course,   $v^\alpha   w_\alpha =0 $  holds, 
and $ v^\alpha   v_\alpha
=   w^\alpha   w_\alpha  = 1$ shall be attained. Then, inserting the second
fundamental tensor cf. section \ref{s47}, this becomes equivalent to
\be\label{4.11}
H = \frac{1}{2} (v^\alpha   v^\alpha
+   w^\alpha   w^ \alpha    )  k_{\alpha \beta }
= \frac{1}{2} (g^{\alpha \beta} + u^\alpha u^\beta)
k_{\alpha \beta }\, . 
\ee
But this latter relation makes sense without 
any reference to condition  eq. (\ref{4.10}). Therefore, we define
 eq. (\ref{4.11}) to  be the general relativistic 
analogue  to the mean curvature of a surface. For the case   
$ H^+ \ne H^- $
we take their arithmetic mean 
\be
 H = (    H^+ + H^- )/2   \,   ,
\ee
 and  then we obtain from  eqs. (\ref{4.9})  and   (\ref{4.11})
\be\label{4.12}
     -2HA = \frac{1}{2} \left(
k_{\alpha \beta }^+ + k_{\alpha \beta }^- \right)
 \tau^{\alpha \beta }     \, . 
\ee
This   choice can  be accepted noting that  
in the Newtonian limit 
\be
 \vert H^+ - H^- \vert \ll  \vert H^+ + H^- \vert  
\ee
anyhow.   To compare this with Kelvin's relation 
we have to relate the right-hand side of equation   (\ref{4.12}) to
the pressure difference $\Delta P$ at $\Sigma$. 
To this end  we investigate the equation of motion for the surface
layer.

\subsection{Equation of motion for the surface layer}\label{s45}

The equation of  motion, $T^k_{i;k}=  0$, contains products 
of $\delta $ distributions and $\theta$-step functions, where  
$\theta(x) =1 $ for $x \ge 0$, and $\theta(x) =0 $ else, 
  at points  where $ \Gamma^i_{jk}$
  has a  jump discontinuity. These products  require special care; 
cf. \cite{cohen} for a discussion of  his point. 
But defining   $ \theta \cdot \delta = \frac{1}{2} \delta $
 we obtain,  cf.  section \ref{s48}
\bea\label{4.13}
\Delta P \equiv \Delta n_i n^k T^i_k =  \frac{1}{2} \left(
k_{\alpha \beta }^+ + k_{\alpha \beta }^- \right)
 \tau^{\alpha \beta }
 \qquad {\rm   and }  \\
  \Delta T^1_\alpha \equiv \Delta n_i e^k_\alpha 
 T^i_k =  - \tau^\beta_{\alpha \| \beta} \, . \label{4.14}
\eea
From equation   (\ref{4.14}) we  see  the following: 
The equation $ \tau^\beta_{\alpha \| \beta} =0$ 
holds only under the additional presumption that the regular, 
i.e., not delta-like,  part of $ T^1_\alpha$ 
has no jump on $\Sigma$. 

\bigskip

This  condition is fulfilled e.g.,  presuming $\Sigma$
 to be such a boundary that the regular energy  flow does
not cross it and the four-velocity is parallel to $\Sigma$
  in both $V_+$  and $V_-$. This we will presume in the
following. Then $\Delta P$   is indeed 
the difference of the pressures on both sides, and together with
equations   (\ref{4.12}) and   (\ref{4.13}) 
we obtain exactly Kelvin's  relation  eq. (\ref{4.1}). That means, it is the definition
of mean curvature used here that enables 
 us to generalize Kelvin's formula to  general relativity.
Furthermore, $O(M/R)$  of equation   (\ref{4.8}) vanishes.

\bigskip

Finally we want to discuss 
the equation $ \tau^\beta_{\alpha \| \beta} =0$. 
Transvection with $u^\alpha$  and 
$ \delta^\alpha_\gamma + u^\alpha u_\gamma $
 yields
\be\label{4.15}
 u^\alpha_{\| \alpha} =0  \qquad {\rm and} \qquad
 u^\alpha u_{\gamma \| \alpha} + (\ln A)_{\| \alpha}
( \delta^\alpha_\gamma + u^\alpha u_\gamma ) =0
\ee
respectively. But $A$ is a constant here, and therefore $u^\alpha$
is an expansion-free  geodesic vector
field in $\Sigma$. {But observe that the  $u^\alpha$ 
lines are geodesics in $V_4$  under additional presumptions only. 

\bigskip

Here, we have only considered a phenomenological theory of 
surface tensions, and, of course, a more detailed theory has to include 
intermolecular forces. But on that phenomenological level
equations   (\ref{4.9}) and   (\ref{4.15}) together with 
Kelvin's relation  eq. (\ref{4.1}), which has been shown to follow from
the equation of motion,  and 
$A =$ const as a solely temperature-dependent equation of state 
complete the usual general relativistic 
Cauchy problem for a thermodynamical system by including
surface tensions.

\subsection{Notation, distributions and mean 
curvature}\label{s46}\label{s47}\label{s48}

To make the equations better   readable, some 
conventions and formulas shall be given.  Let $\xi^\alpha$,  
 $\alpha  =  0,2,3$, be coordinates in $\Sigma $
and $x^i$, $i=0,1,2,3$, those for $V_4$. 
The embedding $\Sigma \subset V_4$  is performed by
functions $x^i(\xi^\alpha)$   whose derivatives
\be\label{4.17}
e^i_\alpha = \partial x^i/\partial \xi^\alpha \equiv x^i_{, \alpha}
\ee
form a triad field in $\Sigma$.  $\Sigma$ 
  divides,  at least  locally, $V_4$  into two connected  components, 
$V_+$  and $V_-$, and the normal $n_i$, defined by
\be\label{4.18}
 n_i n^i =1\, , \qquad n_i e^i_\alpha =0 
\ee
is chosen into the $V_+$ direction, 
which can be thought being the outer region.  Possibly $V_+$
  and $V_-$  are endowed with different 
coordinates $x^i_+$, $x^i_-$  and metrics
 $g_{ik^+}$  and  $g_{ik^-}$, respectively. For
this case all subsequent formulas 
had to be indexed with $+/-$, and only the inner 
metric of $\Sigma$, its first fundamental tensor
\be\label{4.19}
g_{\alpha \beta} = e^i_\alpha e^k_\beta \  g_{ik}
\ee
has to be the same in both cases.  As usual, we require $g_{ik}$ 
to be $C^2$-differentiable except for
jumps of $g_{ij,k}$  at $\Sigma$. Covariant derivatives 
within $V_4$  and $\Sigma$  will be denoted by $ \  ; \ $  and 
 $\  \|  \ $  respectively.

\bigskip

The second fundamental tensor  $k^\pm_{\alpha \beta}$
 on both sides of $\Sigma$  is defined by
\be\label{4.20}
k^\pm_{\alpha \beta}=
\left( e^i_\alpha e^k_\beta n_{i;k}     \right)^\pm
= \left( e^i_{\alpha;k} e^k_\beta n_{i}     \right)^\pm
\ee
and the difference,  $ \Delta k_{\alpha \beta} = k^+{\alpha \beta} -
k^-_{\alpha \beta}$,  $\Delta k = g^{\alpha \beta } \Delta k_{\alpha \beta}$, 
enters the energy-momentum tensor via
equation    (\ref{4.6})  and the relation
\be\label{4.21}
\tau^{ik} \equiv e^i_\alpha e^k_\beta 
\tau^{\alpha \beta} \, , 
 \qquad {\rm  where} \qquad  \kappa \tau_{\alpha \beta} = 
g_{\alpha \beta}\Delta  k - \Delta   k_{\alpha \beta}
\ee
cf. e.g., \cite{israel}. 
From equation    (\ref{4.18}) and the  Lanczos equation    (\ref{4.21})  one obtains 
$n^i\tau_{ik} = 0$, i.e., indeed the absence of a delta-like
 energy flow in the normal direction.

\bigskip

Now take a special coordinate 
system: $x^\alpha  = \xi^\alpha$, and the 
$x^1$ lines are geodesics starting from $\Sigma$ 
into $n_i$ direction with natural parameter $x^1$. 
Then the  line element of $V_4$ reads
\be\label{4.22}
ds^2 =  - \left( dx^1 \right)^2 + g_{\alpha \beta}
 dx^\alpha dx^\beta
\ee
and the only jumps of  $\Gamma^i_{jk}$
 are
$$
\Gamma^{\pm 1}_{\alpha \beta} = - k^{\pm 1}_{\alpha \beta}
= - \frac{1}{2} g^{\pm }_{\alpha \beta , 1}  \, .
$$
The most natural  definition of $\theta \cdot \delta$  is
 $\frac{1}{2} \delta$
 being equivalent to the choice 
\be\label{4.23}
\Gamma^i_{jk} = \frac{1}{2} \left( 
 \Gamma^{+i}_{jk} +  \Gamma^{-i}_{jk}\right)
   \qquad {\rm on} \qquad 
\Sigma \, .
\ee
But cf. Dautcourt 1963 
\cite{dautcourt1963} for another choice of  $\Gamma^i_{jk}$   with
 the consequence that 
$T^k_{i;k} \ne 0 $ at $\Sigma$.

\bigskip
 
 Now the $\delta$ part
 of the equation $T^k_{1;k} =  0 $  reads
\be\label{4.24} 
\Delta T^1_1 \equiv T^1_{+1} - T^1_{-1} =
  \frac{1}{2} \left(
k_{\alpha \beta }^+ + k_{\alpha \beta }^- \right)
 \tau^{\alpha \beta } \, .
\ee
Analogously one obtains for the other
components
\be\label{4.25}
\Delta T^1_\alpha = - \tau^\beta_{\alpha \| \beta}     \, .
\ee
Reintroducing the original 
 coordinate system, the left-hand sides of equations 
   (\ref{4.24})  and     (\ref{4.25}) have to be replaced by 
$ \Delta P_1 = \Delta n_i n^k T^i_k$ 
 and 
$\Delta P_\alpha = \Delta n_i e^k_\alpha T^i_k   $, 
respectively; cf. \cite{maeda}.
 Thereby, $\Delta P_i$  
 is the difference of the energy flows on both  sides 
 of  $\Sigma$.

\bigskip

To deduce equation  eq. (\ref{4.7}) we take 
proper time $\xi^0$  and angular coordinates 
 $\psi = \xi^2$  and $\varphi  = \xi^3$.
Then, inside $\Sigma$,
\be
ds^2 = - \left( d \xi^0 \right)^2  + R^2 d\Omega^2    \, .
\ee
Using equation    (\ref{4.20}) and the exterior Schwarzschild
 solution one obtains 
$k_{\alpha \beta }^+ $
 the non-vanishing components of which are
\be
k_{00}^+   = - M/R^2(l - 2M/R)^{1/2} \, , \qquad
   k_{22}^+ =R(1 - 2M/R)^{1/2} 
\ee
and
\be
k_{33}^+ =  k_{22}^+    \sin^2 \psi
\ee
because of spherical symmetry. To 
avoid long calculations with the metric (\ref{4.3}) one can proceed
as follows. By construction, $\tau_{00} =0$,  and 
together with equation    (\ref{4.21})   and the spherical
symmetry    
$  k_{22}^- =  k_{22}^+$, $k_{33}^- = k_{33}^+$
follows. 
From equation    (\ref{4.24}), equation  (\ref{4.7})  follows then
immediately without the necessity 
of determining the actual value  of $  k_{00}^-  $.

\bigskip

And to be independent of the discussions 
connected with equation     (\ref{4.23}), we deduce equation    (\ref{4.24}) 
another way. First, independent of  surface 
layers, for an arbitrary timelike hypersurface and a
coordinate system such that     eq. (\ref{4.22})  holds, we have
\be\label{4.16}    
\kappa  T_{11} = \frac{1}{2}
\left(
{}^{(3)}R + k^2 -k_{\alpha \beta }k^{\alpha \beta }
\right)
\ee
where $\  {}^{(3)}R$  is the curvature scalar 
within that surface. Now turn to a surface layer with 
$ k^+_{\alpha \beta } \ne  k^-_{\alpha \beta } $. 
Then equation     (\ref{4.16})  splits into a ``$+$''  and a ``$-$''  equation,
having in common solely   $\  {}^{(3)}R$. 
Inserting all this into equation    (\ref{4.21}), one obtains
$$
  \frac{1}{2} \left(
k_{\alpha \beta }^+ + k_{\alpha \beta }^- \right)
 \tau^{\alpha \beta } =
 \frac{1}{2 \kappa }  \left(
k_{\alpha \beta }^+ + k_{\alpha \beta }^- \right)
 \left(
g^{\alpha \beta } \Delta k - \Delta  k^{\alpha \beta }
\right)
$$
\be
= \frac{1}{2 \kappa } \left[ (k^+)^2 - (k^-)^2 
- k_{\alpha \beta }^+ k^{+ \alpha \beta }
+ k_{\alpha \beta }^-  k^{- \alpha \beta } 
\right] = T^+_{11} - T^-_{11}
\ee
i.e., just equation     (\ref{4.24}).

\newpage

\section[The massive scalar field in a 
 Friedmann universe]{The massive scalar field in a closed
 Friedmann universe}\label{Kap5}
\setcounter{equation}{0}
\setcounter{page}{57}

For the minimally coupled scalar field  in Einstein's theory of gravitation we 
look for the space of solutions within  the class of closed Friedmann universe models.
 We prove $D \ge 1$, where $D$ is the dimension 
of the set of solutions which can be integrated up to 
 $t \to \infty$.   $D > 0$ was conjectured by Page in 1984 \cite{page1984}. 
We discuss concepts like ``the probability of the 
appearance of a sufficiently long inflationary phase" 
and argue that it is primarily a probability measure $\mu$
 in the space $V$ of solutions and not in the space of 
initial conditions,  which has to be  applied. The measure 
$\mu$ is naturally defined for Bianchi-type I
 cosmological models because $V$  is a compact cube. 
The problems with the closed Friedmann model,  
which led to controversial claims in the literature, 
will be shown to originate from the fact that $V$ has a 
complicated non-compact non-Hausdorff Geroch topology: 
no natural definition of $\mu$ can  be  given.
 We conclude: the present state of our universe 
can be explained by models of the type discussed, 
but thereby the anthropic principle cannot be fully circumvented.
We consider a closed Friedmann cosmological model,
\be\label{5.1}
ds^2 = 
g_{ij} dx^i dx^j =
dt^2 - a^2(t) [dr^2 + \sin^2 r(d \psi^2  + \sin^2 \psi d\chi^2)] \, ,
\ee
with the cosmic scale factor $a(t)$. We 
apply Einstein's General Relativity Theory 
 and take a minimally coupled scalar field $\phi$ without self-interaction 
as source, i.e., the Lagrangian is
\be\label{5.2}
{\cal L} = \frac{R}{16\pi G} + \frac{1}{2} g^{ij}
\nabla_i \phi  \nabla_j \phi - \frac{1}{2} m^2 \phi^2 \, .
\ee
In this chapter we assume units such that $\hbar  = c =1$; 
$R$ is the scalar curvature, $G$ Newton's constant, and $m$ the mass 
of the scalar field. The variational derivative 
$\delta {\cal L}/\delta \phi=0$ yields the field equation for the scalar field
\be\label{5.3}
     (m^2+ \Box) \phi = 0\,  ,   \qquad \Box \equiv  g^{ij}
\nabla_i \nabla_j
\ee
where $\Box$ denotes  the covariant D'Alembertian. The variation
 of the Lagrangian  with respect to the metric 
\be
\delta {\cal L}\sqrt{ - {\rm det} g_{kl}}  /\delta g^{ij}=0
\ee
yields Einstein's field  equation
\be\label{5.4}
R_{ij } - \frac{R}{2} g_{ij} = 8 \pi G T_{ij} \, , \quad
 T_{ij} \equiv \nabla_i \phi \nabla_j \phi
- \frac{1}{2} \left(\nabla^k \phi \nabla_k \phi - m^2 \phi^2
\right) g_{ij}\, , 
\ee
with Ricci tensor $R_{ij}$. It is the aim of this chapter, which is based on
 \cite{sch90a},   to present  some  rigorous results about the space $V$ of solutions 
of eqs. (\ref{5.3}),     (\ref{5.4})  with metric   (\ref{5.1}). 
This is done in   sections \ref{s52},   \ref{s53} 
and  \ref{s54}. We  discuss them in 
the context of cosmology in  section \ref{s55}: the probability  of  the appearance 
of a sufficiently long inflationary phase and the anthropic principle.

\bigskip

Schr\"odinger  \cite{schr1938}  
already dealt with the massive scalar field in the closed Friedmann 
universe in   1938, but there the back-reaction of the scalar field on 
the evolution of the cosmic scale factor had been neglected. 
35 years later the model enjoyed a renewed interest, especially as a 
semi-classical description of quantum effects, cf. e.g. 
 Fulling and Parker  \cite{fulling},  \cite{fullingparker}, 
Starobinsky  \cite{staro78}, Barrow and Matzner 
 \cite{barrowmatzner},  Gottl\"ober  \cite{gott1984a}  and Hawking and 
Luttrell  \cite{hawkingluttrell}. One intriguing 
property - the possibility of a bounce, i.e., a positive local minimum of the 
cosmic scale factor - made it interesting in connection with a possible 
avoidance of a big bang singularity
 $ a(t) \to   0$. The existence of periodic solutions $(\phi(t), a(t))$  of 
 eqs. (\ref{5.1}),  (\ref{5.3}) and  (\ref{5.4}) 
  became  clear in 1984  by  independent
work of  Hawking   \cite{hawkinga}  and Gottl\"ober
 and Schmidt  \cite{gottschmidt}. 
 The existence of an inflationary phase, which is  
 defined by 
\be
\vert dh/dt \vert \ll  h^2 \, , \qquad {\rm  where} \qquad  h = d (\ln a)/dt
\ee 
   is the Hubble parameter,  in the cosmic evolution is discussed 
in many papers to that model, cf  e.g. 
Belinsky et al.  \cite{belinskyb},  \cite{belinsky}, 
Gottl\"ober and M\"uller  \cite{gottmuell}  and   Page 1987  \cite{page1984}.
Soon it became clear that the satisfactory results obtained for 
the spatially flat model --   the existence of a naturally defined 
measure in the space of solutions and with this measure the
very large probability to have sufficient inflation -- cannot 
be generalized to the closed model easily. We shall turn to that point in 
section \ref{s55}.

\subsection{Some closed-form approximations}\label{s52}

We consider the system   (\ref{5.3}),   (\ref{5.4})   with
 metric  (\ref{5.1}). Sometimes, one takes it as an additional 
assumption that $\phi$ depends on the coordinate $t$ only, 
but it holds, cf. Turner  \cite{turner}: the spatial homogeneity 
of metric    (\ref{5.1})  already implies this property.
 Proof:  For $i \ne j$ we have $R_{ij} =g_{ij} = 0$ and, therefore we get 
 $ \nabla_i \phi \nabla_j \phi= 0$. This means, 
locally  $\phi$ depends on one coordinate $x^i$  only. Supposed $i  \ne  0$, 
then $g^{ij}T_{ij}    - 2T_{00} = m^2 \phi^2$.
      The l.h.s. depends on $t$ only. For $m \ne 0$ this is 
already a contradiction. For $m = 0$  we
have  $R_{ij} = \nabla_i \phi \nabla_j \phi$, 
which is a contradiction to the spatial isotropy of $R_{ij}$, q.e.d.

\bigskip

We always require $a(t) > 0$, for otherwise the 
metric   (\ref{5.1})  is degenerated, leading to a  ``big bang".
 Inserting metric   (\ref{5.1}),  eqs.   (\ref{5.3}),  (\ref{5.4})   reduce to
\be\label{5.5}
 d^2\phi/dt^2 + 3hd\phi/dt + m^2 \phi = 0
\ee
and
\be\label{5.6}
     3\left( h^2 + a^{-2}\right) = 4\pi G \, [m^2\phi^2 + (d\phi/dt)^2] \,  . 
\ee
Eq.   (\ref{5.6})  is the $00$-component of  eq. (\ref{5.4}), 
the other components are  a consequence of it.  To get the analogous equations 
for the spatially flat Friedmann model, the l.h.s. of eq. (\ref{5.6}) has
to be replaced by $3 h^2$. 

\bigskip 

For the massless case   $m= 0$ representing  stiff matter, i.e., pressure equals 
energy density, eqs.   (\ref{5.5}), 
 (\ref{5.6})  can be integrated in closed form. First, for the spatially flat model, we get 
solution   eq. (\ref{2.4}) with $\tau = 1/3$, i.e. $a(t) = t^{1/3}$; 
this means $\a =1 $ in    eq. (\ref{2.5}). 
 Second, for the closed model, let $\psi  = d\phi/dt$, 
then eq.   (\ref{5.5}) implies $\psi a^3 =$ const. Inserting 
this into  eq. (\ref{5.6}), we get 
\be
da/dt = \pm\sqrt{L^4/a^4 -1}
\ee
 with a constant $L > 0$ and
$    0 < a \le L$. We get $a(t)$ via the inverted function up to a $ t$ 
translation from the following equation
\be\label{5.6b}
    \pm  t = \frac{1}{2}L \arcsin (a/L)
          - \frac{a}{2}   \sqrt{1 - a^2/L^2} \, .
\ee
 The minimally coupled  massless scalar field in Einstein's theory 
 is conformally  equivalent to the theory of gravity following from the 
Lagrangian  $L = R^2$, cf. Bicknell \cite{bicknell}, or  \cite{starosch}
 and  \cite{barrowzia}. Details of this Bicknell theorem are presented in chapter
 \ref{Kap7}, especially section \ref{s75}. 

\bigskip

For $a \ll L$ we have from eq. (\ref{5.6b}) $a \sim  t^{1/3}$  and 
 $\phi = \phi_0 + \phi_1 \ln t $. The limit $L \to \infty$ is, only locally of course, 
the limit from the closed to the spatially  flat Friedmann model. Thus we recover,
as it must be the case, $\tau = 1/3$ given above. 
 Eq. (\ref{5.6b}) shows that  the function $a(t)$ has 
  a maximum at  $a = L$. $a(t) > 0$ is fulfilled for a 
$t$-interval of length $\Delta t = \pi L/2$ only, and there is no bounce.
 In which range can one except the massless case to be a 
good approximation for the massive case? To this end we perform 
the following substitutions:
\be
   \tilde  t=t/\e\, , \quad \tilde a=a/\e \, , \quad
\tilde m=m \e \,  , \quad \tilde \phi =\phi \,  .
\ee
They do not change the differential equations. Therefore, to get a 
solution $a(t)$ with a maximum 
$a_{\rm max} = \e \ll 1$ for $m$ fixed, we can transform 
to a solution with  $\tilde  a_{\rm max}= 1$, 
 $\tilde m = \e m \ll m$  and apply solutions  
$\tilde a \left( \tilde t \right)$
 with $\tilde m = 0$ as a good approximation. Then $a(t) > 0$  is fulfilled 
for a $t$-interval of length $\Delta t \approx \pi     a_{\rm max}/2$
 only. This is in quite 
good agreement with the estimate  in eq. (22) of Page 1984 \cite{page1984}. 
But in the contrary to the massless  case it holds: to each $\e > 0$
 there  exist bouncing solutions which possess a local maximum 
$a_{\rm max} = \e$.

\bigskip

Let henceforth be $m > 0$. Then we use   
 $ \sqrt{4\pi G/3} \, \,  \cdot \, \phi$
instead of $\phi$ as scalar 
field and take units such $m = 1$. With a dot denoting 
$d/dt$ we finally get from eqs.   (\ref{5.5}),   (\ref{5.6}) 
\be\label{5.7}
 \ddot \phi  + 3h \, \dot \phi + \phi  = 0 \, ,
\quad   h^2 + a^{-2} = \phi^2 + \dot \phi^2 \, .
\ee
$\phi \to - \phi$ is a 
$Z_2$-gauge transformation, where  $Z_2$ denotes  the two-point group.  
Derivating  eq. (\ref{5.7})  we can express $\phi$ and 
$ \dot \phi$ as  follows 
\be\label{5.8}
\phi = \pm \,  \sqrt{2+2\dot a^2+   a \ddot a} \,  / \sqrt{3a^2}
\, , \quad 
\dot \phi = \pm \, \sqrt{1+\dot a^2 -  a \ddot a} \,  / \sqrt{3a^2} \, .
\ee
Inserting  eq. (\ref{5.8})   into  eq. (\ref{5.7})  we get
\be\label{5.9}
a^2 \frac{d^3 a}{dt^3} = 4 \dot a(1+\dot a^2) - 3 a \dot a \ddot a
 \pm 2a  \sqrt{2+2\dot a^2+   a \ddot a} \, 
\sqrt{1+\dot a^2 -  a \ddot a} \, .
\ee
On the r.h.s. we have $``+"$
 if   $\phi \dot \phi  > 0$ and $``-"$ otherwise. At 
all points $t$, where one of the roots becomes zero,   $``+"$ and $``-"$ 
 have  to be interchanged.
 To get the temporal behaviour for very large 
values $a \gg 1$ but small values $\vert h \vert$,
 we make the ansatz
\be\label{5.10}
     a(t) = 1/\e + \e A(t)\, , \quad  \e>0\, , \quad  \e \approx 0\, .  
\ee
In lowest order of $\e$ we get from  eq. (\ref{5.9}) 
\be\label{5.11}
\frac{d^3 A}{dt^3}\,  = \,  \pm \, 2 \,  \sqrt{2+\ddot A}
 \  \sqrt{1-\ddot A} \, .
\ee
An additive constant to $A(t)$ can be absorbed by a redefinition of $\e$, 
 eq. (\ref{5.10}), so we require $A(0)=0$. After a suitable translation of $t$, 
each solution of  eq. (\ref{5.11}) can be represented as
\be\label{5.12} 
     A(t) = \a t - t^2/4 + \frac{3}{8}   \sin(2t) \, , \quad
     \vert \a \vert \le \pi/4 \, .
\ee
$\a = \pi/4$ and $\a = - \pi/4$   represent the same solution, so 
we have a $S^1$ space of solutions. Here, $S^n$ denotes  
the $n$-dimensional sphere. $a(t) > 0$ is fulfilled for 
$A(t) > - 1/\e^2$,  i.e. $\vert t\vert 
<  2/\e $ only. In dependence of the value $\a$, $A(t)$ has one or two 
maxima and, accordingly, null or one minimum. The corresponding 
intervals for $\a$ meet at two points, $\a \approx  0$ and 
$\vert \a \vert \approx  \pi/4$, where one 
maximum and one horizontal turning point
\be
\dot a = \ddot a =  0, \ \frac{d^3 a}{dt^3}  \ne 0
\ee
exist.

\subsection{The qualitative behaviour}\label{s53}

Eqs.   (\ref{5.7}) represent  a regular system: at $t =  0$ we prescribe 
$a_0$, sgn $\dot a_0$, $\phi_0$, $\dot \phi_0$ 
fulfilling
\be
a_0^2 \left (\phi_0^2 + \dot \phi_0^2 \right) \le 1
\ee
 and then 
all higher derivatives can be obtained by differentiating:
\be
\vert \dot a \vert = \sqrt{a^2(\phi^2 + \dot \phi^2) -1} \, , 
\ddot \phi = - 3 \dot a \dot \phi/a - \phi \, , 
\ee 
the r.h.s. being smooth functions. It follows
\be\label{5.13} 
d/dt \left(h^2 + a^{-2}\right) = -6h(d\phi/dt)^2 \, .
\ee

\subsubsection{Existence of a maximum}\label{t531}

The existence of a local maximum for each
 solution $a(t)$  already follows from the ``closed universe 
recollapse conjecture", but we shall prove it for our model as 
follows: if we start integrating with $ h \ge 0 $; 
otherwise, $t \to   -t$ serves 
to reach that;  then $h^2 + a^{-2}$  is a monotonously decreasing 
function as long as $h \ge 0$ holds, because $h \,  \dot \phi  = 0$
 holds at isolated points $t$ only, 
cf.  eq. (\ref{5.13}). We want to show that after a finite time, $h$ changes 
its sign giving rise to a local maximum of $a(t)$. If this is not the 
case after a short time, then we have after a long time 
$h \ll 1$, $a \gg 1$, and eqs.   (\ref{5.10}),  (\ref{5.12})
 become a good approximation to the exact solution. 
The approximate solution   (\ref{5.10}),  (\ref{5.12}) 
 has already shown to possess 
a local quadratic maximum and this property is a stable one 
within $C^2$-perturbations. So the exact solution has a maximum, too, q.e.d.

\subsubsection{The space of solutions}\label{t532} 

We denote the space of solutions for  eqs. (\ref{5.1}),  (\ref{5.7})  by $V$ 
and endow it with Geroch's topology from 1969, see \cite{geroch} and 
 subsection   \ref{t543}  for further details. 
Applying the result of   subsection \ref{t531}, $V$ can be 
constructed as follows: the set of solutions will 
not be diminished if we start integrating with $\dot a_0 = 0$. We prescribe
\be\label{5.14} 
f= a_0^{-1} \quad {\rm  and } \quad 
 g = (1 - a_0 \ddot a_0) \cdot {\rm sgn} 
(\phi_0 \dot \phi_0) \, ,
\ee
then all other values are fixed. $f$ and $g$ are restricted 
by $f> 0$; $\vert g \vert \le 3$, where $g = 3$ and $ g = -3$ describe the same 
solution. Therefore
\be\label{5.15} 
V    = ({\cal R} \times S^1)/Q \, ,
\ee
where ${\cal R} $ denotes the real line, considered  
as topological space,  and $Q$ is an equivalence  relation 
defined as follows: 
Some solutions $a(t)$  have more  than one, but 
at most countably many,  extremal points, but each
 extremum defines one point in $  {\cal R} \times S^1  $; these are just 
the points being $Q$-equivalent.

\bigskip

Considering eqs.   (\ref{5.7}),  (\ref{5.14})   in more details, we get the 
following: for $\vert g \vert  < 1$ we have a minimum and for 
$\vert g \vert > 1$ 
 a maximum for $a(t)$. 
\be
\frac{  d^3 a_0}{dt^3} > 0
\ee
 holds  for $0 < g < 3$. $a(t)$ is an even 
function for $g = 0$ and $g =\pm 3$ only. For $g = 0$ it is 
 a symmetric minimum of $a(t)$  and $\phi$ is even, too. For $g = \pm 3$
 it is a symmetric
maximum of $a(t)$  and $\phi$ is odd. For $\vert g \vert  = 1$
 one has a horizontal turning point for $a(t)$. Time reversal $t \to   -t$
 leads to $g \to  -g$.

\subsubsection{From one extremum to the next}

To prove our result, $D \ge 1$,  where $D$ is the dimension
 of the set of solutions which can be integrated up to $t \to \infty$, 
we need a better knowledge of the equivalence relation $Q$. 
To this end we define a map 
\be\label{5.15b}
p: {\cal R} \times S^1   \to {\cal R} \times S^1
\ee
 as follows:  For $x = (f, \, g)$ we start integrating at $t = 0$.  Let 
\be
t_1 = {\rm min} \{t \vert t > 0, \,  \dot a(t) = 0\} \, .
\ee
For $t_1 < \infty$  we define 
\be\label{5.15d}
p(x) = \bar x = (\bar f, \, \bar g)   = \left( f(t_1), \, g(t_1) \right) 
\ee
and otherwise $p$ is not defined. 
In words: $p$ maps the initial conditions from one 
extremal point of $a(t)$ to the next one. It holds: $p$ is 
injective and $xQy$ if and only if there exists an 
integer $m$ such that $p^m(x) = y$. 
Let $V_m \subset {\cal R} \times S^1$ be that
 subspace for which $p^m$ is defined. Then the inclusion  
\be
\{(f, \, g) \vert - 1 <g \le 1 \} \subset  V_1
\ee
means:  a minimum is always followed  by a maximum; $\vert g \vert  < 1$
 implies $\vert \bar g \vert \ge 1$. $V_2 \ne \emptyset$  follows from the 
end of section \ref{s52}.  For $x = (f, \,  g)$ we define 
 $-x = (f, \, - g)$. With this 
notation it holds, see the end of subsection \ref{t532},  
\be
p^m(V_m) = V_{-m} = - V_m \, ,
\ee
 and for $x \in V$  we  have
\be\label{5.16}
    p^m \left( -p^m(x)\right) = -x
 \quad {\rm  and} \quad 
 p^{-m}\left( V_m \cap V_{-m} \right)   = V_{2m} \, .
\ee
A horizontal turning point can be continuously deformed   to a pair 
of extrema; such points give rise to discontinuities of the function $p$, 
but for a suitably defined non-constant integer power $m$
 the function $p^m$ is 
a continuous one. If $x \in V_1 \backslash {\rm int}(V_1)$ 
 where \  int \   denotes the topological interior, 
then $a(t)$ possesses a horizontal turning point. To elucidate
 the contents of these sentences we give an example:

\bigskip

For very small values $f$  and $\bar f$ we may apply eqs. 
 (\ref{5.10}),  (\ref{5.12})   to calculate the function $p^m$.
 In the approximation used, no more  than 3 extremal 
points appear, so we have to consider 
\be
 p^m \, , \quad m \in \{-2, \,  - 1, \,  0, \, 1, \,  2 \}
\ee
only. $f$ and $\bar f$  approximately coincide, so  we 
concentrate on the function $\bar g(g)$.
 The necessary power $m$ is sketched at the curve.
 To come from $m$ to $-m$, the curve has to he reflected at the line 
$\bar g = g$. Time reversal 
can be achieved, if it is reflected at $\bar g = -g$. $V_1$
 is the interval $
-1 <g \le  2.8853$. The jump discontinuity of $p$ at $g = 1$ and the 
boundary value $g =2.8853$ \dots  are both connected with  horizontal 
turning points $g, \, \bar g = \pm 1$. The shape of the 
function $\bar g(g)$  for 
$m = 1$ can be obtained by calculating  the extrema of $A(t)$
  eq. (\ref{5.12})  in dependence of $\a$
 and then applying  eq. (\ref{5.14}). For $m \ne 1$ one applies  eq. (\ref{5.16}).

\subsubsection{The periodic solutions}

The periodic solutions $a(t)$ are characterized by 
the fixed points of some $p^m$, $ m \ge 2$, whereas  $p$  itself has no 
fixed points.  The notation $p^m$ means, that the map $p$, eqs. (\ref{5.15b}),
 (\ref{5.15d}), will be applied $m$ times. 
The existence  of the fixed points  can be proved as follows: 
we start integrating at $t = 0$ with  a symmetric minimum,  
 $g = 0$,  of $a(t)$  and count the  number $m$ of zeros of $\phi(t)$
 in the  time interval $0 <t < t_2$, where $a(t_2)$  is the first 1ocal 
maximum of $ a(t)$. The number $m$ depends on $f = 1/a_0$  and has jump 
discontinuities only at points where $\phi(t_2) = 0$.  For initial 
values $f = f_k$, where the number  $m$ jumps from $k$  to $k + 1$, 
 the function $a(t)$  is symmetric 
about $t = t_2$. But a function, which is symmetric about two 
different points, is a periodic one.  We call periodic solutions obtained by 
this procedure, periodic solutions 
of the first type, they represent  fixed points of the map  $p^2$.
 Page 1984  \cite{page1984} gives the numerically obtained result: 
\be
f_1 = 1/a_0 \qquad {\rm  with} \qquad  a_0=0.76207\dots
\ee
 That  $f_k$  exists also  for 
very large values $k$ can be seen as follows: take solution 
 eqs. (\ref{5.10}),  (\ref{5.12})  with  a  very small value $\e$
 and insert it into  eq. (\ref{5.8}); we roughly get  $ m = 1/\e \pi$.

\bigskip

Fixed points of $p^4$, which are not fixed points of $p^2$, 
are called periodic solutions of the second type. They
can be constructed  as follows:  we 
start with $g = 0$ but $f \ne f_k$  and count the number 
$n$ of  zeros of $\dot \phi(t)$ in the interval $0 < t < t_3$, where $ a(t_3) > 0$
 is next local minimum of $a(t)$. 
For $f \approx  f_k$, $t_3$ is defined by continuity reasons.
 For $f = f^l$, the number $n$ jumps from $l$ to $l + 1$, and we have a periodic 
solution. Numerical evaluations yield 
\be
f^1 = 1/a_0 \, ,   \qquad a_0 = 0.74720\dots
\ee
 The   existence of $f^l$ for very large values $l$
 is again ensured by solution   (\ref{5.10}),  (\ref{5.12}).

\subsubsection{The aperiodic  perpetually oscillating solutions}\label{t535}

Now we look at the solutions in the neighbourhood of  the 
periodic ones. Let $x_0 = (f_k, \, 0) \in V_2$ be one of the fixed 
points of $p^2$ representing a periodic solution of the first 
type; $a_0(t)$, the corresponding  solution  
 has no horizontal turning points. Therefore, the function 
$p$ is smooth at $x_0$.  Let us denote the circle with boundary of radius $\e$
 around $x_0$ by $K(\e)$. 
By continuity reasons there exists an $\e > 0$ such that 
$K(\e) \subset {\rm  int} ( V_2 \cap V_{-2})$ and the first two extrema of the functions  
$a(t)$ corresponding to points of $K(\e)$  are either maxima or minima. 
Let   $R_0 = K(\e)$  and for $n \ge 1$
\be
     R_n: = R_0 \cap p^2 \left( R_{n-1}\right) \, ; 
 \quad       R_{-n}: = R_0 \cap p^{-2} \left( R_{1-n}\right)  \, . 
\ee
By assumption, $p^{\pm 2}$
 is defined in $R_0$, so $R_m$  is a well-defined compact 
set with $x_0 \in {\rm int} \, R_m$ for each integer $m$. It
 holds
\be
 R_{n+1} \subset R_n \, , \quad R_{-n} = - R_n \, 
\ee
and 
\be
R_n = \{ x \vert p^{2m}(x) \in R_0 \ {\rm for} \quad 
m = 0, \dots n \} \, .
\ee
Then the set 
\be
R_\infty := \bigcap_{n=0}^\infty R_n
\ee
is  a non-empty  compact set with $x_0 \in  R_\infty$. In words: $R_n$
 represents the set of all these solutions which have $n$ different maxima. 
$R_\infty$ then corresponds to the set of solutions possessing infinitely
many maxima. For each $x \in  R_\infty$,  the 
corresponding solution $a(t)$ can be integrated up to $t \to \infty$. 
The last statement follows from the fact
that the time from one extremum to the next is bounded from 
below by a positive number within the compact set $R_0$, i.e., an 
infinite number of extrema can be covered only by an infinite 
amount of time. Analogous statements hold for   
$R_{- \infty} =  - R_\infty$ and  $t \to - \infty$. 

\bigskip

Let us fix an integer $m \ge 1$. 
We start integrating at $x_\d := (f_k, \, \d ) \in R_0$, $0 < \d \le \e$.
\be
\d(m) := {\rm max} \{ \d \vert \d \le \e, \ p^{2k}(x_\d) \in R_0 \ {\rm for } \
 k=0, \dots m \}
\ee
exists because of compactness, i.e.,  $x_{\d(m)} \in 
 R_m$   and there is an  integer $k(m) \le  m$  such that
\be
 y(m) : = p^{2k(m)} \left(x_{\d(m)} \right) \in \d R_0   \, , 
\ee
 $ \d R_0 $ being the boundary of $R_0$.
\be
    j(m) :=  \left\{
 \begin{array}{c}
  1\,  , \quad   {\rm   if } \quad  k(m) < m/2   \\ 
0  \qquad {\rm  otherwise}\, .
\end{array}
\right.  
\ee
The sequence 
\be
\left(y(m), \,  j(m) \right) \subset  S^1 \times  Z_2
\ee
 possesses a  converging subsequence with 
\be
y(\infty):= \lim_{i \to \infty} y(m_i)\, , \quad 
j(\infty):= \lim_{i \to \infty} j(m_i) \, . 
\ee
Now let us start integrating  from $y(\infty)$ forward in 
time for $j(\infty) = 1$  and backwards in time for 
$j(\infty) =  0$. We consider  only the case $j(\infty)=  1$, the other case will be solved by 
$t \to  - t$. For  each $I$  with $m_i  > 2m$  we have $y(m_i) \in R_m$,
 therefore, $y(\infty) \in  R_m$   for all $m$, i.e., $ y(\infty) \in
 R_\infty$. $y(\infty) \in  \d R_0$  and for all 
 $n \ge 1$, $ p^{2n} \left( y(\infty) \right) \in R_0$.

\bigskip

By continuously diminishing $\e$ we get a one-parameter set 
of solutions $a_\e(t)$, which can be integrated up to $t \to \infty$.
To this end  remember that one solution $a(t)$  is represented by at most 
countably many points of $R_0$.  Supposed $a_\e(t)$  is a periodic function. 
By construction this solution has a symmetric minimum and there exist 
only countably many such solutions. Let the set $M$ of solutions be defined as 
follows:   $M = \{a(t) \vert a(0)$  is a minimum 
parametrized by a point  of $R_0$, $a(t_1)$  is the next minimum, and 
 $0 \le t \le t_1 \}$;   then we get 
 $M=[ a_{\rm min}, \, a_{\rm max}] $
with  $ a_{\rm min} > 0$, $ a_{\rm max} < \infty $, 
and for each $\e$ and each $t \ge 0$ it  holds 
$ a_{\rm min} \le a_\e(t) \le  a_{\rm max}  $. 
So we have proven: in each neighbourhood of the periodic solutions of 
the first type there exists a set of Hausdorff  dimension $D \ge 1$  of 
uniformly bounded aperiodic perpetually oscillating solutions  which can 
be integrated up to $ t \to \infty$.

\subsection{Problems with die probability measure}\label{s54}

To give concepts like ``the probability $p$ of the
 appearance of  a sufficiently long inflationary phase''    a 
concrete meaning, we have to define a probability measure $\mu$  in 
the space $V$ of solutions. Let us suppose we can find a hypersurface 
 $H$  in the space $G$ of initial conditions such that each solution 
is characterized by exactly one point of $H$. Then $V$ and $H$ are 
homeomorphic and we need not to make a distinction between them. 
Let us further suppose that $H \subset G$ is defined by a suitably 
chosen physical quantity $\psi$ to take the Planckian value. Then 
we are justified to call $H$ the quantum boundary. By construction, 
$H$ divides $G$ into two connected components; $\psi \le \psi_{\rm Pl}$ 
defines the classical region. All classical trajectories start their 
evolution at $H$ and remain in the classical region forever. 

\bigskip

Let us remember the situation for the spatially flat Friedmann model 
Belinsky  et al.  \cite{belinskyb}
 subsection \ref{t541},  and for the Bianchi type I model, Lukash   and 
Schmidt  \cite{lukash} subsection \ref{t542},  before  we discuss the closed 
Friedmann model in  subsection \ref{t543}.

\subsubsection{The spatially flat Friedmann model}\label{t541} 

For this case, eq.  (\ref{5.13})  reduces  to $h \dot h = -3h \dot \phi^2$,
 i.e., each solution crosses the surface $h = h_{\rm Pl}$  
exactly once, the only exception is the flat Minkowski spacetime 
 $h \equiv  0$. Reason: For $\dot \phi \ne 0$ we get $\dot h < 0$. 
The corresponding physical quantity
\be\label{5.17} 
\psi = h^2 = \phi^2 + \dot \phi^2
\ee
is the energy density. The space of non-flat spatially flat Friedmann 
models is topologically $S^1$,  and equipartition of initial conditions 
gives a natural probability measure there. With this definition it turned 
out that for $ m \ll m_{\rm Pl}$  it holds
\be
p \approx   1 - 8m/ m_{\rm Pl} \, ,
\ee
 cf. e.g. M\"uller and Schmidt  \cite{muell1985}, i.e., 
inflation becomes quite probable. If,  on the other hand, equipartition is taken at some 
$h_0 \ll h_{\rm Pl} m /m_{\rm Pl}$  then inflation is quite improbable.
The total space $V$  of  solutions  has Geroch 
topology    $ \alpha S^1$,   i.e., $V = S^1 \cup \{ \alpha \}$,
 and the space itself is the only  neighbourhood  around the added point $\alpha$
 which corresponds to $h \equiv 0$, 
because each solution is asymptotically flat  for $t \to \infty$.

\subsubsection{The Bianchi-type I  model}\label{t542}

With the metric
\be
     ds^2 = dt^2   - e^{2 \a} [e^{2(s+ \sqrt 3 r)} dx^2
 + e^{2(s - \sqrt 3 r)} dy^2 + e^{-4s}   dz^2 ], \quad      h = \dot \a
\ee
the analogue to eq.   (\ref{5.17}) is
\be\label{5.18}
\psi  = h^2 = \phi^2 + \dot \phi^2 + \dot r^2 + \dot s^2 \, , 
\ee
and $h =   h_{\rm Pl}$   defines a sphere $S^3$ in  eq. (\ref{5.18}). Here
 all solutions cross this sphere exactly once, even the flat Minkowski 
spacetime: it is represented as (0 0 1)-Kasner solution $\a$,
 so the space of solutions is $V = S^3/Q$, where $Q$  is a 
12-fold cover of $S^3$ composed  of the $Z_2$-gauge transformation 
 $\phi \to - \phi $   and of the six permutations of 
the three  spatial axes.
Eq.   (\ref{5.18}) induces a natural probability measure on the space 
 $\psi = \psi_{\rm Pl}$ 
by equipartition,  and the equivalence relation $Q$ does not essentially 
influence this. As in subsection \ref{t541}, $\a \in V$  has only one neighbourhood: 
$V$ itself. Up to this exception, $V$ is  topologically a 3-dimensional 
cube with boundary. One diagonal line through it represents the 
Kasner solution $\phi \equiv 0$.
The boundary $\delta V$ of $V$  has topology $S^2$  and 
represents the axially symmetric solutions and one great
 circle of it the isotropic 
ones. As usual, the solutions with higher symmetry form the 
boundary of  the space of solutions. $p$ turns out 
to be the same as in   subsection \ref{t541} 
and how $\psi$ eq.   (\ref{5.18}) can be invariantly defined, is discussed 
in \cite{lukash}.

\subsubsection{The closed  Friedmann  model}\label{t543}

Now  we  come to the analogous questions concerning
 the closed Friedmann model.  Before defining a  measure, 
one should  have a topology  in a set. I  feel it should be a variant of  
Geroch  \cite{geroch}. The Geroch topology, cf.  \cite{sch87b},  
   is defined as follows: let $x_i = (a_i(t), \, \phi_i(t))$
  be a sequence of  solutions and  $x = (a(t), \, \phi(t))$  a further solution. 
 Then $x_i \to  x$ in Geroch's topology, if there exist suitable gauge and 
coordinate transformations after which $a_i(t) \to a(t)$ 
 and $\phi_i(t) \to \phi(t)$  converge uniformly 
together with all their derivatives in the interval 
 $t \in  [- \varepsilon, \, \varepsilon ]$ 
for some $\varepsilon > 0$. Because  of the validity of the field 
equations, ``with all their derivatives" may be substituted 
by ``with their first derivatives".

\bigskip

With this definition one gets just the same space $V$ as in subsection \ref{t532}, 
eq.   (\ref{5.15}). The existence of aperiodic 
perpetually oscillating  solutions, which go right across the region
 of  astrophysical interest subsection \ref{t535}  shows that 
for a subset of dimension $D \ge 1$ of $R \times S^1$, $Q$
 identifies  countably many points. All these points lie in a compact 
neighbourhood of the corresponding periodic  point
 $ (f_k, \,  0)$ 
 and possess therefore at least one accumulation point $z$. 
At these points $z$, $V$  has a highly non-Euclidean  topology. 
Further, $V$  has a non-compact non-Hausdorff topology. 
So there is no chance to define a probability measure  in a natural 
way and no possibility to define a continuous hypersurface in the 
space of initial conditions which each trajectory crosses exactly once.
     
\subsection{Discussion of the inflationary phase}\label{s55}

Supposed, we had obtained a result of the 
type: ``Each solution $a(t)$  has at least one but at most seven 
local maxima." Then  one could define  --- up to a factor $7 =O(1)$
 ---  a probability measure. So it is just the existence of the 
perpetually bouncing  aperiodic solutions which gives the problems. 
We conclude: it is not a lack of mathematical knowledge but 
an inherent  property of the closed Friedmann model which hinders 
to generalize the convincing results obtained for the spatially 
flat  model. So it is no wonder that different trials led 
to controversial  results, cf. Belinsky  et al. \cite{belinskyb}, \cite{belinsky}
 and Page 1987  \cite{page1984}.  One of these results reads ``inflationary 
and non-inflationary solutions have both infinite measure", hence, nothing is clear.  

\bigskip

Let us now discuss the results of Starobinsky  \cite{staro78}
and Barrow and Matzner  \cite{barrowmatzner}  concerning the 
probability of a  bounce. They have obtained a very low 
probability  to get a bounce, but they used equipartition at 
some $h_0 \ll h_{\rm Pl} m/ m_{\rm Pl}$. 
As seen in subsection \ref{t541}  for the spatially flat Friedmann model 
concerning inflation, this low probability does not hinder to 
get a considerable large probability if equipartition is applied at 
 $h = h_{\rm Pl}$.
 We conclude, the probability of bouncing solutions is not 
a well-defined concept up to now. Well-defined is, on the other 
hand, some type of conditional probability. If we suppose that 
some fixed value $a$, say $10^{28}$ cm or so, and there some fixed value
 $ h$, say 50 km/sec $\cdot$ Mpc or so, appear within the cosmic evolution, then 
the remaining degree of freedom is just the phase of
 the scalar field, which 
 is the compact set $S^1$  as configuration space. But this solves 
not all problems, because the perpetually bouncing solutions discussed 
in subsection \ref{t535} cross this range of astrophysical interest infinitely 
often.  Calculating conditional probabilities instead of absolute 
probabilities, and, if this condition is related to our own 
human existence, then we have already applied the anthropic principle. 
Cf.  similar opinions in Singh and Padmanabhan  \cite{singh}
 concerning  the so far proposed explanations 
of the smallness of the cosmological constant.

\bigskip

The massive scalar field in  a closed Friedmann 
model with Einstein's  theory of gravity cannot explain 
the long inflationary stage of cosmic evolution as an absolutely  
probable event and so some type of an anthropic principle has to 
be applied; see the well-balanced monograph by  Barrow and Tipler   
 \cite{barrowtipler}  to this theme.
We have discussed the solutions for the minimally 
coupled scalar field, but many results for the conformally 
coupled one are similar, 
see e.g. Turner and Widrow  \cite{turnerwidrow}; 
this fact in turn can be  explained by the existence of a conformal transformation 
relating between them,  see   \cite{sch88c}.
The problems in the case of defining a probability 
measure in the set of  not necessarily spatially flat  Friedmann models 
are also discussed in Madsen  and Ellis  \cite{madsen}. 
They conclude that   inflation need not to solve 
the flatness problem.
 The Gibbons-Hawking-Stewart approach \cite{gibbons}
gives approximately the same probability measure as the equipartition of 
initial conditions used here.
The Wheeler-de Witt equation for the 
massive scalar field in a closed Friedmann universe 
model is also  discussed by   Calzetta   \cite{calzetta}.
 Possibly, this approach is the route 
out of the problems mentioned here.

\newpage

\section[The superspace of Riemannian
metrics]{The metric in the superspace of Riemannian
metrics}\label{Kap6}
\setcounter{equation}{0}

The space of all Riemannian metrics is  infinite-dimensional. Nevertheless 
a great deal 
of usual Riemannian geometry can be carried over. The superspace of all Riemannian 
metrics shall be endowed with a class of Riemannian metrics;
 their curvature and  invariance properties are  discussed. Just one of this class
 has the property to bring the
 Lagrangian of General Relativity into the form of a classical particle's motion. 
The signature of the superspace metric depends  on the signature  of  the 
original metric in a non-trivial manner, we derive the corresponding formula. 
Our approach, which is based on \cite{sch90c}, 
 is  a local  one: the essence  is a metric in the space of all 
symmetric rank-two tensors,  and then  
 the space becomes a warped product of the real line with an Einstein space.

\subsection{The superspace}

Let $n \ge 2$, $n$ be the dimension of   the basic 
 Riemannian spaces. Let $M$ be an 
$n$-dimensional differentiable manifold with an 
atlas $x$ of coordinates $x^i$, 
 $i = 1, \dots, n$. The signature $s$ shall be fixed;  $s$ is the 
number of negative  eigenvalues of the metric. Let $V$  be the space of all Riemannian 
metrics $g_{ij}(x)$  in $M$  with signature $s$,
 related to the coordinates $x^i$. 

\bigskip

This implies that isometric  metrics in $M$ are different points in $V$  in general. 
The $V$ is called superspace,  its points are the Riemannian metrics. The tangent 
space in $V$  is the vector space
\be\label{6.1}
     T = \{ h_{ij}(x) \vert   x \in M, \  h_{ij}= h_{ji} \}   \, , 
\ee
the space  of all symmetric  tensor fields of rank 2. 
All considerations are local ones, so we may have in mind 
one single fixed coordinate system in $M$.

\subsection{Coordinates in superspace}

Coordinates should possess one contravariant index, 
so we need a transformation of the type
\be\label{6.2}
 y^A     = \mu^{Aij} g_{ij}(x)
\ee
such that the $y^A$  are the coordinates for $V$.
To have a defined one--to--one correspondence
 between the index pairs $ (i, \, j)$ 
 and the index $A$   we require 
\be
A = 1, \dots N = n(n + 1)/2 \, ,
\ee
 and $A = 1, \dots  N$  corresponds to the pairs
\bea\label{6.3}
(1,1), \ (2,2),  \dots (n,n), \ (1,2), \ (2,3),
 \dots (n-1,n), \ (1,3),   \nonumber \\ 
  \dots (n-2,n), \dots (1,n)
\eea 
consecutively. $(i, \, j)$ and 
$(j, \, i)$  correspond to the same $A$. We make the ansatz
\be\label{6.4}
    \mu^{Aij} = \mu_{Aij} = 
 \left\{
 \begin{array}{c}
 b  \quad {\rm for} \quad  i \ne j \\
 c \quad {\rm for} \quad  i = j \\
          0 \ {\rm if} \  (i,j) \ {\rm does \ not \ correspond \  to \ } A
\end{array}
\right.   
\ee
with certain real numbers $b$ and $c$ to be fixed later  
 and require the usual inversion relations
\be\label{6.5}     
 \mu^{Aij} \mu_{Bij} = \d^A_B
\quad {\rm  and}  \quad 
\mu^{Aij}\mu_{Akl}
= \d_k^{(i} \d_l^{j)} \, . 
\ee
Bracketed indices are to be symmetrized, 
which is necessary because of symmetry 
of the metric $g_{ij}$.
 Inserting ansatz   (\ref{6.4}) into eq. (\ref{6.5})  gives $c^2 = 1$,
 $b^2 = 1/2$. Changing the sign of $b$ or $c$ only changes the sign 
of the coordinates, so we may put
\be\label{6.6}
     c=1, \qquad  b=1/\sqrt 2   \, .
\ee
The object $ \mu^{Aij}$ is
 analogous to the Pauli spin matrices relating two 
spinorial indices to one vector index.

\subsection{Metric in superspace}\label{s63}

The metric in the superspace shall be denoted by $H_{AB}$,  it holds
\be\label{6.7} 
    H_{AB} = H_{BA}
\ee
and the transformed metric is
\be\label{6.8}
G^{ijkl} = H_{AB}     \mu^{Aij}    \mu^{Bkl} \, , \quad
H_{AB} =      \mu_{Aij}    \mu_{Bkl}
G^{ijkl} \, .
\ee
From  eqs. (\ref{6.4}) and    (\ref{6.7}) it follows that
\be\label{6.9}
G^{ijkl} = G^{jikl} = G^{klij} \, .
\ee
The inverse to $H_{AB}$  is $H^{AB}$, and we define
\be\label{6.10}
G_{ijkl} = H^{AB}     \mu_{Aij}    \mu_{Bkl}
\ee
which has the same symmetries as  eq. (\ref{6.9}). We  require 
$G^{ijkl}$  to be a tensor and use only the metric $g_{ij}(x)$  to define it. 

\bigskip

Then the ansatz
\bea\label{6.11}
G^{ijkl}
= z \,  g^{i(k} \,  g^{l)j} + \a \,  g^{ij} \,  g^{kl}
\\
 G_{ijkl}    = v \, g_{i(k} \,  g_{l)j} + \b \,  g_{ij} \,  g_{kl} \label{6.12}
\eea
where $v$, $z$, $\a$ and $\b$ are constants,  is 
the most general one to fulfil  the symmetries  eq. (\ref{6.9}). One 
should mention that also curvature-dependent constants 
could have heen introduced. 

\bigskip

The requirement 
that $H_{AB}$  is the inverse to $H^{AB}$  leads via 
  eqs. (\ref{6.8}) and  (\ref{6.10})  to
\be\label{6.13}
G_{ijkl} \,  G^{klmp}  =  \d_i^{(m} \,  \d_j^{p)} \, . 
\ee
The requirement that  $G^{ijkl}$  is a tensor can be 
justified as follows: Let 
a curve $y^A(t)$, $0 < t < 1$  in $V$ be given, 
then its length is 
\be
\sigma  = \int_0^1 \left(   H_{AB} 
\frac{dy^A }{dt}\frac{dy^B }{dt}  \right)^{1/2} \, dt
\ee
i.e.,  with   eqs. (\ref{6.2}) and  (\ref{6.8}) 
\be\label{6.14}
\sigma  = \int_0^1 \left(
G^{ijkl} \,  \frac{d g_{ij} }{dt}\,  \frac{d g_{kl} }{dt}
 \right)^{1/2} \, dt    \, .
\ee
A coordinate transformation in $M: x^i \to \varepsilon x^i$ 
changes   $g_{ij} \to \varepsilon^{-2} g_{ij}$.
 
\bigskip

We now require that $\sigma$ shall not 
be changed by such a transformation. 
Then $\a$ and $z$ are constant real numbers. Inserting 
  eqs. (\ref{6.11}) and  (\ref{6.12})
 into  eq. (\ref{6.13})  gives $v \, z = 1$, hence $z \ne 0$. 
By a  constant rescaling we get
\be\label{6.15}
v = z = 1
\ee
and then  eqs. (\ref{6.11}),  (\ref{6.12}) and  (\ref{6.13}) yield
 the conditions 
\be\label{6.16}
\a \ne \frac{-1}{n} \, , \qquad \b = \frac{- \a}{1 + \a n}   \, .
\ee
So we have got a one-parameter 
set of metrics in $V$. 

\bigskip

Eq.   (\ref{6.16})  fulfils the 
following duality relation: with 
\be
f(\a) = - \a/(1 + \a n) \, ,
\ee
the equation  $f(f(\a)) = \a$ holds for all $\a \ne -1/n$.

\bigskip

Inserting eq. (\ref{6.15}) into eqs. (\ref{6.11}) and  (\ref{6.12}), we finally
 get for the superspace metric  
\bea\label{6.17} 
G^{ijkl}
=   g^{i(k} \,  g^{l)j} + \a \,  g^{ij} \,  g^{kl} \qquad {\rm and}
\nonumber  \\
 G_{ijkl}    =  g_{i(k} \,  g_{l)j} + \b \,  g_{ij} \,  g_{kl} \,. 
\eea
It holds: For $\a = -1/n$, the metric $H_{AB}$  is not invertible.

\bigskip

 Indirect proof: $G^{ijkl}$   depends continuously on $\a$, so it must 
be the case with the inverse. But 
$$
\lim_{\a \to -1/n}
$$
 applied to $ G_{ijkl}    $     gives no finite result. Contradiction.

\subsection{Signature of the superspace metric}

Let $S$ be the superspace signature, i.e., the number of negative eigenvalues  
of the superspace metric $H_{AB}$.
 $S$ depends on $\a$ and $s$, but $\a = -1/n$ is excluded.  
 For convenience we define
\be \label{6.18}
\Theta = \left\{
\begin{array}{c}
  0, \qquad   \a>-1/n \\
 1, \qquad   \a<-1/n
\end{array}
\right.
\ee
From continuity reasons it follows 
that $S$  is a function of $\Theta$ and $s$: $S = S(\Theta, s)$. 
If we transform $g_{ij} \to - g_{ij}$
 i.e., $s \to  n - s$, then $H_{AB}$  is not changed, i.e.,
\be\label{6.19}
     S(\Theta,s) = S(\Theta,n - s) \, .
\ee
We transform $g_{ij}$  to diagonal form as follows
\be\label{6.20}  
g_{11} = g_{22} = \dots = g_{ss} 
 = -1, \quad g_{ij} = \d_{ij} \quad {\rm otherwise}\, .
\ee

\bigskip  

Let us now give the signature for $\Theta = 0$. 
 To calculate $S(0,s)$ we may put $\a = 0$  and get with 
  eqs. (\ref{6.8}),  (\ref{6.11}) and  (\ref{6.15})
\be\label{6.21}
    H_{AB} =      \mu_{Aij} \,   \mu_{Bkl} \, 
g^{ik} \,  g^{jl} 
\ee
which is a diagonal matrix. It holds $H_{11} = \dots   = H_{nn}    = 1$  and 
the other diagonal components are $\pm 1$. 
A full estimate gives in agreement with eq. (\ref{6.19})
\be\label{6.22}
     S(0,s) = s(n - s) \, . 
\ee

\bigskip

Now, we look for the signature for $\Theta = 1$. 
 To calculate $S(1,s)$  we may put $\a = -1$ and get
\be\label{6.23} 
    H_{AB} =      \mu_{Aij} \,   \mu_{Bkl} \, \left(
g^{ik} \,  g^{jl}  -   g^{ij} \,  g^{kl}    \right) \, .
\ee
For $A \le  n < B$, $H_{AB} = 0$, i.e., 
the matrix $H_{AB}$  is composed of two blocks. 
For $A, \, B \le  n$ we get
\be
H_{AB} =       \left\{ \begin{array}{c}
  0 \quad {\rm for}   \quad  A=B \\
 1    \quad {\rm for}   \quad  A \ne B
\end{array}
\right.
\ee
a matrix which has the $(n - 1)$-fold
 eigenvalue 1 and the single eigenvalue $1 - n$.
For $A,B > n$ we have the same
 result as for the case $\a = 0$, i.e., we get
 $ S(1,s) = 1 + s(n - s)$.

\bigskip

Result:  The signature of the superspace metric is
\be\label{6.24}
S = \Theta  + s(n - s) \, .
\ee

\subsection{Supercurvature and superdeterminant}

We use exactly the same formulae  as for finite-dimensional Riemannian 
geometry to define Christoffel affinities  $\Gamma^A_{BC}$
 and Riemann tensor $R^A_{BCD}$ in superspace. We even omit the
 prefix ``super{}" in the following.  Using  eq. (\ref{6.4}) we write all
equations with indices $i,j=   1,\dots n$.

\bigskip

 Then each pair 
of covariant indices $i,j$ corresponds
 to one contravariant index $A$. The following formulae appear:
\bea\label{6.25}
\frac{\partial g^{ij}}{\pa g_{km}} = - g^{i(k} \, g^{m)j}
\\
\Gamma^{ijklmp} = - \frac{1}{2}g^{i(k} g^{l)(m} g^{p)j}
- \a g^{ij} g^{k(m} g^{p)l}- \frac{1}{2}
 g^{j(k} g^{l)(m} g^{p)i}
\eea
and, surprisingly independent of $\a$  we get
\be\label{6.27}
\Gamma^{klmp}_{ij}
=      -    \d^{(k}_{(i} \, g^{l)(m}
 \, \d^{p)}_{j)} \, .
\ee
Consequently, also Riemann- and Ricci tensor do not depend 
on $\a$:
\be\label{6.28}  
R_{rs}^{klmpij} = \frac{1}{2} \left(
   \d^{(k}_{(r} \,g^{l)(m}g^{p)(i}
 \, \d^{j)}_{s)}
-    \d^{(k}_{(r} \,g^{l)(i}g^{j)(m}  \, \d^{p)}_{s)}
\right) \, .
\ee
Summing over $r=  m$ and $s = p$ we get
\be\label{6.29}
R^{klij}  = \frac{1}{4} \left( g^{ij} g^{kl} - n g^{k(i} g^{j)l}
\right) \, .
\ee
The Ricci tensor has one eigenvalue 0.  Proof: It is not invertible because it 
is proportional to the metric  for the degenerated case 
$\a = -1/n$, cf. eq. (\ref{6.17}) in  section \ref{s63}.

\bigskip

The co--contravariant Ricci tensor reads
\be\label{6.30}
R^{ij}_{kl} = G_{klmp} R^{mpij} 
= \frac{1}{4} \left(
g^{ij} g_{kl} - n \d_k^{(i} \d^{j)}_l
\right) \, , 
\ee
and the curvature scalar is
\be\label{6.31} 
R=   - \frac{1}{8} n(n-1)(n+2) \, .
\ee
The eigenvector to the eigenvalue 0 of the Ricci 
tensor is $g_{ij}$. All other eigenvalues equal $-n/4$,
 and  the corresponding eigenvectors can be parametrized 
by the symmetric  traceless matrices, i.e.
the multiplicity of the eigenvalue $-n/4$ is $(n - 1)(n + 2)/2$.
 This is another view to the well-known split of metric perturbations 
into conformal transformations,
 i.e., one bulk degree of freedom on the one hand, and the remaining 
volume-preserving  degrees of freedom on the other.   

\bigskip 

We define the superdeterminant 
\be\label{6.32}
     H = \det H_{AB} \, .
\ee
$H$  is a function of $g$, $\a$ and 
 $n$ which  becomes zero for $\a = -1/n$, cf.  section \ref{s63}.
We use eqs. (\ref{6.8})  and  (\ref{6.17})  to look in more 
details for the explicit 
 value of $H$. Though we assume $n \ge 2$, here 
the formal calculation for $n = 1$ makes sense; it leads to 
\be\label{6.32a}
H=H_{11} =G^{1111} =g^{11}  g^{11} + \a  g^{11}  g^{11}  
= (1+\a) g^{-2} \, .
\ee
Let us return to the general case $n \ge 2$. 
Multiplication of $g_{ij}$  with $\e$ gives $g \to \e^n g$,
 $ H_{AB} \to \e^{-2} H_{AB} $ and 
 $H \to \e^{-n(n+1)} H$.  

\bigskip

So we get in an intermediate step
\be\label{6.33}
     H = H_1 \, g^{-n-1}
\ee
where $H_1$ is 
the value of $H$ for $g= 1$. $H_1$ depends on $\a$ 
and $n$ only. To calculate $H_1$ we put 
 $g_{ij } = \d_{ij}$   and get via $H_{ij} = \d_{ij} + \a$,
 $ H_{Ai} =  0$ for $A> n$, and $H_{AB} = \d_{AB}$  for 
$A, B > n$  finally
\be\label{6.34}
 H_1=1+\a n \, .
\ee
This result  is in agreement with the $n = 1$-calculation eq. (\ref{6.32a})
 and also with the fact that $\a = -1/n$ gives $H=0$.

\subsection{Gravity and quantum cosmology}

Now, we come to the main application: The action for gravity shall be expressed 
by the metric of superspace. The purpose is to explain the mathematical background 
of quantum cosmology, see [9], [59], [73], [101], [126], [148], [158], [179], [182] 
and [198] for new papers on that topic.

\bigskip

We start from the metric
\be\label{6.35} 
     ds^2 = dt^2 - g_{ij} \, dx^i \, dx^j
\ee
$i,j =1, \dots n$  with positive definite $g_{ij}$  and $x^0 = t$.  
 We define the second fundamental form $K_{ij}$  by
\be\label{6.36} 
   K_{ij}  = \frac{1}{2} \, g_{ij,0} \, .
\ee
The  Einstein action for metric  (\ref{6.35}) is 
\be\label{6.37}
 I = - \int \        \frac{1}{2} \     {}^{\ast}R \  \,  \sqrt g \, d^{n+1}x 
\ee
where $g  = \det g_{ij}$  and $ {}^{\ast}R$  is the 
 $(n + 1)$-dimensional curvature 
scalar for eq. (\ref{6.35}). Indices at $K_{ij}$  will be 
shifted with $g_{ij}$, and $K = K^i_i$. With  eq. (\ref{6.36}) we get
\be\label{6.38}
     (K \sqrt g)_{,0} = (K_{,0} + K^2) \sqrt g \, . 
\ee
This divergence can  be added to the 
integrand of eq. (\ref{6.37})  without changing the 
field equations.  It serves to cancel the term $K_{,0}$
 of $I$. So we get
\be\label{6.39}
 I =  \int \   \frac{1}{2} \left(  K^{ij} K_{ij} - K^2 + R 
 \right)   \sqrt g \, d^{n+1}x 
\ee
where $R$ is the $n$-dimensional curvature 
scalar for $g_{ij}$.  

\bigskip
 
Applying eq. (\ref{6.17}), we now make the 
ansatz for the kinetic energy
\be\label{6.40}
W =    \frac{1}{2} \,  G^{ijmp} K_{ij} K_{mp}
=    \frac{1}{2} \left(   K^{ij} K_{ij} + \a  K^2  
 \right) \, .
\ee
Comparing  eq. (\ref{6.40})  with eq. (\ref{6.39}) we see that 
for $\a = -1$
\be\label{6.41}
 I =  \int \  \left( W +    \frac{R}{2} 
  \right)    \sqrt g \, d^{n+1}x 
\ee
holds.   

\bigskip

 Surprisingly,  
this value for $\a$ does not depend on $n$. 
Because of $n \ge  2$  this value $\a$  
gives a regular superspace metric. For 
 $n = 1$, eq. (\ref{6.37}) is a divergence, 
and $\a = -1$ does not  give  an invertible superspace metric. This is another
 form of the result that  Einstein gravity does not lead to a local
field equation in 1+1-dimensional spacetime. 

\bigskip

Using the $\mu^{Aij}$  and the notations 
$z^A = \mu^{Aij} \, g_{ij } \, / \, 2$
 and $v^A =  dz^A/dt $ we get from  eqs. (\ref{6.40}) and  (\ref{6.41})
\be\label{6.42}
 I =  \int \   \frac{1}{2} \left( 
 H_{AB} v^A v^B  + R(z^A)   \right)   \sqrt g \, d^{n+1}x 
\ee
i.e., the action has the classical form of  kinetic plus 
potential energy. The signature of the metric 
 $H_{AB}$  is $S = 1$. This can be seen from eqs. 
  (\ref{6.18}) and   (\ref{6.24}).

\subsection{The Wheeler-DeWitt equation}

In  eq. (\ref{6.42}), Einstein gravity is given in a  form to allow canonical quantization:
 The momentum $v^A$ is  replaced  by   $- i \pa / \pa z^A $ \ in units 
where  $\hbar =  1$,  and then the Wheeler-DeWitt equation for the world 
function $\psi(z^A)$  appears as Hamiltonian constraint 
in form of  a wave equation:
\be\label{6.43} 
\left(  \Box - R(z^A)  \right) \psi   =0 \, .
\ee
After early attempts in \cite{arno},  the  Wheeler - DeWitt equation 
has often been discussed, especially for 
cosmology, see e.g. \cite{bleyerliebscher},   \cite{gibbons}, 
 \cite{grishchuk} and   \cite{halliwell}.  Besides curvature, 
matter fields can be inserted as potential, too. It
 is remarkable  that exactly for Lorentz and 
for Euclidean signatures in eq. (\ref{6.35}), i.e., positive and
 negative definite $g_{ij}$  respectively,
  the usual D'Alembert operator with $S = 1$
 in eq. (\ref{6.43})  appears. For other 
signatures in eq. (\ref{6.35}), eq. (\ref{6.43}) has at least two timelike axes.

\newpage

\section[Scalar fields and $f(R)$ for  cosmology]{Comparison of 
scalar fields and $f(R)$ for  cosmology}\label{Kap7}
\setcounter{equation}{0}
\setcounter{page}{87}

Following \cite{sch87a},
 we generalize the well-known analogies between $m^2 \phi^2$ 
 and $R + R^2$  theories to include   the self-interaction 
$ \l \phi^4$-term for the scalar field. It turns out to be the $R + R^3$
 Lagrangian which gives an appropriate model for it. Considering a spatially 
flat Friedmann cosmological model, common and different properties 
of  these models are discussed, e.g., by linearizing around 
a ground state the masses of the corresponding  spin 0-parts coincide. 
 Then  we prove   a general conformal equivalence
 theorem between a Lagrangian
$ L = L(R)$, $ L'L'' \ne  0$, and a minimally coupled scalar
 field in a general potential in section \ref{s75}.
 This theorem was independently deduced 
 by several persons, and it is now known as Bicknell theorem \cite{bicknell}.
In the final section \ref{s76}, which is based on \cite{sch88b}
we discuss Ellis' programme, on 
 which length scale the Einstein field equation is valid, 
on microscipic  or  on cosmic distances? 

\subsection{Introduction to scalar fields}

For the gravitational Lagrangian
\be\label{7.1}
     L = (R/2 + \b R^2)/8\pi G \,  ,    
\ee
where $\b$ is some free but constant  parameter, the value  
\be\label{7.1f}
R=R_{\rm  crit} = - 1/4\b 
\ee 
 is the critical value of the curvature scalar,  
cf.  Nariai \cite{nariai1973} and Schmidt  
 \cite{sch86a}, \cite{sch86b}. It is  defined by 
\be\label{7.1a}
\pa L/ \pa R = 0 \, .
\ee 
 In regions where 
\be
R/ R_{\rm  crit} < 1
\ee
 holds,  we can define 
\be
 \psi  = \ln (1 - R/R_{\rm  crit})
\ee
  and 
\be\label{7.1d}
 \tilde g_{ij}   =   \left(  1 - R/R_{\rm  crit} \right)   g_{ij} \, .
\ee
 In units where  $8\pi G = 1$ we now obtain from the Lagrangian
 eq. (\ref{7.1}) via the conformal transformation eq. (\ref{7.1d})
 the transformed Lagrangian 
\be\label{7.2}
 \tilde   L = \tilde R/2  - 3 \tilde g^{ij} \psi_{;i} \psi_{;j}/4 - \left(1
 - e^{-\psi}\right)^2 / 16 \b
\ee
being equivalent to $L$,   cf. Whitt   \cite{whitt}; see  \cite{sch86b} 
for the version of this equivalence used here.

\bigskip

For $\b < 0$, i.e., the absence of 
tachyons in $L$ eq. (\ref{7.1}), we have massive  gravitons of mass 
\be\label{7.2a}
m_0 = (-12 \b)^{-1/2}
\ee
 in $L$, cf. Stelle  \cite{stelle77}. For the weak field limit, the potential
 in  eq. (\ref{7.2}) can be simplified to be $\psi^2/ (16\cdot \b )$,
 i.e., we have got a minimally coupled scalar field whose mass 
is also $m_0$.  The superfluous factor 
3/2 in  eq. (\ref{7.2}) can be absorbed by a redefinition of $\psi$.
Therefore, it is not astonishing, that all results concerning 
the weak field limit for both $R + R^2$-gravity without 
tachyons and Einstein gravity with a minimally coupled 
massive scalar field exactly coincide. Of course, one cannot 
expect this coincidence to hold for the non-linear region, too, 
but it is interesting to observe which properties hold there also.

\bigskip

We give only one example here: we consider a cosmological 
model of the spatially flat Friedmann type, start integrating 
at the quantum boundary, which is obtained by  
$$
R_{ijkl}R^{ijkl}
$$
 on the one hand, and $T_{00}$
 on the other hand, to have Planckian values, 
 with uniformly distributed initial conditions and look 
whether or not an inflationary phase of 
the expansion appears. In both  cases we get the following result: The probability $p$
 to have sufficient inflation  is about $p = 1 - \sqrt{\l} m_0/m_{\rm Pl}$,  i.e., 
$p = 99.992 \% $ if we take $m_0 = 10^{-5} m_{\rm Pl}$ 
 from GUT and $\l = 64$, where $e^\l$ is the linear multiplication factor of 
inflation. Cf.  Belinsky  et al.  \cite{belinskyc}
for the scalar field and Schmidt   \cite{sch86b} for $R + R^2$, respectively.

\bigskip

From Quantum field theory, however, instead of the massive 
scalar field, a Higgs field with self-interaction turns out to 
he a better candidate for describing effects of  
the early universe. One of the advances of 
the latter is its possibility to describe a 
spontaneous breakdown of symmetry. In the following, we 
try to look for a purely geometric model for this Higgs field which is analogous to the above 
mentioned type where $L = R + R^2$  modelled a massive scalar field.

\subsection{The Higgs field}\label{s72}

For the massive scalar Field $\phi $ we have the mater Lagrangian 
\be\label{7.3a}
L_m = - \left(  \phi_{;i} \phi^{;i} - m^2 \phi^2
\right) /2 \, , 
\ee
and for the Higgs field to be discussed now,
\be\label{7.3b}
     L_\l = - \left(
 \phi_{;i} \phi^{;i} + \mu^2 \phi^2 - \l \phi^4/12 
\right) /2 \, . 
\ee
The ground states are defined by $\phi =$ const. and
$\pa L/\pa \phi =0$. This means  $\phi = 0$  for the scalar field, and 
the three ground states $\phi = \phi_0 = 0$, and
\be\label{7.4a}
\phi = \phi_\pm = \pm \sqrt{6 \mu^2/\l}
\ee
  for the Higgs field. 

\bigskip

The expression 
\be\label{7.3}
     \left(  \pa^2 L/ \pa \phi^2 \right)^{1/2}
\ee
represents  the effective mass at these points. This gives the
 value $m$ for the scalar field  eq. (\ref{7.3a}), so justifying the notation. 
 Further, eq. (\ref{7.3}) give mass  $\, i \, \mu$  at $\phi = 0$  and 
 $\sqrt{2} \mu $  at   $\phi = \phi_\pm$ 
 for the Higgs field  eq. (\ref{7.3b}). The imaginary value of the mass 
at the ground state $\phi = 0$ shows the instability met there, and in 
the particle picture,  this gives rise to a tachyon. 

\bigskip

To have a vanishing 
Lagrangian at the ground state     $ \phi_\pm$ eq. (\ref{7.4a})
we add  a   constant
\be\label{7.4}
\L  = -3 \mu^4/2 \l 
\ee
to the Lagrangian  eq. (\ref{7.3b}).  The final Lagrangian reads
\be\label{7.5}
L = R/2 + L_\l + \L 
\ee
 with $L_\l$  eq. (\ref{7.3b})  and $\L$  eq. (\ref{7.4}).

\subsection{The non-linear gravitational Lagrangian}

Preliminarily  we direct the attention to the 
following fact: on the one hand, for Lagrangians   (\ref{7.3a}),   (\ref{7.3b})
 and  (\ref{7.5})  the transformation $\phi \to - \phi$
 is a pure  gauge transformation, it does not change any 
invariant or geometric objects. On the other hand,
\be\label{7.6}
 R_{ijkl} \to -  R_{ijkl}
\ee
or simpler
\be\label{7.7}
R \to  -R
\ee
is a gauge transformation at the linearized level only: taking
\be
g_{ik} = \eta_{ik} +  \e h_{ik} \,   ,
\ee
where
\be
\eta_{ik} = {\rm diag} (1, -1, -  1, - 1) \, , 
\ee
then $\e \to - \e $ implies curvature inversion  eq. (\ref{7.6}) 
 To be strict  at the linearized level in $\e$.
 On the other hand, the curvature inversion eq. (\ref{7.6}), 
 and  even its simpler version eq. (\ref{7.7}), fails to  hold quadratic  in $\e$. 
This corresponds to  fact that the $\e^2$-term in   eq. (\ref{7.2}), which   
corresponds  to the $\psi^3$-term in the development
 of $\tilde L$ in powers of $\psi$,  
  is the first one to break  the $\psi \to - \psi$ symmetry in  eq. (\ref{7.2}).

\bigskip

In fact, the potential is essentially 
\be
 \left( 1 - e^{( - x)} \right)^{2} \, .
\ee
Calculating this to $Order=14$, the mapleresult reads 
\begin{eqnarray}
x^{2} - x^{3} + {\displaystyle \frac {7}{12}} \,x^{4} - 
{\displaystyle \frac {1}{4}} \,x^{5} + {\displaystyle \frac {31}{
360}} \,x^{6} - {\displaystyle \frac {1}{40}} \,x^{7} +  \nonumber \\
{\displaystyle \frac {127}{20160}} \,x^{8} - {\displaystyle 
\frac {17}{12096}} \,x^{9} + {\displaystyle \frac {73}{259200}} 
\,x^{10} - {\displaystyle \frac {31}{604800}} \,x^{11} +  \nonumber \\
  {\displaystyle \frac {2047}{239500800}} \,x^{12} - 
{\displaystyle \frac {1}{760320}} \,x^{13} + {\rm O}(x^{14})
\end{eqnarray}

\bigskip

Now, let us introduce the general non-linear Lagrangian $L = L(R)$  which 
we at the moment  only assume to be an analytical function of $R$.
 The  ground states are defined by $R =$ const., i.e.,
\be\label{7.8}
     L'R_{ik}- g_{ik} L/2   = 0    \, .
\ee
Here, $L' = \pa L/ \pa R$.

\subsubsection{Calculation of the ground states}\label{t731}

From  eq. (\ref{7.8}) one immediately sees that $\pa L/ \pa R =0 $ 
defines critical values of the curvature scalar. 
For these values $ R = R_{\rm  crit}$    it holds: 
For $ L(R_{\rm  crit})  \ne  0 $  no such ground state exists,  and for 
$ L(R_{\rm  crit}) =  0 $, we have  only one equation 
 $ R = R_{\rm  crit}$  to be solved with  10 arbitrary functions $g_{ik}$. 
 We call these ground states degenerated ones.  For $L = R^2$, 
 $ R_{\rm  crit}  = 0$, this has been discussed by Buchdahl  \cite{buchdahl}. 
 Now, let us concentrate on the case $\pa L/ \pa R  \ne  0$. 
 Then $R_{ij} $  is proportional to $g_{ij}$  with a constant 
proportionality factor, i.e., each ground 
state is an Einstein space
\be\label{7.9}
R_{ij}= R \, g_{ij} /4 \,    , 
\ee    
 with a prescribed constant value $R$. 
Inserting  eq. (\ref{7.9}) into  eq. (\ref{7.8})  we get as condition for ground states
\be\label{7.11a}
            RL' = 2L \, .
\ee
     As an example, let $L$  be a third order polynomial
\be\label{7.10}
            L = \L  + R/2 + \b R^2 + \l R^3/12 \,  . 
\ee
We consider only Lagrangians with a positive 
linear term as we wish to reestablish Einstein gravity in 
the $\L \to   0$  weak field limit, and $\b  <  0$  to exclude tachyons there.

\bigskip

We now solve eq. (\ref{7.11a}) for the Lagrangian eq. (\ref{7.10}).
For $\l = 0$ we have,  independently of $\b$,  the only 
ground state $R = -4\L$. It is a degenerated one 
if and only if   $\b \L = 1/16$. That implies that for 
usual $R + R^2$ gravity eq. (\ref{7.1}), i.e. 
$\l = \L = 0$, we get  \, $ R = 0 $ \, as the 
only ground state;  it is a non-degenerated one.

\bigskip

Now, let $\l \ne 0$ and $\L =  0$. To get non-trivial 
ground states we need the additional assumption  
 $\l  > 0$. Then, besides $R = 0$, the ground states are
\be\label{7.11}
     R=R_\pm  = \pm \sqrt{6/\l}
\ee
being quite analogous to the ground states eq. (\ref{7.4a}) of the Higgs 
field  eq. (\ref{7.3b}). The ground state $R = 0$  is  not 
degenerated. Of course, this statement  is 
independent of $\l$  and holds true, as one knows, 
for $\l = 0$.  To exclude tachyons, we require $\b < 0$, then $R_-$  is not degenerated and 
$R_+$  is degenerated if and only if   $\b  = -\sqrt{6/\l}$.
 The case $\l \L  \ne  0$  will not be considered here.

\subsubsection{Definition of the masses}\label{t732}

For the usual $R + R^2$  theory  eq. (\ref{7.1}), the mass is 
 calculated at the level of the linearized theory. Then the equivalence to the 
 Einstein field equation with a scalar field applies, and we use 
 eqs. (\ref{7.3}) and (\ref{7.1f}) to calculate 
\be
m_0 = (R_{\rm  crit}/3)^{1/2} =  (- 12 \b)^{-1/2} \, .
\ee
 Thus, we recover the value eq. (\ref{7.2a}).  But how to define the graviton's masses for the
 Lagrangian  eq. (\ref{7.10})? To give such a definition a 
profound meaning one should  do the following: 
linearize the full vacuum field equation around 
the ground state, preferably de Sitter- or anti-de Sitter spacetime, 
 respectively,  decompose its solutions with respect to 
a suitably chosen orthonormal system, which is a 
kind of higher spherical harmonics,  and look for the 
properties of its single modes. For $L$  eq. (\ref{7.1})  this procedure just gave 
 $m_0$.

\bigskip

A little less complicated way to look at this mass problem 
is to consider a spatially flat Friedmann  cosmological model 
and to calculate the frequency with which the scale factor 
oscillates around the ground state, from which the 
mass $m_0$  turned out to be the graviton's mass for $L$   eq. (\ref{7.1}), too.

\bigskip

Keep in mind, 1. that all things concerning a linearization around flat vacuous 
spacetime do not depend on the parameter $\l$ neither for the 
Higgs field nor for the $L(R)$  model, and 2. 
that a field redefinition $R \to   R^\ast + R_\pm$  is {\it  not} 
possible like   $\phi \to \phi^\ast + \phi_\pm$  because curvature remains  absolutely present.

\subsection{The cosmological model}\label{s74x}

Now we take as Lagrangian eq.   (\ref{7.10}) and as line element
\be\label{7.12}
 ds^2= dt^2 - a^2(t) (dx^2 + dy^2 + dz^2)   \, .
\ee
The dot denotes $d/dt$  and $h = \dot a/a$. We have
\be\label{7.13}
          R =  -6 \dot h - 12h^2 \,   ,
\ee
and the field equation will be obtained as follows.

\subsubsection{The field equation}

For $L = L(R)$  the variation
\be
\d \left( L \,  \sqrt{-g} \,  \right) / \d g^{ij} =0
\ee
 gives with $L' = \pa L/ \pa R $ the following fourth-order gravitational
 field equation 
\be\label{7.14}
L' R_{ij} - g_{ij} L/2 + g_{ij } \Box L' - L'_{;ij}
=0                  \, ,
\ee
cf. e.g., Novotny (1985) \cite{novotny}; see
also the ideas presented by  Kerner  \cite{kerner}.  It holds 
\be\label{7.15}
L'_{;ij} = L'' R_{;ij} + L'''\,  R_{;i} R_{;j} \, .
\ee
With  eq. (\ref{7.15}), the trace of  eq. (\ref{7.14})  reads
\be\label{7.16}
     L'R - 2L + 3L'' \Box  R + 3L'''\, R_{;k}R^{;k} = 0 \,   , 
\ee
i.e., with $L$   eq. (\ref{7.10}) 
\be
-2 \L  -R/2 + \l R^3/12 + 6 \b \Box R
 + \frac{3 \l}{2}  (R \Box R + R_{;k}R^{;k} )  =0 \,  .  
\ee
Inserting eqs.  (\ref{7.12}),  (\ref{7.13}) and  (\ref{7.15})
into the $00$-component of   eq. (\ref{7.14})  we get the equation
\bea\label{7.17}
0    = h^2/2 - \L /6 - 6 \b (2h \ddot h - \dot  h^2 + 
6h^2 \dot h ) \nonumber \\  + 3 \l (\dot h  + 2h^2)
 (6h \ddot h  + 19h^2 \dot h - 2\dot h^2 - 2h^4) \,  . 
\eea
The remaining components are a consequence of this one.

\subsubsection{The masses}
Linearizing the trace equation   (\ref{7.16})  around the flat
 spacetime, hence  $\L = 0$,  gives independently of 
$\l$  of course, $ R = 12 \b \Box R$, and the oscillations around 
the flat spacetime indeed correspond to a mass $m_0   = (-12 \b) ^{ -1/2}$. 
 This once again confirms the evaluation eq. (\ref{7.2a}). 

\bigskip

Now, let us linearize around  the ground 
states  eq. (\ref{7.11})  by inserting $\L = 0$
 and $R =\pm\sqrt{6/\l} + Z$  into  eq. (\ref{7.16}). It gives
\be\label{7.19a}
Z = \left( -6\b \mp \sqrt{ 27 \l /2} \right) \Box Z \,  ,
\ee
and, correspondingly, comparing eq. (\ref{7.19a}) with 
 the equation $(\Box \, + \,  m_\pm) Z=0$, we get 
\be\label{7.18}
     m_\pm = \left(  6 \b \pm \sqrt{ 27 \l /2} \right)^{-1/2}   \, .
\ee
For $\b \ll  - \sqrt \l$, \   $m_\pm$ 
is imaginary, and its absolute value differs by a factor $\sqrt 2$ from $m_0$. 
This is quite analogous to the $\l \phi^4$-theory, cf.  section \ref{s72}. 
 Therefore, we concentrate on discussing  this range of  parameters.

\bigskip

For the ground state for $\L  \ne  0$, $\l = 0$ we get with 
$R = -4 \L  + Z $  just $ Z = 12\b \Box  Z$, i.e., mass $m_0$ eq. (\ref{7.2a})
 just as in the case $\l  = \L = 0$.

\bigskip

Let us generalize this estimate to  $L = L(R)$; according to eq. (\ref{7.11a}), 
$R = R_0 = $ const. is a ground  state if    
\be
 L'(R_0) \,  R_0 \,  = \,  2 \, L(R_0)
\ee
  holds. It is degenerated if  $L'(R_0) = 0$. Now, we linearize around 
$R = R_0$: $R = R_0 + Z$. For $L''(R_0) = 0$, 
only $Z = 0$ solves the linearized equation, and 
 $R = R_0$  is a singular solution. For $L''(R_0)  \ne  0$ we get the mass
\be\label{7.19}
m = \Bigl( R_0/3 - L'(R_0)/3L''(R_0 ) \Bigr)^{1/2} 
\ee
meaning the absence of tachyons 
for real values $m$. Eq.   (\ref{7.19})  is the analogue to  eq. (\ref{7.3}) for the
 general Lagrangian $L(R)$.

\subsubsection{The Friedmann   model}

Here we only consider the spatially flat 
Friedmann  model  eq. (\ref{7.12}). Therefore, we can 
discuss only de Sitter  stages with $R < 0$, especially  the ground state 
 $R_+$  eq. (\ref{7.11}) representing an anti-de Sitter spacetime 
  does not enter our discussion, but $R_-$ does. 

\bigskip

Now, let $\L =  0$. Solutions of   eq. (\ref{7.17})  with 
constant values $h$ are $h=  0$ representing 
flat spacetime  and in the case that $\l > 0$ also 
\be
h = \frac{1}{\sqrt[4]{24 \l} }
\ee
representing the de Sitter spacetime. These are  the non-degenerated ground states $R = 0$ 
and $ R = R_-   = - \sqrt 6/\l$,  respectively. Eq.   (\ref{7.17})  can be written as
\bea\label{7.20}
     0 = h^2(1 - 24 \l h^4)/2 + h \ddot h \left(
 1/m_0^2 + 18 \l (\dot h + 2 h^2) \right) \nonumber \\
 - 6 \l \dot h^3 + \dot h^2 (45 \l h^2 - 1/2m_0^2 )
+ 3 h^2 \dot h (1/m_0^2 + 36 \l h^2 ) \, .
\eea
First, let us consider the singular curve 
defined by the vanishing of the coefficient of   $\ddot h$  in  eq. (\ref{7.20})  in 
the $h - \dot h$-phase plane. It is, besides $h = 0$, the curve
\be\label{7.21}
\dot h =    -2h^2 - 1/18\l m_0^2
\ee
 i.e., just the curve 
\be
R = 1/3\l m_0^2 = -4 \b /\l
\ee
 which is  defined by $L'' = 0$, cf.   eq. (\ref{7.16}). This value coincides with 
 $R_+ $ if $\b  = - \sqrt{3 \l /8}$,  this value we need  not discuss here. Points 
of the curve  eq. (\ref{7.21})  fulfil  eq. (\ref{7.20})  for
\be
h = \pm \, 1 \, / \, 18\l m_0^3 \sqrt 3 \,  \sqrt{1-1/18\l m_0^4} 
\ee
only, which is not real because of $\l \ll m_0^4$. 
Therefore, the space of solutions is composed of at least two  connected 
components. 

\bigskip

Second, for $h = 0$ we have $\dot h = 0$ or
\be\label{7.22}
\dot h = - 1/12 \l m^2   \, .
\ee
From the field equation we get: $h = \dot h =0$
 implies $h \ddot h \ge 0$,  i.e. $h$ does not 
change its sign.  We know such a behaviour  already 
from the calculations in  \cite{muell1985}, where the same model with 
 $\l = 0$ was discussed.  In a neighbourhood of  eq. (\ref{7.22}) 
 we can make the ansatz
\be
h = -t/12 \l m_0^2 + \sum_{n=2}^\infty \, a_n \, t^n
\ee
which has solutions with  arbitrary values $a_2$. This  means:  one can 
change from expansion to subsequent recontraction,  
but only through the ``eye of a needle"  eq. (\ref{7.22}). On the other hand, 
a local minimum of the scale factor 
never appears. Further,  eq. (\ref{7.22}) does not 
belong to the connected component of  flat spacetime.

\bigskip

But we are especially interested in the latter one, and therefore, we restrict to the subset 
$\dot h > \dot h$( eq. (\ref{7.21})) and need 
only to discuss expanding solutions $h \ge 0$. Inserting $\dot h = 0$,
\be
\ddot h =    h(24 \l h^4 - l)/(2/m_0^2 + 72\l h^2)
\ee
turns out, i.e., $\ddot h > 0$ for $h > 1/ \sqrt[4]{24 \l }$  only. 
All other points in the $h - \dot h$ phase plane are regular ones, and one can 
write $d \dot h / dh \equiv \ddot h / \dot h  = F(h, \dot h)$
 which can be calculated by  eq. (\ref{7.20}).

\bigskip

For a concrete discussion let $ \l \approx 10^2 l^4_{\rm Pl} $ 
 and $m_0 = 10^{-5}m _{\rm Pl} $.  Then both 
conditions  $\b \ll - \sqrt \l$  and $\vert  R_- \vert < l^{-2}_{\rm Pl}$
 are fulfilled.  Now the qualitative behaviour of  the solutions can be summarized: There 
exist two special solutions which approximate the 
ground state $R_-$ for $t \to - \infty$. All other 
solutions have a past singularity $h \to \infty$.  
Two other special solutions approximate 
the ground state $R_-$ for $t \to   + \infty$. 
Further solutions have a future  singularity $h \to \infty$, and all other solutions 
have a power-like behaviour for $t \to \infty$,
 $ a(t)$  oscillates around the  classical dust model $a(t) \sim t^{2/3}$. 
But if we restrict the initial conditions to lie in 
a small neighbourhood of the unstable ground 
state $R_-$, only one of the following three cases appears:

\noindent 
1.   Immediately one goes with increasing values $h$ to a singularity.

\noindent 
2.  As a special case:  one goes back to the de Sitter stage $R_-$.

\noindent
3.  The only interesting one: One starts with a finite 
$l_{\rm Pl}$-valued inflationary era, goes 
over to a GUT-valued second inflation and ends with a power-like 
Friedmann  behaviour.

\bigskip

In the last case to be considered here, let $\l = 0$, $\L > 0$ and $\b < 0$. 
The analogue to  eq. (\ref{7.20})   then reads
\be
0    = h^2/2 - \L /6 + (2h \ddot h - \dot  h^2 +
 6h^2 \dot h)/2m_0^2 \,  .
\ee
Here, always $h \ne 0$  holds, we consider 
only expanding solutions $h > 0$. For $\dot h = 0$ we have
\be
\ddot h =    (\L  m_0^2/3 -m_0^2 h^2)/2h \, .
\ee
For $\ddot h = 0$ we have $\dot h >  m_0^2/6$ and

\be
h = (\L /3 + \dot  h^2/m_0^2)^{1/2}  (1 + 6 \dot h/m_0^2)^{-1/2} \, .
\ee
Using the methods of   \cite{muell1985}, where the case $\L = 0$
 has been discussed,  we obtain the following result:
All solutions approach the de Sitter phase $h^2 = \L /3$
 as $t \to \infty$.   There exists one special solution  approaching 
$ \dot h = -m_0^2/6$ for $ h \to \infty $,
 and all solutions have a past singularity $h \to \infty$.
 For a sufficiently small value $\L$
 we have again two different inflationary eras  in most of all models.

\subsection{The generalized Bicknell theorem}\label{s75}

In this section we derive a general equivalence theorem between a non-linear 
Lagrangian $ L(R)$  and a minimally coupled  scalar field $\phi$ with a general 
potential with Einstein's theory. Instead of $\phi$  we take
\be\label{7.24f}
\psi = \sqrt{2/3} \  \phi\, .
\ee
This is done to avoid square roots in the exponents. 
Then the Lagrangian for the scalar field reads
\be\label{7.23}
 \tilde   L = \tilde R/2 - 3 \tilde g^{ij} \psi_{;i} \psi_{;j}/4 
 + V(\psi) \, .
\ee
At ground states $\psi = \psi_0$, defined by
 $\pa V/ \pa \psi = 0$  the effective mass is
\be\label{7.24}
          m= \sqrt{2/3}  \sqrt{ \pa^2 V/ \pa \psi^2   }     \,  ,   
\ee
cf.  eqs. (\ref{7.3}) and (\ref{7.24f}). The variation $ 0 = \d \tilde L/ \d \psi$ gives
\be\label{7.25} 
         0 =\pa V/ \pa \psi + 3 \, \tilde g^{ij} \tilde \nabla_i \tilde  \nabla_j
\,   \psi / 2
\ee
and Einstein's equation is
\be\label{7.26}
\tilde E_{ij} = \kappa \tilde T_{ij}
\ee
with
\be\label{7.27}
\kappa \tilde T_{ij} =  3 \psi_{;i}\psi_{;j} /2 +
   \tilde g_{ij} \left(  V(\psi) - \frac{3}{4} \tilde g^{ab} \psi_{;a}\psi_{;b}
\right)        \, .
\ee
Now, let
\be\label{7.28}
   \tilde g_{ij} =e^\psi  g_{ij} \, .
\ee
The conformal transformation  eq. (\ref{7.28}) shall 
be inserted into eqs.  (\ref{7.25}),  (\ref{7.26}) and  eq. (\ref{7.27}).
 One obtains from  eq. (\ref{7.25})  with 
\bea\label{7.29}
\psi^{;k} := g^{ik} \psi_{;i}       \nonumber \\
\Box \psi + \psi^{;k}\psi_{;k} = - 2 (e^\psi \pa V/\pa \psi )/3
\eea
and from   eqs. (\ref{7.26}),  (\ref{7.27})
\be\label{7.30}
E_{ij} = \psi_{; ij} + \psi_{;i}\psi_{;j} +g_{ij} 
\left(  e^\psi V(\psi) - \Box \psi   - \psi_{;a} \psi^{;a}  \right) \, .
\ee
Its trace reads
\be\label{7.31}
        -R = 4 e^\psi  V(\psi) - 3 \Box \psi   - 3 \psi_{;a} \psi^{;a} \, .
\ee
Comparing with  eq. (\ref{7.29}) one obtains
\be\label{7.32}
        R = R(\psi) =   - 2e^{-\psi} \pa \left( e^{2\psi} V(\psi) \right) /
 \pa \psi \, .
\ee
Now, let us presume $\pa R/\pa \psi  \ne  0$,
 then  eq. (\ref{7.32}) can be inverted as
\be\label{7.33}
 \psi = F(R) \,   . 
\ee 
In the last step,  eq. (\ref{7.33}) shall be inserted into eqs. 
   (\ref{7.29}),  (\ref{7.30}),  (\ref{7.31}).  Because of
\be
F(R)_{; ij}
= \pa F/ \pa R \cdot  R_{; ij} + \pa^2 F/ \pa R^2 \cdot R_{; i}R_{;j}
\ee  
and $\pa F/\pa R  \ne  0$,   eq. (\ref{7.30})   is a fourth-order 
equation for the metric $g_{ ij}$.   We 
try  to find a Lagrangian    $L = L(R)$ such 
that the equation $ \d L \sqrt{-g} / \d g^{ij} =  0$ 
becomes just  eq. (\ref{7.30}). For $L' = \pa L/\pa R  \ne  0$,
  eq. (\ref{7.14})  can be solved to be
\be\label{7.34}
E_{ij} = - g_{ij}R/2 + g_{ij}L/2 L'  - g_{ij}
\Box L'/L'    - L'_{;ij}/L' \, . 
\ee
We compare     the coefficients of the $R_{;ij}$  terms  
in eqs.   (\ref{7.30})  and   (\ref{7.34}), this gives 
\bea
\pa F/\pa R = L''/L' \,  , \qquad {\rm  hence} \nonumber   \\
L(R) = \mu \int_{R_0}^R e^{F(x)} dx + \L_0   \label{7.35}
\eea 
with suitable constants $\L_0$, $\mu$, and $R_0$, $\mu  \ne  0$.
 We fix them as follows: We are interested in a neighbourhood of 
$R = R_0$   and require $L'(R_0) = 1/2$.  Otherwise $L$
 should be multiplied by a constant factor.  Further, a constant translation
of $\psi$  can be used to obtain $F(R_0) = 0$,
 hence $\mu = 1/2$, $L(R_0) = \L_0$, and
\be
L' (R_0)  = \pa F/\pa R(R_0)/2  \ne  0\,  .
\ee
With  eq. (\ref{7.35}) being fulfilled, the traceless 
parts of eqs.   (\ref{7.30}) and   (\ref{7.35})  
 identically coincide. Furthermore, we have
\be
\Box L'/L' = \Box F + F^{;i} F_{;i}
\ee
and it suffices to test the validity of the relation
\be
e^F \,  V(F(R)) = -R/2 + L/2L' \, . 
\ee
It holds
\bea
2L' = e^F\,  , \qquad {\rm i.e.,} \nonumber \\
 e^{2F} V(F(R)) = L - R e^F/2 \, . \label{7.36}
\eea
At $R = R_0$, this relation reads $V(0) = \L_0 - R_0/2$.  
Applying $\pa /\pa R$  to  eq. (\ref{7.36}) gives 
just  eq. (\ref{7.29}), and, by  the way, $V'(0) = R_0/2 -    2\L_0$. In sum,
\be
L(R) = V(0) + R_0/2 +   \int_{R_0}^R e^{F(x)} dx/2 \, ,
\ee
where $F(x)$ is defined via $F(R_0) = 0$,
\be
\psi   = F\left( -2 e^{-\psi}  \pa(e^{2\psi} V(\psi))   / \pa \psi 
 \right) \, .
\ee

\bigskip

Now, let us go the other direction:  Let $L = L(R)$ be given 
such that  at $R = R_0$,  $ L'L''  \ne  0$. By a constant change of   $L$ let 
$L'(R_0) =     1/2$. Define $\L_0 = L(R_0)$, $\psi = F(R) = \ln (2L'(R))$  and 
consider the inverted function $R = F^{-1}(\psi)$.  Then
\be\label{7.37}
V(\psi) = (\L_0 - R_0/2) e^{-2\psi}
- e^{-2\psi} \int_0^\psi   e^x \ F^{-1}(x) dx/2
\ee
is the potential ensuring the  above mentioned conformal equivalence. 
This procedure is possible at all $R$-intervals 
where $L' \, L''  \ne  0$ holds. For analytical 
functions $L(R)$, this inequality can be violated 
for discrete values $R$  only, or one has simply 
a linear function $L(R)$ being Einstein gravity with $\L$-term.

\bigskip

Eq. (\ref{7.37}) is  given here  in  the form published in \cite{sch87a}. Later it
turned out that this integral can be evaluated in closed form as follows: 
\be
V(\psi) = L \bigl( F^{-1}(\psi) \bigr) \cdot e^{-2\psi} 
- \frac{1}{2} \,  F^{-1}(\psi)   \cdot e^{-\psi}    \, .
\ee
\bigskip

Examples: \  1. Let $L = \L + R^2$, $R_0 = 1/4$,   then $4R = e^\psi$  and
\be\label{7.38}
V(\psi ) = \L  e^{-2\psi} - 1/16\, .
\ee
For $\L = 0$, this is proven in Bicknell  \cite{bicknell}
and Starobinsky  and Schmidt   \cite{starosch}.

\bigskip

2.   Let $L = \L + R/2 + \b R^2 + \l  R^3/12$, \  $R_0=0$, hence 
$\b  \ne  0$ is necessary. We get
\bea
e^\psi -    1 = 4 \b R + \l R^2/2 \qquad {\rm   and} \nonumber \\
V(\psi) = \L e^{-2\psi} +  \nonumber \\ 
 2\b \l^{-1} e^{-2\psi}  \left( e^\psi - 1 - 
16\b^2(3\l)^{-1}
 ((1 + \l(e^\psi - l)/8\b^2)^{3/2} - 1) \right) \, . \label{7.39}
\eea
The limit $\l \to  0$  in  eq. (\ref{7.39})  is  possible and leads to
\be
V(\psi) = \L e^{-2\psi} -  (1 -  e^{-\psi})^2/16\b \, , 
\ee
so we get for $\L =0$ again the potential from   eq. (\ref{7.2}).

\bigskip

Now, let $R_0$ be a non-degenerated ground state, hence
\be
L(R) = \L_0 + (R - R_0)/2 + L''(R_0) (R - R_0)^2/2 +\dots 
\ee
with $L''(R_0)     \ne  0$  and $\L_0 = R_0/4$, 
cf. subsection \ref{t731}. Using  eq. (\ref{7.37})  we get $V'(0) = 0$  and 
\be
V''(0) = R_0/2 - 1 \, / \bigl(  4L''(R_0) \bigr) \, .
\ee
Inserting this into  eq. (\ref{7.24})  
we exactly reproduce  eq. (\ref{7.19}). 
This fact once again confirms  the estimate   eq. (\ref{7.19})  and, moreover, shows 
it to be a true analogue to  eq. (\ref{7.3}).  To understand this coincidence 
one should note that at ground states,  the conformal factor becomes a constant = 1. 

\subsection{On Ellis' programme within homogeneous world models}\label{s76}

For the non-tachyonic curvature squared action we show that the 
expanding Bianchi-type I models tend to the dust-filled 
Einstein-de Sitter model for $t$ tending to infinity if 
the metric is averaged over the typical oscillation period. 
Applying a conformal equivalence between curvature squared 
action and a minimally coupled scalar field, which holds for all 
dimensions $>$ 2,  the problem is solved by discussing 
a massive scalar field in an anisotropic cosmological model.

\subsubsection{Ellis' programme}\label{t761}

Ellis   \cite{ellis} has asked in 1984 on
 which length scale the Einstein field equation is valid. 
On laboratory or on cosmic distances? Here 
we extend this question to cover microscopic 
scales, too. This is a real problem as one knows:

\noindent
1.   An averaging procedure does not commute with a non-linear 
differential operator as the Einstein tensor is.

\noindent
2.   Einstein's theory is well tested at large distances $\gg$ 1 cm.

\noindent 
3. The ultraviolet divergencies of Einstein gravity 
can be removed by adding curvature squared terms 
to the action, see Weinberg   \cite{wein}; this is a microscopic 
phenomenon and forces  to prefer a curvature squared action 
already on the level of classical field theory, as we are concerned here.

\bigskip

Now, we propose a synthesis of 1., 2., and 3. as follows: microscopically, we take
\be\label{8.1}
     L_{\rm g} = (R/2 - l^2 R^2)/8\pi G \, 
\ee
the minus sign before the $R^2$-term shows that we
 consider the  non-tachyonic case only. By an averaging procedure 
we get Einstein gravity on large scales $ \gg l \approx   10^{-28}$  cm. 
As we are dealing with homogeneous cosmological models, the average is taken 
over the time coordinate only: We will show: 
 the curvature squared contribution represents effectively dust in 
the asymptotic region $t \to \infty$.  For spatially flat Friedmann 
models  this was already proven in M\"uller 
 and Schmidt   \cite{muell1985}, here we generalize to models 
with less symmetry. In the present paper we consider only the 
vacuum case
\be\label{7.41}
\delta L_{\rm g} \sqrt{- \det g_{ij}} \, 
 \delta g^{kl}   = 0 \,   ,
\ee
 and  therefore, we interpret the effectively obtained dust  
as invisible gravitating matter necessary to get a spatially flat universe.

\bigskip

We conjecture that additional matter contributions, that means usual 
dust plus radiation,  do not alter the result qualitatively. 
The general expectation for Lagrangian eq. (\ref{8.1}) 
is the following: starting at die Planck era
\be\label{7.42}
R_{ijkl}R^{ijkl}  \approx  10^{131} \, {\rm  cm}^{-4} \, ,
\ee
the $R^2$-term is dominant,  the inflationary de 
Sitter phase is an attractor; its appearance becomes very 
probable, see  \cite{sch86b} and  \cite{starosch}; 
the $R$-term in  eq. (\ref{8.1}) yields a 
parametric decay of the value of $R$ and one turns over to the
 region $t \to \infty$ where $R^2$  gives just dust in the mean.

\bigskip

In Carfora and Marzuori   \cite{carfora}
another approach to Ellis' programme was initiated: a 
smoothing out of die spatially closed 3-geometry in the
  direction of die spatial Ricci-tensor leads to different values of mean mass density 
before and after the smoothing out procedure.

\subsubsection{The massive scalar field 
in a Bianchi-type I  model}\label{t762}

Consider  a minimally coupled scalar 
field $\phi$  in a potential $V(\phi)$, where $8\pi G = 1$ is assumed, 
\be\label{7.43}
 L=R/2-   \frac{1}{2} \phi_{;i} \phi^{;i}     + V(\phi)    \, .
\ee
We suppose  $V(\phi)$  to be a $C^3$-function and 
$V(0) = 0$ to be   a local quadratic minimum of the potential $V$,
 i.e., $V'(0) = 0, V''(0) =m^2$, $m > 0$. To describe the 
 asymptotic behaviour $\phi \to  0$ as $t \to \infty$
 it suffices to use  $V(\phi) =  m^2 \phi^2/2 $ because the higher
order terms do not affect it.  This statement can be proved  as follows: 
For each $\e > 0$, $\e < m$, there exists a 
$\phi_0 > 0$ such that for all $\phi $ with $\vert \phi \vert < \phi_0 $ it holds
\be\label{7.44}
(m - \e)^2 \phi^2 \le 2V(\phi) \le  (m + \e)^2 \phi^2 \, .
 \ee
Then all further development is enclosed by inequalities with 
$m \pm \e$, $\e \to  0$.

\bigskip

Here, we concentrate on a cosmological model of Bianchi-type I. 
In Misner parametrization it can be written as
\be\label{8.2}
ds^2=   dt^2 - e^{2\a } [ e^{2(s+\sqrt{3}r)} dx^2
 + e^{2(s- \sqrt{3}r)} dy^2 + e^{-4s} dz^2]    \, .
\ee
The functions $\phi$, $\a$, $r$, $s$  depend on $t$ only. 
$\dot \a = d\a /dt = h$  is the Hubble parameter and $\eta = u/h$, 
$ u =   (\dot r^2 + \dot s^2)^{1/2}$  the anisotropy parameters. 
It holds $0 \le \eta \le 1$, $\eta = 0$
 represents the isotropic model, and $\eta = 1$
 gives $\phi \equiv 0$, therefore, we consider only 
the case $0 < \eta < 1$ in the following. $h = 0$
 is possible for Minkowski spacetime only, and 
the time arrow is defined by $h> 0$,  i.e., we restrict to expanding solutions. 

\bigskip
 
In Lukash and  Schmidt  \cite{lukash} it was shown that all such solutions 
tend to Minkowski  spacetime for $t \to \infty$.
 More detailed: At $t = 0$, let $\a = r = s = 0$ by a coordinate transformation. 
Then for each prescribed quadruple
 $ (\dot r, \, \dot s, \, \phi, \, \dot \phi)$
  at $t = 0$  the integration of the relevant system
   up to $t \to \infty$  is 
possible with $r$, $s$, $\phi$, $\dot \phi$, $h$, $\dot h$
  tending  to zero in that limit. The relevant equations are
\bea\label{8.3}
     (\phi^2  + \dot \phi^2 + u^2)^{1/2} = h \\
\ddot \phi  +  3h \dot \phi + \phi = 0 \label{8.4}       \\
\dot r = C_r e^{-3\a} \, , \qquad    \dot s  = C_s e^{-3\a} \, . \label{8.5}
\eea
We put the mass $m = 1$; $C_r$, $C_s$  are constants. Now, we consider 
this limit in more details. Up to now the 
following is known: For the isotropic 
models $u \equiv 0$, $e^\a = a$ one knows that the 
asymptotic behaviour is given by oscillations around 
$a \sim  t^{2/3}$  i.e., we get the Einstein-de Sitter model 
in the mean, and the effective equation of 
state is that of dust, cf. e.g. Starobinsky  \cite{staro78}. 
In Gottl\"ober 1977  \cite{gott1984a} this  is generalized to a special class
 of inhomogeneous, nearly isotropic models.
After averaging over
 space and oscillation period one gets as effective equation of state
\be\label{8.6} 
p \sim \rho/a^2
\ee
i.e., also dust in the limit $t \to \infty$. 

\bigskip

The question for the Bianchi-type I model 
is now: does one get the energy-momentum tensor 
of an ideal fluid in the mean, 
or will there appear a strongly anisotropic pressure? 
The first question to be  answered is about  the averaging procedure:
From  eq. (\ref{8.4})  and the fact that $h \to  0$  as $t \to \infty$  one 
gets a fixed oscillation period: let $t = t_n$ be  the $n$-th
 local extremum of $\phi(t)$, then
\be\label{7.45}
\lim_{n \to \infty} t_{n+1} - t_n = \pi   \, .
\ee
Therefore, we average the metric about this period of time. 
Let us denote $r(t_n)$  by $r_n$, \dots \, 
 For a fixed finite time it is ambiguous how to perform this average, 
but in the limit $t  \to \infty$  one can describe the system by adiabatic 
or parametric deviations from pure $\phi \sim   \cos t$-oscillations, and 
the averaging procedure consists of 
constructing   monotonous smooth curves $\bar \a (t)$
 fulfilling $\bar \a(t_n) = \a_n$, \dots ,  and the effective 
energy-momentum tensor is obtained 
from  eq. (\ref{8.1}) with $\a = \bar \a (t)$, \dots  using the Einstein equation.

\bigskip

Derivating  eq. (\ref{8.3})  one gets with   eq. (\ref{8.4}) 
\be\label{8.7}
0 \ge \dot  h = 3\phi^2 - 3h^2 \ge  -3h^2 
\ee
and therefore,
\be\label{7.46}
h_n \ge h_{n+1} > h_n  - 3\pi h_n^2 \, .
\ee
Let 
\be\label{7.47}
h   \ge \frac{1}{3\pi (n+n_0)} 
\ee
be valid for one value $n = n_1$  then, by induction, 
 this holds true also for all larger values $n$, i.e.,
\be\label{8.8}
h(t) \ge \frac{1}{3(t+t_0)}
\ee
for some $t_0$,  let $t_0 = 0$ subsequently. 
Derivating   eq. (\ref{8.5})  we get with   eq. (\ref{8.3})  
\be\label{8.9} 
\dot \eta 
= - 3 \eta (1- \eta^2 ) h \cdot 
\frac{\phi^2}{\phi^2 + \dot \phi^2} \  .
\ee
The last factor can be substituted by its mean value, which equals $1/2$.
 From  eq. (\ref{8.8})  one
 can see that for initial conditions $0 < \eta < 1$
as we met here,  eq. (\ref{8.9})  leads to $ \eta \to 0$  as $t \to \infty$.
 Knowing this, we can perform a stronger estimate for $h$, 
because in the mean,  $\dot h = -3h^2/2$, i.e., in the leading order 
we have $h(t) = 2/3t$  and, using   eq. (\ref{8.9}), we get
\be\label{7.48}
\dot \eta =- \eta/t \, , \qquad    \eta = \eta_0/t \, .
\ee
Remember: the inflationary era diminishes $\eta$  exponentially with
increasing time, but  
here we consider only the asymptotic region. In sum  we get for metric 
 (\ref{8.2}) 
\be\label{8.10}
e^\a  = t^{2/3}\, , \qquad r=-  C_r/t \,  ,  \qquad    s = -C_s/t \,   . 
\ee
Inserting   eq. (\ref{8.10})  into  eq. (\ref{8.2})  we get, via the 
Einstein equation, a diagonal energy-momentum tensor 
with $\rho  = T_{00} = 4/3t^2$  and pressure of the 
order $ (C_r^2 + C_s^2)/t^4$. The effective equation of 
state is $p \sim \rho/t^2 \sim \rho/a^3$, i.e., also dust in the 
limit $t \to \infty$  as  eq. (\ref{8.6}). From the details of the 
paper   \cite{lukash} one can see that for almost all models the end of the 
inflationary stage is just the beginning of the Einstein-de Sitter stage.
 A similar result occurs if not the metric, 
but $\rho$ and $p$ will be averaged, cf.  Gottl\"ober  \cite{gott1984a}.

\bigskip

Result: For the minimally coupled scalar field in a potential $V$ which has  a  single 
quadratic minimum at $V = 0$  all expanding Bianchi-type I 
cosmological solutions tend to the Einstein-de Sitter model 
for $t \to \infty$  if the metric is averaged over the oscillation period.

\subsubsection{The generalized equivalence}\label{t763}

Some types of  a conformal equivalence  theorem between fourth-order gravity 
and minimally coupled scalar fields are obtained 
in  \cite{sch87a},  \cite{sch88a}. This theorem was also independently obtained 
by Ferraris   \cite{ferraris}, see also Jakubiec  
and Kijowski  \cite{jakubiec1988},  
 Goenner  \cite{goenner}. All of them are 
restricted to 4-dimensional spacetimes. On the other hand, 
both $R^2$-terms and scalar fields have been discussed 
for higher-dimensional spacetimes, cf.  e.g. Ishihara  \cite{ishihara}. 

\bigskip
 
Therefore, it is worth mentioning that this conformal equivalence 
theorem can be formulated for arbitrary dimensions $n > 2$: Let
\be\label{8.11}
\tilde{\cal L}
 = \tilde  R/2-   \frac{1}{2}\tilde g^{ij} \phi_{\vert i}
 \phi_{\vert j}     + V(\phi)    
\ee
and
\be\label{8.12}
\tilde g_{ij}= e^{\l \phi} g_{ij} \, , \qquad \l = \frac{  2}{ (n-1)(n-2)}
\ee
be the conformally transformed metric. Then 
the solutions of the variation of   eq. (\ref{8.11})  are transformed 
by  eq. (\ref{8.12})  to the solutions of the variation of $ L = L(R)$, where
\be\label{8.13}
 R = - 2e ^{\l \phi} \left(  \frac{nV}{n-2} + \mu \frac{dV}{d \phi}
\right) \, , \qquad  \mu = \sqrt{\frac{n-1}{n-2}}
\ee
is supposed to  be locally, i.e. near  $R = R_0$,   invertible as
\bea
 \phi=F(R) \, ,      \qquad F(R_0)=0 \, , \qquad F'(R_0) \ne 0   \nonumber \\
{\cal L}(R) = \frac{1}{2} R_0 + V(0) + 
 \frac{1}{2} \int_{R_0}^R e^{F(x)/\mu}  dx    \, . \label{8.14}
\eea
The inverse direction is possible  provided 
 $L'(R) \cdot  L''(R) \ne 0$, cf.    \cite{sch87a},  \cite{sch88a} for details 
with $n = 4$. Other cosmological models with higher dimensions are in 
[32], [119], and [190]. Two-dimensional models are discussed in [154] and [229]. 

\subsubsection{The fourth-order gravity  model}\label{t764}

Now we come to the question posed  in the 
introduction: Let $L(R)$  be a $C^3$-function 
fulfilling $L(0) = 0$, $L'(0) \cdot L''(0) < 0$. Then we can write
\be\label{8.15}
{ \cal L} (R) = \frac{R}{2 } + \b R^2 + O(R^3) \,  ,   \qquad \b <0 \,  .  
\ee
We consider the Bianchi-type I vacuum 
solutions which start in a neighbourhood of the  Minkowski spacetime and ask for the 
behaviour as $t   \to \infty $. Applying the
 equivalence theorem cited in subsection \ref{t763} we arrive at the
 models discussed in subsection \ref{t762}, and this is applicable for
$\vert \, R \,  \vert $  being small enough. The conformal factor 
depends on $t$ only, and therefore, the space of Bianchi-type I models
 will not be leaved, and we can formulate the 
following: In a neighbourhood of Minkowski spacetime, all 
Bianchi-type I  models which represent a stationary point
 of the action eq. (\ref{8.15}), can be integrated 
up to $t   \to \infty $ or  $-  \infty $, let it be $ + \infty $. One singular solution 
is the Kasner solution and all other solutions undergo 
isotropization and have an averaged equation of state $p = 0$ for $t  \to \infty $.

\bigskip

Let us conclude: Anderson  \cite{anderson} discussed the 
possibility that curvature squared terms  give an effective contribution to the 
energy density  of a Friedmann model.  This could explain the discrepancy between the 
observed mean mass density   of about  1/10 the critical one and the predicted one,
from inflationary cosmology,  nearly the critical one. Here, we have shown that also 
for a large initial anisotropy the oscillating curvature
 squared contributions give just dust in the mean 
and not an equally large anisotropic pressure as one could have expected. The next 
step would be to look for a generalization 
of this fact to inhomogeneous cosmological models.
 Further fourth-order gravity models can be seen in [11], [48], [49], [61],
 [74], [110], [111], [115], [124], [128], [139], [193], [214] and [246]. 
 Another approach to averaging procedures, i.e., to Ellis' programme,
 is due to Zalaletdinov [275].

\newpage

\section[Models with De Sitter and power--law inflation]{Friedmann
 models with de Sitter \\ and  power--law inflation}\label{Kap9}
\setcounter{equation}{0}
\setcounter{page}{113}

Following \cite{sch90b}, we consider the spatially flat Friedmann model
\be
ds^2=dt^2  - a^2(t) (dx^2+dy^2+dz^2) \, .
\ee
For $a \approx   t^p$, especially, if 
$p \ge 1$, this is called power-law inflation. 
For the Lagrangian 
\be\label{9.3}
L = R^m \qquad {\rm  with} \qquad 
 p=   - (m - 1) (2m - 1)/(m - 2)
\ee
 power-law inflation is an exact solution, as it
 is for Einstein gravity with a 
minimally coupled scalar field $\Phi$ in an exponential potential 
\be
V(\Phi) = \exp (\mu \Phi) 
\ee
 and also for the higher-dimensional Einstein 
equation with a special Kaluza-Klein ansatz. The synchronized coordinates 
are not adapted to allow a closed-form solution, so we write
\be
ds^2=a^2 \Bigl( Q^2 ( a) \,  da^2 - dx^2 -dy^2-dz^2 \Bigr) \ .
\ee
The general solutions reads 
\be
Q(a) =   (a^b + C)^{f/b}
\ee
with free integration  constant $C$, where $C = 0$ gives exact power-law inflation,
 and $m$-dependent values $b$
and $f$: 
\be
f = -2 +1/p \, , \qquad b =   (4m - 5)/(m   - 1) \, .
\ee 
Further, special solutions for the closed and open Friedmann model are found.

\bigskip

The de Sitter spacetime
\be\label{9.1}
ds^2=dt^2  - e^{2Ht} (dx^2+dy^2+dz^2) \, , \qquad H \ne 0
\ee
is the spacetime being mainly discussed to represent 
the inflationary phase of cosmic evolution. 
However,  a spacetime defined by
\be\label{9.2}
ds^2=dt^2  - \vert t \vert ^{2p} (dx^2+dy^2+dz^2) \, , \qquad p \ne 0
\ee
enjoys increasing interest for these discussions, too. 
Especially, eq. (\ref{9.2}) with $p \ge 1$, $ t > 0$ is called 
power-law inflation; and with $p < 0$, $t < 0$ it is called polar inflation.

\bigskip

We summarize some differential-geometrical properties 
of both de Sitter and power-law/polar inflation 
in section \ref{s92} and show, from which kind of 
scale-invariant field equations they arise in section \ref{s93}; 
we give the complete set of solutions for 
the spatially flat and special solutions for the closed and open Friedmann 
models in closed form for field equations following from the Lagrangian
$ L = R^m $
in section \ref{s94}, and discuss the results in  section \ref{s95}  under the 
point of view that power-law inflation is an attractor solution of the 
corresponding field equations.

\subsection{Differential--geometrical properties}\label{s92}

Eq.   (\ref{9.2}) defines a self-similar spacetime: if we 
multiply the metric $ds^2$ by an arbitrary positive 
constant $a^2$, then the resulting  $d \hat s^2 = a^2  ds^2$ 
 is isometric to $ds^2$. 
Proof: We perform a coordinate transformation 
$\hat t = at$, $\hat  x = b(a, p) x$ \dots 
On the other hand, the de Sitter spacetime  eq. (\ref{9.1})  is not 
self-similar, because it has a constant non-vanishing curvature scalar.  

\bigskip

Power-law inflation is intrinsically time-oriented. 
Proof: The gradient of the curvature scalar defines a temporal 
orientation. \ 
On the other hand, the expanding ($H > 0$) and the contracting ($H < 0$) 
de Sitter spacetime can be transformed into each other by a 
coordinate transformation, because both of them can be transformed 
to the closed Friedmann universe with scale factor $\cosh(Ht)$,
 which is an even function of $t$. \ 
This property  is connected with the fact 
that  eq. (\ref{9.2}) gives a global description, whereas 
 eq. (\ref{9.1}) gives only a  proper subset of the full de Sitter spacetime.

\bigskip

For $p \to \infty$,  eq. (\ref{9.2}) tends to   eq. (\ref{9.1}). Such a statement has 
to be taken with care, even for the case with real functions. Even more 
carefully one has to deal with spacetimes. The most often used 
limit ---  the Geroch-limit \cite{geroch} of spacetimes --- has the property that a 
symmetry, here we take it to be self-similarity, of all the elements of the sequence must 
also be a symmetry of the limit.

\bigskip

From this it follows that the Geroch limit 
of spacetimes  eq. (\ref{9.2})  with $p \to \infty$  cannot be unique, 
moreover, it is just the one-parameter set   eq. (\ref{9.1})  
 parametrized by  arbitrary values $H > 0$.

\subsection{Scale--invariant field equations}\label{s93}

A gravitational field equation is called scale-invariant, if 
 to each solution $ds^2$ and to each positive constant $c^2$  
the resulting homothetically equivalent metric $d \hat s^2 = c^2  ds^2$
is also a solution. 
One example of such field equations is that one following 
from  eq. (\ref{9.3}). 
Moreover, no Lagrangian $L = L(R)$ gives
 rise to a scale-invariant field 
equation which is not yet covered by  eq. (\ref{9.3})  already.

\bigskip

Secondly, for
\be\label{9.4}
     L = R/16 \pi G -    \frac{1}{2} g^{ij} \Phi_{, i}  \Phi_{, j} 
 + V_0 \exp(\mu \Phi) \, , 
\ee
we assume  $8 \pi G = 1$ henceforth,  the homothetic transformation has 
to be accompanied
 by a suitable translation of $\Phi$ to ensure scale-invariance. 
For $\mu \ne  0$, the value of $V_0$ can be normalized  to 1, 0 or $- 1$.

\bigskip

A third example is the following: for the Kaluza-Klein ansatz   eq. (\ref{9.5})  
the $N$-dimensional Einstein equation $R_{AB} = 0$
 is scale-invariant. We take the ansatz as 
\be\label{9.5}
 dS^2     = ds^2(x^i) + W(x^i) d\tau^2(x^\alpha) \, ,  
\ee
 where $i,j = 0, \dots 3$; $\a, \b = 4, \dots N- 1$;
 $A, B = 0 \dots N- 1$, and  
restrict to the warped product of the $(N - 4)$-dimensional internal 
space $d\tau^2$ with  
 4--dimensional spacetime $ds^2$.

\bigskip

Let us consider the limits $m \to \infty$  and 
$m \to 0$ of   eq. (\ref{9.3}). One gets 
$L = \exp (R/R_0)$  and $ L = \ln (R/R_0)$, respectively. 
Both of them give rise to a field equation which is 
not scale-invariant: a homothetic transformation 
changes also the reference value $R_0$.  For the second case 
this means a change of the 
cosmological constant. So we have a similar result as before: the limits 
exist, but they are not unique.

\subsection{Cosmological  Friedmann models}\label{s94}

We consider the closed  and open model in subsection \ref{t941}. 
and the spatially flat model in subsection  \ref{t942}. 

\subsubsection{The closed and open models}\label{t941}

Let us start with the closed model. We restrict 
ourselves to a region where one has  expansion, so we may use the cosmic 
scale factor as timelike coordinate:
\be\label{9.6}
ds^2=a^2 \left( Q^2 ( a) \,  da^2 - d\sigma^2 \right) \, ,  \qquad Q(a) > 0\, ,
\ee
where $d\sigma^2$, defined by 
\be\label{9.7}
d\sigma^2  = dr^2 + \sin^2 r d \Omega^2 \, , \qquad
    d\Omega^2 = d\theta^2 + \sin^2 \theta  d\psi^2 
\ee
is the positively curved 3-space of constant curvature. 
It holds: if the 00-component of  the field equation, here:   eq. (\ref{9.10})  below,
 is fulfilled, then all other components 
are fulfilled, too. Such a statement holds true for all 
Friedmann models and ``almost all sensible field 
theories". With ansatz   eqs. (\ref{9.6}),  (\ref{9.7})   we get via
\be\label{9.8}
R^0_0  = 3(a^{-4} Q^{-2} + a^{-3} Q^{-3} dQ/da)  
\ee
and
\be\label{9.9}
R = 6(a^{-3}Q^{-3}dQ/da - a^{-2}) 
\ee
the result: the field equation following from the 
Lagrangian   (\ref{9.3})   is fulfilled for metric 
 (\ref{9.6}),   (\ref{9.7}), if and only if
\be\label{9.10}
     mR  R^0_0 -R^2/2     + 3m(m -  1) Q^{-2} a^{-3} dR/da = 0  
\ee
holds. For $m = 1$, which represents Einstein's theory, no solution exists. 
For all other  values $m$, eqs.  (\ref{9.8}),  (\ref{9.9}) and  (\ref{9.10})  
 lead to a second order equation for $Q(a)$.

\bigskip

We look now for solutions with vanishing $R^{\, 0}_0$\, , 
 i.e., with   eq. (\ref{9.8})   we get
\be\label{9.11}
     Q= C/a \,  , \qquad   C= {\rm  const.} >0 \, . 
\ee
By the way, metric  (\ref{9.6}),  (\ref{9.7})   is self-similar if and only 
if    eq. (\ref{9.11})   holds. From  eqs. (\ref{9.9}),  (\ref{9.11})  we get
\be\label{9.12}
     R=D/a^2\,  , \qquad     D=-6(l +1/C^2) \, .   
\ee
We insert   eqs. (\ref{9.11}),  (\ref{9.12}) into   eq. (\ref{9.10})  and get
\be\label{9.13}
C =  (2m^2 - 2m - 1)^{1/2}   \, ,
\ee
which fulfils    eq. (\ref{9.11})  if
\be\label{9.14} 
     m >(1 + \sqrt 3)/2 \quad {\rm  or} \quad     m < (1 - \sqrt 3)/2
\ee
holds. Inserting    eq. (\ref{9.11}) and
   eq. (\ref{9.13})   into   eq. (\ref{9.6})  and introducing synchronized 
coordinates we get as a result: if    eq. (\ref{9.14})  holds, then 
\be\label{9.15}
ds^2 = dt^2  - t^2 d\sigma^2 /  (2m^2 - 2m - 1)
\ee
is a solution of the fourth-order field equation 
following from Lagrangian   (\ref{9.3}). It is a 
self-similar solution, and no other self--similar solution 
describing a closed Friedmann model exists.

\bigskip

For the open Friedmann model all things are 
analogous, one gets for
\be
 (1 - \sqrt 3)/2 < m < (1 + \sqrt 3)/2
\ee
and with sinh $r$ instead of sin $r$ in   eq. (\ref{9.7})  the only self-similar open 
solution, which is flat for $m \in \{0, 1\}$,
\be
ds^2 = dt^2  - t^2 d\sigma^2 /  (- 2m^2 + 2m + 1) \, .
\ee

\subsubsection{The spatially flat model}\label{t942}

The field equation for the spatially flat model 
can be deduced from that one of a closed model 
by a limiting procedure as follows: we insert the transformation $r \to \e r$, 
$  a \to   a/\e$, $  Q \to  Q\e^2$   and 
apply the limit $\e \to  0$ afterwards. One gets via
\be
\lim_{\e \to 0} \  \sin (\e r) = r  
\ee
the metric
\be\label{9.16}
ds^2=a^2 \left( Q^2 ( a) da^2 - dx^2 -dy^2-dz^2\right)
\ee
with unchanged eqs.   (\ref{9.8}) and  (\ref{9.10}), whereas   eq. (\ref{9.9})  yields
\be\label{9.17}
     R = 6a^{-3}Q^{-3} dQ/da \, .  
\ee
The trivial solutions are the flat Minkowski spacetime 
and the model with constant value 
of $Q$, i.e., $R = 0$, a valid solution for $m > 1$ only, 
which has the geometry of Friedmann's  radiation model.

\bigskip

Now, we consider only regions with non-vanishing curvature 
scalar. For the next step we apply the fact that the spatially flat
 model has one symmetry more than the closed one: 
the spatial part of the metric is self-similar. In the coordinates 
  eq. (\ref{9.16})  this means that each 
solution $Q(a)$  may be multiplied by an arbitrary constant. 
To cancel this arbitrariness, we define a new function
\be\label{9.18} 
P(a) =  d(\ln  Q)/da \,  .
\ee
We insert   eqs. (\ref{9.8}), (\ref{9.17})  into   eq. (\ref{9.10}),
 and then   eq. (\ref{9.18})  into the resulting 
second order equation for $Q$. We get the first order equation for $P$
\be\label{9.19}
0    = m(m - 1) dP/da + (m - 1) (1 - 2m) P^2 + m(4 - 3m) P/a \, .
\ee
As it must be the case, for $m = 0,1$,  only $P = 0$  
is a solution. For $m = 1/2$, 
\be
P \sim  a^5 \qquad {\rm   and} \qquad  Q = \exp(c \, a^6) \, ,
\ee
where $c$ denotes an integration constant.
 For the other values $m$ we define $ z =  aP$
 as new dependent and $t=\ln a$   as new independent variable.
 Eq.  (\ref{9.19})  then becomes 
\be
0=  dz/dt + gz^2 - bz\, , \qquad 
 g = l/m - 2 \, , \qquad b = (4m -5)/(m-1) \, .
\ee 
For $m = 5/4$ we get 
\be
 z = - 5/(6t - c) \, , \quad  {\rm  i.e.}, \quad P = - 5/(6a \ln (a/c)) \, , \quad 
Q=( \ln (a/c) )^{-5/6}  \, .
\ee
 For the other values  we get
\bea
    z =f/ (\exp(-bt + c) - 1) \, , \quad
f= -b/g\,  , \quad  P =f/(e^c a^{1-b} - a)
\nonumber \\   Q = (\pm a^b + c)^{1/g} \, . \label{9.20}
\eea
For $m \to    1/2$  we get $b \to  6$  and 
$1/g \to \infty$; for $m \to 5/4$  we get $b \to  0 $  and $1/g \to    -5/6$, so  the
 two special cases could also have 
been obtained by a limiting procedure from   eq. (\ref{9.20}).

\bigskip

Metric   (\ref{9.16})  with   eq. (\ref{9.20})  can be explicitly written in synchronized 
coordinates for special examples only, see e.g. Burd  and Barrow  \cite{burd}.

\subsection{Discussion of power-law inflation}\label{s95}

We have considered  scale--invariant field equations. The three 
examples mentioned in section \ref{s93} can 
be transformed into each other by a conformal transformation of the 
four--dimensional spacetime metric. 
The parameters of eqs.   (\ref{9.3})   and  (\ref{9.4})   are related 
by $ \sqrt 3 \,  \mu = \sqrt 2 (2 -m)/(m      - 1)$,  cf.  \cite{sch89}, 
a similar relation exists to the internal  dimension in
  eq. (\ref{9.5}), one has 
\be
m = 1 + 1/ \left\{   1  + \sqrt 3  \left[  1 + 2/(N - 4)
\right]^{\pm 1/2}  \right\}  \, , 
\ee
for details see \cite{bleyerschmidt}. The 
necessary conformal factor is a suitable power 
of the curvature scalar   eq. (\ref{9.3}). A further conformal transformation in 
addition with the field redefinition $\theta  = \tanh \Phi $ 
 leads to the conformally coupled scalar field $\theta$ 
in the potential
\be
 (1 +  \theta )^{2 +  \mu } \, (1 -  \theta )^{2 -  \mu } \, , 
\ee
the conformal factor being $\cosh^4 \Phi$, 
 cf. \cite{sch88c}. So, equations stemming from quite different physical foundations 
are seen to be equivalent. We have looked at 
them from the point of view of  self-simi1ar 
solutions and of limiting processes changing the type of symmetry.

\bigskip

The general solution for the spatially flat Friedmann models in fourth-order 
gravity   eq. (\ref{9.3}), eqs.   (\ref{9.16}),  (\ref{9.20})  
 can be written for small $c$ in synchronized coordinates as follows
\be
ds^2=dt^2  - a^2(t) (dx^2+dy^2+dz^2) \, , \quad
 a(t) = t^p ( 1 + \e t^{-bp} + O(\e^2)) \, . 
\ee
$\e = 0$ gives exact power-law inflation 
with $p = (m - 1) (2m - 1)/(2 - m)$ and $b = (4m - 5)/(m - 1)$.
 We see: in the range $1 < m < 2$, power-law
 inflation is an attractor solution within the set 
of spatially flat Friedmann models for $m \ge  5/4$ only, because otherwise
$b$ would become negative.  The generalization to 
polar inflation with $m > 2$ and  $p <0$  is similar.

\subsection{The cosmic no hair theorem}\label{s96}

Following \cite{k}, we now discuss the cosmic no hair theorem, which 
tells under  which circumstances the de Sitter spacetime represents an
attractor solution within the set of other nearby solutions. 
  This property ensures the inflationary model to be  a typical solution.  The notion
``cosmic no hair theorem" is chosen because of  its analogous
properties  to the ``no hair theorem" for black holes, for which one can cf. \cite{x71}.

\bigskip

After a general introduction we restrict our consideration to 
spatially flat Friedmann  models. In this section, we choose  
gravitational Lagrangians 
\be
R  \Box \sp k R \sqrt{-g}
\ee
 and linear combinations of them. They are  motivated from trials how to 
overcome the non--renormalizability \cite{stelle77} of 
 Einstein's theory of gravity. Results are: For arbitrary  $k$, i.e., 
 for arbitrarily large order $2k+4$ of the 
field equation, one can always find examples where the
attractor property  takes place. Such examples necessarily need a non-vanishing
$R\sp 2$-term.  The main formulas  do not depend on the dimension, so one gets
similar results also for 1+1-dimensional gravity and for Kaluza-Klein cosmology.

\subsubsection{Introduction to no hair theorems }

Over the years, the notion ``no hair conjecture'' drifted to  
``no hair theorem'' without possessing a generally accepted formulation  or even a
complete proof. Several trials have been made to formulate and prove it at
least for certain  special cases. They all have the overall structure: ``For a
geometrically  defined class of spacetimes and physically motivated
properties of the energy-momentum tensor, all the solutions of the gravitational
field equation  asymptotically converge to a space of constant curvature.'' 

\bigskip

The paper Weyl \cite{x90} is cited in \cite{x6}
with the phrase: ``The behaviour of every world satisfying certain natural
homogeneity conditions in the large follows the de Sitter solution asymptotically.''
to be the  first published version of the no hair conjecture. 
 Barrow and G\"otz  \cite{x6} apply the formulation: ``All
ever-expanding universes with $\Lambda >0$ approach the de Sitter spacetime locally.''
The first proof of the stability of the de Sitter solution,
in this case  within the steady-state theory, is due to Hoyle and Narlikar 1963 [112].

\bigskip

 The probability of inflation is large if 
the no hair theorem is valid, cf. Altshuler [1].
Peter, Polarski,  Starobinsky [181]  and Kofman, Linde, Starobinsky [130]
compared the double--inflationary  models with cosmological  observations. 
 Brauer, Rendall, Reula [40]
discussed the no hair conjecture within  Newtonian cosmological models.
H\"ubner and Ehlers [113] and Burd [44] considered 
inflation in an open Friedmann  universe and have noted that 
inflationary models need not to be spatially flat.

\bigskip

Gibbons and Hawking 1977 [88] have found two of the earliest 
strict results on the no hair conjecture for Einstein's theory, 
cf. also Hawking, Moss [109] and Demianski [62].
Barrow 1986 [12] gave examples that the no hair conjecture fails 
if the  energy condition  is relaxed and points out, that
this is necessary to solve the graceful exit problem. He uses the formulation of
the no hair  conjecture ``in the presence of an effective cosmological
constant, stemming  e.g. from viscosity,  the de Sitter spacetime is a stable
asymptotic solution''.

\bigskip

 Usually, energy inequalities are presumed for formulating 
the no hair conjecture. Nakao et al. [168] found some cases
where it remains valid also for negative Abbott-Deser mass. 
The latter goes over to the well-known ADM-mass \cite{arno}  
for $\Lambda \longrightarrow 0$. In Murphy [165],
  viscosity terms  as source are considered to get a
singularity-free cosmological model. 
In the eighties, these non-singular models with viscosity
where reinterpreted as inflationary ones, cf. Oleak \cite{x62}. 

\bigskip

In the three papers [186],  Prigogine et al.  developed a
phenomenological model of particle and  entropy creation.
It allows particle creation from spacetime curvature, but the inverse procedure,
i.e. particle decay into spacetime curvature is forbidden. 
This breaks the  $t\longrightarrow -t$-invariance of the model. Within 
that model, the expanding de Sitter spacetime is an attractor
solution independently of the initial fluctuations; this
means, only the expanding de Sitter solution is thermodynamically possible.

\bigskip

Vilenkin 1992 [262] and  Borde [33], [34] discussed  future-eternal inflating
universe models; they must have a singularity if the condition
D: ``There is at least one point $p$ such that for some point $q$ to the future of $p$
the volume of the difference of the pasts of $p$ and $q$ is finite'' is fulfilled.
Mondaini and Vilar [157] have considered  recollapse and
the no hair conjecture in closed higher-dimensional Friedmann  models.

\bigskip

 In Shiromizu et al. [237], Shibata et al. [235]  and Chiba,
Maeda [52]  the following argument is given: If the matter 
distribution is too clumpy, then 
a large number of small black holes appears. Then one should
look for an inflationary scenario where these black holes are harmless.
They cannot   clump together to one giant black hole because of the
exponential expansion of the universe; this explains the existing upper bound of the
mass of black holes in the quasi-de Sitter  model: above 
\be
M_{\rm  crit} = \frac{1}{3\sqrt\Lambda}
\ee
there do not exist  horizons; this  restriction is called  cosmic  hoop conjecture.
The paper [52] from 1994  is entitled
``Cosmic hoop conjecture?'',  and that conjecture is formulated as:
when an apparent horizon forms in a gravitational collapse,
the matter must be sufficiently 
compactified such that the circumference $C$ satisfies the
condition 
\be
C < 4 \pi M < 4 \pi M_{\rm crit} \,  ,
\ee
 where $M$ is the  Abbott--Deser  mass of the collapsed body.

\bigskip

 Xu, Li and Liu [272]  proved 
the instability of the anti-de Sitter spacetime  against gravitational waves and dust 
matter perturbations in 1994.  Coley and Tavakol [54]  discussed the robustness of the
cosmic no  hair conjecture under using the concept of the structural
stability.  Sirousse-Zia [239] considered the Bianchi type IX model in
Einstein's theory with a positive $\Lambda $-term and got an
asymptotic  isotropization of  the mixmaster model. She uses methods of
Belinsky et al. [22].  M\"uller [162]  used $L=R^2$ and discussed the
power-asymptotes of Bianchi models. 
Barrow and Sirousse-Zia [16]  discussed the  mixmaster
$R\sp 2$-model  and the question, 
under which conditions  the Bianchi type IX 
model becomes asymptotic de Sitter. 
In  Yokoyama, Maeda [273] and Cotsakis et al. [57]
 the no hair conjecture for Bianchi type 
IX models and Einstein's theory with a cosmological term
are discussed; they get  $R\sp 2-$ inflation in anisotropic universe models 
and  typically, an initial anisotropy helps to enhance inflation. 
They got some recollapsing solutions besides those 
converging to the de Sitter solution. 
 In Spindel [241]  also general Bianchi type I 
models in general dimensions are considered.

\bigskip

Breizman et al. 1970 [41] considered 
\be 
L=R+\alpha R\sp{4/3}
\ee
 to  get  a singularity-free model, the solutions are of a quasi de Sitter type. 
The papers Barrow [12] from 1987, Pollock [185], M\"uller et al. [164]
and Mijic et al. [155]  consider the  no hair conjecture
for $R\sp 2$ models, they use the formulation ``asymptotical
de Sitter, at least on patch''. The restriction ``on patch'' is not strictly
defined but  refers to a kind of local validity of the statement, e.g., in
a region being covered by one single synchronized system of  reference
in which the  spatial curvature is non-positive and the energy conditions
are fulfilled. The Starobinsky model is  one of those which  does not need
an additional  inflaton field to get the desired quasi de Sitter stage. 

\bigskip

One should  observe a notational change: Originally, 
\be 
L=R+ a R\sp 2 \ln R
\ee
 was called Starobinsky model, whereas
\be
L=R+ a R\sp 2 
\ee
 got the name ``improved Starobinsky model'' -
but now the latter carries simply the name ``Starobinsky
model'',  see e.g. Starobinsky 1983 [244], Hwang [114],  Gottl\"ober et al. [95]
 and  Amendola et al. [3] and [4].
 For  the inflationary phase, both versions are quite similar. 
 A  further result of  the papers   [12] and [185]   
 is that by the addition of a cosmological  term, the 
Starobinsky model leads naturally to double inflation. Let us
comment  this result: It is correct, but one should add that this is
got at the  price of getting a ``graceful exit problem'',  by this phrase
there is meant the problem of how to finish the inflationary phase
dynamically - in the 
Starobinsky model this problem is automatically solved by the fact that 
the quasi de Sitter phase is a transient attractor only.

\bigskip

 The papers  Maeda [150], Maeda et al. [152] and Barrow, Saich
 [15]  discuss the no hair conjecture within $R\sp2$-models and 
found inflation as a transient attractor in fourth-order
gravity.  The papers   Feldman [75], Rogers, Isaacson [192]  
and \cite{x75} discuss the stability of  inflation in $R\sp 2$-gravity.
The papers Bi\v c\'ak, Podolsky [27],  Cotsakis,
Flessas [58], Borzeszkowski, Treder [36]
and Contreras et al. [56]  discuss generalized cosmic no hair theorems
for quasi exponential expansion.

\bigskip

In Starobinsky 1983 [244] the no hair theorem for Einstein's
theory   with a positive $\Lambda$-term is tackled by using a power
series  expansion as ansatz to 
describe a general spacetime. This includes a definition 
of an asymptotic de Sitter spacetime even for   inhomogeneous models. 
However, the convergence of the sequence is not rigorously proven. 
 In \cite{starosch}, the ansatz [244]  
 was generalized  to consider also the cosmic no hair theorem for $L=R\sp 2$.
Shiromizu et al. [237] discussed an inflationary inhomogeneous  scenario and 
mentioned the open problem  how to define asymptotic  de
Sitter spacetimes. 

\bigskip

 Jensen, Stein-Schabes [122]  consider the no hair theorem for a special class 
of inhomogeneous models and give partial proofs.
Morris [161] considers inhomogeneous models for $R+R\sp
2$-cosmology.  In Calzetta, Sakellariadou [47]  inflation in inhomogeneous
but spherically symmetric
cosmological models is obtained only if the Cauchy data are
homogeneous  over several horizon lengths. The  analogous problem is considered in 
Nakao et al. [166], [167]; in Shinkai, Maeda [236]  also the 
  inclusion of colliding plane gravitational waves is considered, they give a
numerical  support of the no hair conjecture by concentrating on the dynamics of
gravitational waves.  Berkin [24] gets as further result, that for $L=f(R)$, a
diagonal Bianchi metric is always possible.  
 Barrow and Sirousse-Zia [16]  and Spindel [241]  also 
worked on  the   diagonalization problem. They apply the diagonalization
condition of  MacCallum et al. [146], [147].

\bigskip

  Kottler 1918 [131]  found the   spherically symmetric
vacuum solution for Einstein's theory with $\Lambda$-term  
\begin{equation}
ds\sp 2 = A(r) dt\sp 2 - \frac{dr\sp 2}{A(r)}
- r\sp 2 (d\theta \sp 2 + \sin \sp 2 \theta \, d \phi \sp 2 )
\end{equation} 
with 
\be
A(r)=1 - \frac{2m}{r} - \frac{\Lambda}{3} r\sp 2 \,  . 
\ee
At the horizon $A(r)=0$ the Killing vector changes its sign
and one  gets by interchanging the coordinates $t$ and $r$ the
corresponding   Kantowski-Sachs model. The cosmic no hair conjecture within 
Kantowski-Sachs models and $\Lambda > 0$ is discussed by 
Moniz [159].  To get a theory renormalizable to one-loop order 
one needs at least curvature squared terms
in the Lagrangian, and they lead to a fourth-order theory, Stelle [248]. 
 But there exist several further motivations to  consider  fourth-order theories; 
to get an overview for early papers on fourth--order gravity see e.g. [199].

\bigskip

Let us now discuss  sixth--order equations.  Stelle [248] 1977  considers 
mainly fourth-order $R\sp 2$-models; in the introduction he
mentioned  that in the next order, terms like 
\be
R\sp 3 + R_{ij;k} R\sp{ij;k}
\ee
become admissible, but  the pure $R\sp 3$-term is not  admissible. 
The paper  Buchdahl 1951 [42] deals with Lagrangians of  arbitrarily high 
order. Its results are applied in [218], Quandt et al. [187], 
 and  Battaglia Mayer et al. [18]   to general  Lagrangians $F(R, \Box)$.
 In the paper  Vilkovisky [263], 
the  Sacharov-approach  was generalized. The original idea of
A. Sakharov [195] from 1967, see the commented reprint from 2000,    
was to define  higher order curvature corrections to the Einstein action to
get a kind  of elasticity of the vacuum. Then the usual
breakdown of  measurements at 
the Planck length,  such a short de Broglie wave length
corresponds to such  a large mass which makes the measuring apparatus to a black
hole, is softened. 

\bigskip 

Vilkovisky [263] discusses the effective gravitational
action in the form $R f(\Box )R$, where
\be
f(\Box )= \int \frac{1}{\Box - x } \rho(x)dx \, . 
\ee
Martin and Mazzitelli [153] discuss the non-local Lagrangian 
$R\frac{1}{\Box }R$ as conformal anomaly in two dimensions.
Treder  1991 [257] used higher-order Lagrangians, especially $R\sp
2$-terms, and he mentioned that for 
\be
R+R_{,i}R_{,k}g\sp{ik}
\ee
 a sixth-order field equation appears. Remark: This Lagrangian leads to the same
field equation as  $R-R\Box R$.
 Lu and Wise [144] consider the gravitational Lagrangian as a
sequence  $S=S_0+S_1+S_2+ \dots$ ordered with respect to physical
dimension. So, $S_0=R$ and $S_1$ sums up the $R\sp 2$-terms. 
 Kirsten et al. [127] consider the effective Lagrangian for
self-interacting scalar fields; in the renormalized action, the term 
\be
\frac{\Box R}{c+R}
\ee
appears. Wands [266]  classifies Lagrangians of the type
$F(R, \Phi)\Box R$ and  mentions that not all 
of them can be conformally transformed
to Einstein's   theory. Amendola [2] 
 considers the Lagrangian $\Phi \sp 2 \Box R$, Gottl\"ober et al. [97]
 the Lagrangian  $R \Box R$,  Gottl\"ober et al. [94]  and Schmidt [219]
 discuss double inflation from $\Phi $ and $R\sp 2$-terms. 
Besides $R\Box R$,  Berkin [24]  and Berkin, Maeda [25]  
 consider the de Sitter  spacetime as 
attractor solution for field equations where the variational
derivative of the term $C_{ijkl}C\sp{ijkl}$ is included.

\bigskip

This section  is organized as follows: subsection \ref{t962} compares several
possible definitions of an asymptotic de Sitter spacetime. 
Subsection \ref{t963} deals with the
Lagrangian and corresponding field equations for higher--order
gravity, and \ref{t964} gives the no hair theorem for it.
 In subsection \ref{t965}, we determine under which circumstances the
Bianchi models in higher--order gravity can be written in
diagonal form without loss of generality, and in 
subsection \ref{t966} we discuss and summarize the results.

\subsubsection{Definition  of  an   asymptotic  de Sitter  
 spacetime}\label{t962}

In this subsection we want to compare some possible definitions
of an asymptotic de Sitter spacetime. To this end let us consider the metric
\begin{equation}\label{y3}
ds^2=dt^2 - e^{2\alpha(t)} \sum_{i=1}^n d(x^i)^2
\end{equation}
which is the  spatially flat Friedmann model in $n$
spatial dimensions. We consider all values $n \ge 1$, but then
 concentrate on the usual case $n=3$. The Hubble parameter is
\be 
H= \dot \alpha \equiv \frac{d\alpha}{dt}\, .
\ee
 We get 
\begin{equation}
R_{00}= - n \left( \frac{dH}{dt} + H^2 \right) \, , \qquad
 R = - 2n \dot H   - n (n+1) H^2 \, .
\end{equation}

Then it holds:  The following 4 conditions for metric  (\ref{y3}) are equivalent. 
1: The spacetime  is flat. \ 2: It holds $R=R_{00}=0$. \ 
3: The curvature invariant  $R_{ij} R^{ij} $ vanishes. \ 
4: Either $\alpha = {\rm const.}$ or  
\be 
n=1 \quad {\rm and } \quad \alpha = \ln \vert t  - t_0 \vert + {\rm  const.}
\ee
For the last case with $n=1$ one has to observe that 
\be
ds^2= dt^2 - (t-t_0)^2dx^2
\ee
 represents flat spacetime in
polar coordinates. For the proof we use  the identity
\begin{equation}
R_{ij} R^{ij} = (R_{00})^2 + \frac{1}{n} (R-R_{00})^2 \, .
\end{equation}

An analogous characterization is valid  for the de Sitter
spacetime.  The following 4 conditions for metric   (\ref{y3})  are
equivalent. 1: It is a non--flat spacetime of constant
curvature. \  2: $R_{00}= R/(n+1)= {\rm  const.}  \ne 0$. \ 
3: $(n+1) R_{ij} R^{ij} = R^2 = {\rm  const.}  \ne 0  $. \
 4: Either $H = {\rm  const.}  \ne 0$ or 
\be
n=1  \ {\rm and } \ \Bigl(  
ds^2= dt^2 - \sin^2(\lambda t)dx^2 \ 
{\rm  or} \  ds^2= dt^2 - \sinh^2(\lambda t)dx^2 \Bigr)  \, .
\ee
For $n=1$, the de Sitter spacetime and anti-de Sitter spacetime differ by 
the factor $-1$  in front of the metric only. For $n>1$, 
 only the de Sitter spacetime,  having 
$R<0$, is covered, because the anti-de Sitter spacetime cannot be represented 
as spatially flat Friedmann model. Our result   shows that within the class of
spatially flat Friedmann models, a characterization of the de Sitter 
spacetime using polynomial  curvature  invariants only, is possible. 

\bigskip

Next, let us look for isometries leaving the form of the
metric    (\ref{y3})   invariant. Besides spatial isometries, 
the map $\a \to \tilde \a$ defined by 
\begin{equation}
\tilde \alpha(t) = c + \alpha ( \pm t +t_0), \qquad c, \ t_0 \, = \,  {\rm const.}
\end{equation}
 leads to an isometric spacetime. The simplest expressions
being  invariant by such a transformation are $H^2$ and $\dot H$. We
take $\alpha $
as dimensionless, then $H$ is an inverse time and $\dot H$ an
inverse time squared. Let $H \not= 0$ in the following. The expression 
\begin{equation}                                              
\varepsilon := \dot H H^{-2}
\end{equation}
is the simplest dimensionless quantity defined for the 
spatially flat Friedmann models  and being invariant with respect to this
 transformation. Let 
$n>1$ in the following, then it holds: Two metrics of type    (\ref{y3})   are
isometric if and only 
if the corresponding functions $\alpha $ and $\tilde \alpha $
are related  by this transformation. 
It follows: Metric     (\ref{y3})    with $H\not= 0$ represents  the de
Sitter spacetime  if and only if  $\varepsilon \equiv 0$. 

\bigskip

All dimensionless invariants containing at
most second order derivatives of the metric can be expressed as 
$f(\varepsilon )$,  where $f$
is any given function. But if one has no restriction to the order, one gets 
a sequence of further invariants
\begin{equation}                                              
\varepsilon _2 = \ddot H H^{-3}, \quad  \ldots, \  \varepsilon _p = 
                 \frac{d^pH}{dt^p} H^{-p-1} \, .
\end{equation}  
 Let $H>0$ in metric (\ref{y3}) with $n>1$. We
call it an asymptotic de Sitter spacetime if  
\begin{equation}
\lim_{t\to \infty} \frac{\alpha(t)}{t} = {\rm const. } \ne 0
\end{equation}
or 
\be
\lim _{t \to \infty}\,  R^2 = {\rm const.} >0 \quad {\rm  and} \quad 
         \lim _{t \to \infty} \, (n+1)R_{ij}R{ij}-R^2=0
\ee
or for some natural number $p$ one has
\be
 \lim_{t \to \infty} \ \varepsilon _j =0 \quad {\rm  for} \quad 1\leq j\leq p \, .
\ee
In general, all these definitions are different. Using the identity
\be 
R_{ij}R^{ij}=n^2(\dot H + H^2)^2 + n (\dot H+nH^2)^2
\ee
 we will see that  all these definitions lead to the same 
result if we restrict ourselves to the set of solutions of the
higher--order  field equations.    The
problem is that none of the above definitions can be
generalized  to inhomogeneous models. One should find a
polynomial curvature  invariant which equals a positive
constant if and only if the  spacetime is locally the de
Sitter spacetime. To our  knowledge, such an invariant
cannot be found in the literature,  but also the
non--existence of such an invariant has not been   proven up
to now. For the anti-de Sitter spacetime, a partial result is given in
 section \ref{s32}.

\bigskip

This situation is quite different for the positive definite 
case: For signature $(++++)$ and 
$S_{ij} = R_{ij}  - \frac{R}{4} g_{ij} $ it holds:
\be
I \equiv  (R-R_0)^2 + C_{ijkl} C^{ijkl}  + S_{ij} S^{ij}  =0
\ee
if and only if  the $V_4$ is a space of constant curvature $R_0$. So 
$I \longrightarrow 0$ is a suitable definition of an
asymptotic  space of constant curvature.
 One possibility exists, however, for the Lorentz signature case, 
 if one allows additional  structure as follows: An ideal
fluid has an energy--momentum  tensor
\be
T_{ij} = ( \rho + p)u_i u_j - p g_{ij} 
\ee
where $u_i$ is a continuous vector field with
$u_i u^i \equiv 1$. For  matter with equation of state $ \rho = - p$, 
the equation  $T^{ij}_{\ \ ;j} \equiv 0$ implies $p=$ const., and so every
solution of  Einstein's theory with such matter is isometric 
to a vacuum  solution of Einstein's theory with a cosmological
term. The  inverse statement, however, is valid only locally: 

\bigskip

Given a  vacuum solution of Einstein's theory with a 
$\Lambda $--term, one  has to find  continuous timelike unit
vector fields which need  not to exist from topological
reasons. And if they exist, they  are not at all unique.  So,
it becomes possible to define an  invariant $J$ which vanishes
if and only if the spacetime is de Sitter by  transvecting the
curvature tensor with $u^i u^j$ and/or $g^{ij} $  and suitable
linear and quadratic  combinations of such terms.  Then time
$t$ becomes defined by the streamlines of the vector  $u^i$.
If one defines the asymptotic de Sitter spacetime by 
$J \longrightarrow 0$ as $t \longrightarrow \infty$, then it
turns out, that this definition is   not  independent of
the vector field $u^i$. 

\subsubsection{Lagrangian $F(R, \Box R, \Box \sp 2R,
\dots ,  \Box \sp k R )$ }\label{t963}

Let us consider the Lagrangian density $L$ given by
\begin{equation}
L = F(R, \Box R, \Box \sp 2R,
\dots  ,  \Box \sp k R) \sqrt{-g}
\end{equation}
where $R$ is the curvature scalar,  $\Box $ the D'Alembertian
and  $g_{ij}$  the metric of a  Pseudo-Riemannian $V_D$ of 
dimension   $D\ge   2$  and  arbitrary  signature;  
 $g  =  - \vert  \det  \, g_{ij}  \vert  $. 
 The main application will be $D=4$ and  metric
signature $(+---)$.  $F$ is supposed to be  a  sufficiently  smooth 
function of its arguments, preferably a polynomial.
 Buchdahl \cite{buch1}  already  dealt with such kind of  Lagrangians
 in 1951,  but then it became quiet of them for decades. 

\bigskip

The  variational  derivative of $L$ with respect  to  the  metric yields the tensor 
\begin{equation}
P\sp{ij}  \  =  - \frac{1}{\sqrt{-g}} \ 
\frac{\delta  L  }{\delta g_{ij} }
\end{equation}
The components of this tensor  read
\begin{equation}
P_{ij} \ = \ G R_{ij} \ - \ \frac{1}{2} F g_{ij} \ - \ G_{;ij}
 \ + \ g_{ij} \Box G \  + \ X_{ij}
\end{equation}
where the semicolon denotes the covariant derivative,
 $R_{ij}$  the Ricci tensor, and
\begin{equation}
 X_{ij} \ = \ \sum_{A=1}\sp k \ \frac{1}{2} g_{ij}
[F_A(\Box\sp{A-1}  R)\sp{;m}  ]_{;m} \  - \  F_{A(;i}[\Box\sp{A-1} 
R]_{;j)}
\end{equation}
having  the round symmetrization brackets in its last  term. 
For $k=0$, i.e. $F = F(R)$, a case considered in subsection \ref{t964}, the
tensor $  X_{ij}$  identically  vanishes.   It  remains  to 
define  the expressions $F_A$, $A=0, \dots,k$ .
\begin{equation}
F_k \ = \ \frac{\partial F}{\partial \Box \sp k R }
\end{equation}
and for $A=k-1, \dots,0$
\begin{equation}
F_A \ = \ \Box F_{A+1} \ + \ 
 \frac{\partial F}{\partial \Box \sp A R }
\end{equation}
and finally $G \ = \ F_0$. 
The brackets are essential, for any scalar $\Phi $ it holds 
\begin{equation}
\Box(\Phi _{;i}) \ - \ (\Box \Phi)_{;i}  \ = \ R_i \sp { \ j}
\ \Phi_{;j}
\end{equation}
Inserting  $\Phi  =  \Box\sp m R$ into this  equation,  one 
gets  identities to be applied in the sequel without further notice.
 The covariant form of energy-momentum conservation reads 
\begin{equation}
P\sp i _{\ j;i} \ \equiv \ 0
\end{equation}
and $P_{ij}$ identically vanishes  if and only 
if $F$ is a divergence, i.e., locally there can be found a
vector  $v\sp i$ such that $F \ = \ v\sp i _{\ ;i}$ holds.  Remark:
Even for compact manifolds without boundary the restriction 
``locally''  is unavoidable, for  example, let  $D=2$ and  $V_2$ be the
Riemannian  two-sphere $S\sp 2$ with arbitrary positive definite metric. 
$R$ is  a divergence,  but there do not exist  continuous
 vector fields $v\sp i$   fulfilling  $R \ = \ v\sp i _{\ ;i}$  
 on the whole  $S\sp 2$. Example: for $m,n \ \ge \ 0$ it holds
\begin{equation}
\Box \sp m  R \ \Box \sp n R \ - \   R \ \Box \sp{m+n} R \ = \
 {\rm  divergence} \, .
\end{equation}
So, the terms $\Box \sp m  R \ \Box \sp n R $ with naturals
$m$ and $n$ can be restricted to the case $m=0$ without loss
of generality. 

\subsubsection{No hair theorems for higher-order gravity}\label{t964}

 For $n>1$, the $n+1$-dimensional de Sitter spacetime  is an attractor 
solution for the field equation derived from the Lagrangian
\be
R^{(n+1)/2} \, . 
\ee
 It  is not an attractor solution for the Lagrangian $R \Box^k R$
and $k>0$. There exist  combinations of coefficients $c_i$, such that 
the de Sitter spacetime is an attractor solution for the field 
equation derived from the more general Lagrangian 
\be\label{y4}
L=c_0 \, R^{(n+1)/2} \  + \  \sum\limits_{k=0}^{m} \ 
c_k \,  R \,  \Box ^k  R \, . 
\ee 
 Idea of Proof: Concerning fourth-order gravity this method 
was previously applied e.g. by Barrow 1986 [12].
  The de Sitter spacetime is an exact solution for the field equation,
if and only if $2RG=(n+1)F$. If we make the ansatz 
\be
\dot  \alpha(t)=1+\beta(t)
\ee 
we get the linearized  field equation $0=\ddot \beta+n\dot \beta$ for the Lagrangian
$R^{\frac{n+1}{2}}$. For the Lagrangian $L=R \Box^k R$ we get the linearized field
equation $\Box^k R=(\Box^k R)_{,0}$. For the characteristic
 polynomial we get a recursive formula such that the next order 
is received from the previous one by multiplying with 
$ \, \cdot \,  x \cdot (x+n)$.  We get the roots $x_1=-n-1$, $x_2=-n$  
($k$-fold), $x_3=0$ ($k$-fold)  and $x_4=+1$. Because of 
the last root the de Sitter spacetime is not an attractor
solution.  For the  Lagrangian (\ref{y4})
we get the characteristic polynomial 
\be                                  
P(x)  =  x(x+n)\left[c_0 + \sum_{k=1}^mc_kx^{k-1}(x + n)^{k-1}(x
- 1)(x + n + 1)\right]
\ee
for the linearized field equation. The transformation
$z=x^2+nx+\frac{n^2}{4}$ gives a polynomial $Q(z)$ which 
can be solved explicitly. So one can find those combinations of the 
coefficients such that  the de Sitter spacetime is an attractor
solution.

\bigskip
 
It turned out that all the variants of the definition of an
asymptotic de Sitter solution given in subsection \ref{t962} lead to
the same class of solutions, i.e., the validity of the
theorems written below does not depend on which of the variants of definition of 
an asymptotic  de Sitter spacetime listed   is applied. 
For the 6th--order case we can summarize as follows:  Let  
\be
L \ = \ R^2 \ + \ c_1 \ R  \  \Box R
\ee
and 
\be
L_{\rm E} \ = \  R \ -  \ \frac{l^2}{6} \ L 
\ee
 with length  $l>0$. Then the following statements are equivalent.

\medskip

1. The Newtonian limit of $L_{\rm E}$ is well--behaved, and the
potential $\phi$ consists of terms 
$ e^{-\alpha r}/r$ with $\alpha \ge 0$ only.

\medskip

2. The de Sitter spacetime with $H= 1/l $ is an
attractor solution for $L$ in the set of spatially flat
Friedmann models, and this can already be seen from the
linearized field equation.

\medskip

3. The coefficient $c_1 \ge 0$, and the graceful exit problem 
is solved for the quasi de Sitter phase $H\le 1/l$ of $L_{\rm  E}$.

\medskip

4. $l^2= l_0^2+l_1^2$ such that  $ l^2 \,  c_1 = l_0^2 \,  l_1^2$ has a
solution with $0\le l_0<l_1$. 

\medskip

5. $0\le c_1 < {l^2}/{4}$. 

\bigskip

A formulation which includes also the marginally well-behaved cases 
reads as follows: Let  $L$ and $L_{\rm E}$ as in the previous result, then the 
following statements   are equivalent: 

\medskip

1. The Newtonian limit of $L_{\rm  E}$ is well--behaved, for the
potential $\phi$ we allow ${1}/{r}$ and terms like
\be
\frac{P(r)}{r} e^{-\alpha r} \qquad {\rm  with} \qquad  \alpha > 0
\ee
 and a  polynomial $P$.

\medskip

2. The de Sitter spacetime with $H= {1}/{l}$ cannot be
ruled out to be an attractor solution for $L$ in the set of
spatially flat Friedmann models if one considers the linearized
field equation only. 

\medskip

3. $L_{\rm  E}$ is tachyonic--free.

\medskip

4. $l^2= l_0^2+l_1^2$ with  $l^2 \,  c_1 = l_0^2 \,  l_1^2$ has a
solution with $0\le l_0 \le l_1$. 

\medskip

5. For the coefficients we have $0\le c_1 \le \frac{l^2}{4}$.

Of course, it would be interesting what happens in the region
where the linearized equation does not suffice to decide.

\subsubsection{Diagonalizability of Bianchi
models}\label{t965}
A Bianchi model can always be written as
\begin{equation}
ds\sp 2 = dt\sp 2 - g_{\alpha \beta} (t) \sigma \sp{\alpha}
\sigma \sp{\beta}
\end{equation}
where $g_{\alpha \beta}$ is positive definite and $\sigma
\sp{\alpha}$ are the characterizing one-forms. It holds
\begin{equation}
d\sigma \sp{\gamma} = - \frac{1}{2} C\sp{\gamma}_{\alpha
\beta} \sigma \sp{\alpha} \wedge \sigma \sp{\beta}
\end{equation}
with structure constants $C\sp{\gamma}_{\alpha \beta}$ of the
corresponding Bianchi type. It belongs to class A if
$C\sp{\beta}_{\alpha \beta} = 0$. The Abelian group, Bianchi
type I, and the rotation group, Bianchi type IX,  both belong
to class A. 
 In most cases, the $g_{\alpha \beta}$ are written in diagonal
form; it is a non-trivial problem to decide under which
circumstances this can be done without loss of generality. 
 For Einstein's theory, this problem is solved. One of
its results reads: If a Bianchi model of class A, except Bianchi 
types I and II, has a diagonal energy-momentum tensor, then the 
metric $g_{\alpha \beta} (t)$ can be chosen in diagonal form. Here, the 
energy-momentum tensor is called diagonal, if it is diagonal
in the basis $(dt,\ \sigma \sp 1,\ \sigma \sp 2,\ \sigma \sp 3)  $.

\bigskip

This result rests of course on Einstein's theory and cannot be
directly applied to higher-order gravity. 
 For fourth-order gravity following from a Lagrangian 
$L=f(R)$ considered in an interval of $R$-values
where 
\be
\frac{df}{dR} \ \cdot \ \frac{d \sp 2f}{dR \sp 2} \ \ne \ 0
\ee
one can do the following: The application of the conformal
equivalence theorem, cf. e.g. [29] or [218], is possible, the
conformal factor depends
on $t$ only, so the diagonal form of metric  does not
change. The conformal picture gives Einstein's theory with a
minimally coupled scalar field as source; the energy-momentum
tensor is automatically diagonal. So, in this class of 
fourth-order theories of gravity, we can apply the above cited
theorem of MacCallum et al. [146], [147], where 
the initial-value formulation of General Relativity is
applied.  As example we formulate: All solutions of Bianchi type IX of
fourth-order gravity following from $L=R\sp 2$ considered in a
region where $R\ne 0$ can be written in diagonal form. 
 Consequently, the ansatz used by Barrow and  Sirousse-Zia [16]
for this problem is already the most general one, cf.   Spindel [241].

\bigskip

For fourth-order gravity of a more complicated structure,
however, things are more involved; example: Let
\be
L=R+a R\sp 2 + b C_{ijkl} C\sp{ijkl}
\ee
with $ab \ne 0$. Then there exist Bianchi type IX models which
cannot be written in diagonal form. This is a non-trivial
statement because its proof needs a careful analysis
of the correspondingly allowed initial values in the Cauchy
problem.  To understand the difference between the cases $b = 0$ and $b
\ne 0$ it proves useful to perform the analysis independently
of the above cited papers [46]. For simplicity, we restrict to
Bianchi type I. Then the internal metric of the hypersurface
$[t=0]$ is flat and we can choose as initial value $g_{\alpha
\beta} (0)=\delta_{\alpha \beta} $. Spatial rotations do not
change this equation, and we can take advantage of them to
diagonalize the second fundamental form $\frac{d}{dt}g_{\alpha
\beta}(0)$. 

\medskip

First case: $b=0$. As additional initial conditions one has
only $R(0)$ and  $\frac{d}{dt}R(0)$. The field equation
ensures  $g_{\alpha \beta}(t)$  to remain diagonal for all times.

\medskip

Second case: $b\ne 0$. Then one has further initial data
\be 
\frac{d\sp 2}{dt\sp 2}g_{\alpha \beta}(0) \, . 
\ee
In the generic case, they cannot be brought to diagonal form simultaneously
with $\frac{d}{dt}g_{\alpha \beta}(0)$. This excludes a
diagonal form of the whole solution. To complete the proof,
one has of course to check that these initial data are not in 
contradiction to the constraint equations. This case has the
following relation to the Bicknell theorem: Just for
this case  $b\ne 0$, the conformal relation to Einstein's
theory breaks down, and if one tries to re-interpret the 
variational derivative of $C_{ijkl}C\sp{ijkl}$ as 
energy-momentum tensor then it turns out to be non-diagonal
generically, and the theorem cannot be applied. 

\bigskip

For higher-order gravity, the situation becomes even more
involved. For a special class of theories, however, the
diagonalizability condition is exactly the same as in  
Einstein's theory: If 
\be 
L=R + \sum_{k=0}\sp m \  a_k \,  R \,  \Box \sp k R\, , \qquad a_m \ne 0\, , 
\ee
 then in a region    where $2L \ne R$
the Cauchy data are the data of General Relativity, $R(0)$,
and the first $2m+1$ temporal derivatives of $R$ at $t=0$. All
terms with the higher derivatives behave as an energy-momentum
tensor in diagonal form, and so the classical theorem applies.
Let us comment on the restriction  $2L \ne R$ supposed above:
 $F_0 = G=0$ represents a singular
point of the differential equation; and for the
Lagrangian given here $G={2L}/{R} - 1$. For fourth--order
gravity defined by a 
non--linear Lagrangian $L(R)$ one 
has
$$
G= \frac{dL}{dR}\, , 
$$
  and $G=0$ defines the critical value of the curvature scalar.

\subsubsection{Discussion of no hair theorems}\label{t966}

In subsection \ref{t964} we have shown: The results of the Starobinsky model   
  are structurally stable with respect to
the addition of  a sixth--order term $ \sim   R \Box R$, if the coefficients fufil 
certain inequalities. For the eighth-order case we got:  For  
\be
L=R^2 + c_1 R \Box R + c_2 R \Box \Box R ,  \qquad c_2 \ne
0
\ee  
and the case $n>1$ the de Sitter spacetime with $H=1$ is  an
attractor solution in the set of spatially flat 
Friedmann  models if and only if the following inequalities are 
fulfilled: 
\be
0<c_1< \frac{1}{n+1}, \qquad 0<c_2< \frac{1}{(n+1)^2}
\ee
and
\be
c_1> - (n^2+n+1) c_2 + \sqrt{ (n^4+4n^3+4n^2)c_2^2 + 4 c_2 }
\ee
These inequalities define  an open region in the $c_1-c_2-$plane whose 
boundary contains the origin; and for the other boundary
points  the linearized equation does not suffice to decide the
attractor  property. 

\bigskip

This situation shall be called
``semi--attractor'' for simplicity. In a general context this notion is used to
describe a situation where all Lyapunov coefficients have non-positive real parts and at
least one  of them is purely imaginary.    

\bigskip

In contrary to the   6th--order case, here we do not have  a one--to--one
correspondence, but a non--void open intersection with that
parameter set  having the Newtonian limit for $L_E$  well--behaved.   

\bigskip

To find out, whether another de Sitter spacetime with an 
arbitrary Hubble parameter $H > 0$ is an attractor solution
for  the eighth--order field equation following from the 
above  Lagrangian, one should remember that $H$ has the
physical  dimension of an inverted time, $c_1$ is a time squared,
$c_2$ is  a time to  power 4. So, we have to replace $c_1$ by
 $c_1 H^2$ and   $c_2$ by $c_2 H^4$ in the above dimensionless inequalities 
to  get  the correct conditions. Example: 
$$
0<c_1 H^2 < \frac{1}{4} \, .
$$ 

\bigskip

Let us summarize: Here  for a theory 
of gravity of order higher than four the Newtonian limit and
the attractor  property of the de Sitter spacetime are systematically
compared. It should be noted that the details of the theory sensibly
depend  on the numerical values of the corresponding coefficients.
So, no general overall result about this class of theories is
ever to be expected. 

\bigskip

The facts,  contradicting each other from the first glance, field
equations of higher than second order  follow 
from quantum gravity considerations  Mirzabekian et al. [156]  on the one hand 
but such equations are known to be unstable 
in general and are therefore unphysical
on the other hand,  have now considered in more details:
It depends on the special circumstances
whether such theories are unstable. 

\bigskip

We have found out
that for the class of theories considered here, 
one of the typical indicators of instability - 
cosmological runaway--solutions -  need not to exist, even 
for an arbitrarily high order of the field equation. 
It is an additional satisfactory result that  
this takes place in the same range of parameters where the Newtonian limit is well behaved.

\vfill

\eject

\newpage

\section{Solutions of the Bach-Einstein  equation}\label{Kap11}
\setcounter{equation}{0}
\setcounter{page}{145}

For field equations of  4th order, following from a 
Lagrangian ``Ricci scalar plus Weyl scalar", 
 it is shown  using methods of non-standard analysis 
that in a neighbourhood of Minkowski spacetime  
there do not exist regular static spherically symmetric solutions. 
With that, besides the known local expansions about $r = 0$ and 
$r = \infty$ respectively,   
 a global statement on the existence of such solutions is given, see \cite{sch85}. 
 Then,  this result will be discussed in connection with Einstein's particle
 programme. Finally, in section \ref{s116} it will be shown, that cosmological 
 Bianchi type I solutions of the Bach equation exist which fail to be conformally
 related to an Einstein space.  

\subsection{Introduction to the Bach equation} 

General Relativity Theory starts from the Einstein-Hilbert Lagrangian 
\be
L_{\rm EH} =  R/2 \kappa 
\ee
 with the Ricci scalar $R$  which leads to the Einstein  vacuum field
 equation $R_i^k= 0$   being of second order in the metrical tensor $g_{ik}$. Its
 validity is proven with high accuracy
 in spacetime regions where the curvature is small only.
 Therefore, the additional presence of a term being 
quadratic  in the curvature is not excluded by the standard weak field experiments.

\bigskip

Lagrangians with  squared curvature have already been discussed 
by Weyl in  1919 \cite{weyl19}, by Bach in 1921 \cite{bach} 
 and Einstein in 1921.  They were guided by ideas about conformal invariance,
 and  Einstein  \cite{einstein1} proposed to look seriously to such alternatives. 
In {Bach} \cite{bach}, the Lagrangian $L_{\rm W}$  with the 
the Weyl scalar $C$,
\be
L_{\rm W} = C/2 \kappa \, , \quad {\rm with} \quad 
C = \frac{1}{2} \cdot C_{ijkl} \,  C^{ijkl}
\ee
  was of special interest. Then it  became quiet of them for a couple 
of decades because of   the brilliant results of General Relativity 
 but not at least because of mathematical difficulties.

\bigskip
 
 Later, new  interest in such equations arose  with arguments coming from
 quantum gravity, cf. e.g.  \cite{treder75}, 
Borzeszkowski, Treder, Yourgrau  \cite{bty}, Stelle  \cite{stelle78}, 
Fiedler, Schimming   \cite{fiedler1983}.
 In the large  set of possible quadratic modifications a linear combination
\be\label{11.1c}
 L = L_{\rm EH} + l^2 L_{\rm W}
\ee
  enjoyed a special interest, cf. e.g.  Treder 1977 \cite{treder75}. 
The coupling constant $l$  has to be a length for dimensional reasons 
and it has to be a small one to avoid conflicts with 
observations: One often takes it to be the Heisenberg length, which is the   Compton wave
length of the proton, \,  $= 1.3 \cdot 10^{-18}$ cm \  or the  Planck length 
$= 1.6 \cdot 10^{-33}$  cm.

\bigskip

In this chapter  we consider the  Lagrangian $L$ eq. (\ref{11.1c})
 in connection with Einstein's particle programme,
 see Einstein,  Pauli  1943 \cite{einsteinpauli}:
 One asks for spherically symmetric singularity-free 
asymptotically flat solutions of  the vacuum field equations 
which shall be interpreted as particles, 
  but this cannot be fulfilled within General Relativity  itself; 
  it is still a hope,  cf.  \cite{hhvb},   to realize it in such 4th order field equations. 

\bigskip
 
Two partial answers have already been given: 
Stelle  \cite{stelle78} has  shown the gravitationa1 potential of  the linearized 
equations  to be
\be\label{11.1}
\Phi(r) =  -m/r + \exp (-r/l) c/r \qquad {\rm for} \qquad  r \ll   l \ ;
\ee
he also gave an expansion series about $r = 0$, and
 Fiedler, Schimming  \cite{fiedler1983}  proved its convergence and smoothness
 in a certain neighbourhood
 of $ r =  0$, but said nothing about the convergence radius of this expansion.

\bigskip

Now we want to join  these two local expansions. 
To this end we write  down the field equations in 
different but equivalent versions section \ref{s112},
 calculate some linearizations section \ref{s113}   and 
prove the statement of the summary in  section \ref{s114}
which will be followed by  a discussion
 of Einstein's particle programme in section \ref{s115}.

\subsection{Notations and field equations}\label{s112} 

We start from a spacetime metric with signature $(+ - - -)$,
 the Riemann and Ricci tensor being  defined by 
$R^i_{ \ jkl} = \Gamma^i_{ \ jl,k} - \dots$,  
 $ R_{ik} =  R^j_{ \  ijk}$
respectively. The Weyl tensor $C^i_{ \ jkl}$
 is the traceless part of the Riemann  tensor and the
Weyl scalar is given by   $2C = C_{ijkl} C^{ijkl}$. 
 Light velocity is taken to be 1    and $G = \kappa/8 \pi$
 is Newton's constant; $g = \det g_{ij}$. 

\bigskip
 
 Then we consider the Lagrangian
\be\label{11.2}
{\cal L}  = \sqrt{-g}  L = \sqrt{-g}  (R + l^2 C)/ 2 \kappa + 
{\cal L}_{\rm mat} \, , 
\ee
where ${\cal L}_{\rm mat}$  is the matter Lagrangian. 
For writing down the corresponding field equation it is convenient 
to introduce the Bach tensor $B_{ij}$,
  cf. Bach \cite{bach} and W\"unsch  \cite{wunsch},  beforehand.
\be\label{11.3}
\frac{1}{2} \,  B_{ij} =
 \frac{  \kappa \,  \delta (\sqrt{-g} \,  L_{\rm W})}
{\sqrt{-g} \,  \delta \,   g^{ij}   }     =  C^{a \ \  b}_{\ ij \ \,  ;ba}
 +     \frac{1}{2}    \,  C^{a \ \  b}_{\ ij } \,  R_{ba} \, .
\ee

\bigskip
 
It holds $ B^i_i =0$, $B^j_{i;j} =0$, $B_{ij} = B_{ji}$,
  and $B_{ij}$ is  conformally invariant of weight -1. 
Variation of eq.  (\ref{11.2})  with respect to $g_{ij}$  leads to the field equation
\be\label{11.4}
R_{ij} - \frac{1}{2} g_{ij} R + l^2 B_{ij}= \kappa T_{ij} \,  .
\ee
Now let us  consider the vacuum case $T_{ij}=  0$.
 The trace of  eq.  (\ref{11.4})  then simply  reads $R=0$, and  eq.  (\ref{11.4})
 becomes  equivalent to the simpler one
\be\label{11.5}
R_{ij}  + l^2 B_{ij}= 0 \, .
\ee
Writing $k = l^{-2}$
 and    $\Box \,  \cdot =  g^{ij} \left( \cdot \right)_{;ij} $ 
 one gets also the equivalent system 
\be\label{11.6}
R = 0 \, ;   \qquad   kR_{ij} + \Box R_{ij} =
 2R_{iabj} R^{ab} + \frac{1}{2} g_{ij} R_{ab}R^{ab} \, .
\ee

\bigskip
 
The static spherically symmetric  line element in Schwarzschild coordinates reads 
\be\label{11.7}
ds^2 = (1 + \beta) e^{-2 \lambda} dt^2 - (1 + \beta)^{-1} dr^2 - r^2d\Omega^2 \, , 
\ee
where $\beta$  and $\lambda$ depend on $r$ only. 
The dot means differentiation with respect to $r$, and defining
\bea
\alpha  = 2 \beta - 2r \dot \beta + 2r \dot \lambda 
(1 + \beta) + 2r^2(\dot \lambda^2 - \ddot \lambda) (1 + \beta) + r^2
(\ddot \beta - 3 \dot \lambda \dot \beta) \, ,   \\
\zeta = r \dot \alpha \, , \quad \eta_1 = \alpha r^{-3}
 \, , \quad \eta_2 = 3 \beta r^{-1}  \, , \quad \eta_3 = \zeta r^{-3} \, , 
\eea
the field equations    (\ref{11.6})  are just eqs. (2.15.a-c)
 and (2.16) of the paper  \cite{fiedler1983}. To avoid the products
 $ r \dot \eta_i$  we define a new 
independent variable $x$  by $r = le^x$.
 Then  $ r \dot \eta_i  = \eta_i ' =  d\eta_i/dx $ and we obtain
\be\label{11.8}
0  = \eta_1' + 3\eta_1 - \eta_3 \, , 
\ee
\be\label{11.9}
0    = (k + r \eta_3 /6) \eta_2 ' + \eta_1 + \eta_3 + r^2k\eta_1
 + r^3 \eta_1\eta_3/6 + r \eta_2\eta_3/2 + r^3\eta_1^2/4 \, , 
\ee
\bea\label{11.10}
0 = (1+r \eta_2/3) \eta_3' - 2 \eta_1 + 2k\eta_2 + 2 \eta_3 - r^2k\eta_1 -
 \nonumber \\
r^3 \eta_1^2 /2 - r^3 \eta_1\eta_3/6 + 2r \eta_2 \eta_3/3 \, .
\eea
Conversely,  if the system eqs.  (\ref{11.8}),  (\ref{11.9}),  (\ref{11.10})
 is solved, the   metric can be obtained by
\be\label{11.11}
\beta= r \eta_2/3 \, , \quad \dot \lambda = 
( 2r\dot \eta_2 + 2 \eta_2 + r^2 \eta_1)/(6 + 2r \eta_2  ) \, , 
\quad \lambda(0)=0 \, .
\ee

\subsection{Linearizations}\label{s113} 

Now we make some approximations to obtain the rough 
behaviour of the solutions. First,  for $r \to \infty$, 
 Stelle  \cite{stelle78}  has shown the following: Linearization about Minkowski 
spacetime   leads, using our notations  eqs.  (\ref{11.1}) and  (\ref{11.7}), 
to $\lambda   = 0$ and $\beta =   2 \Phi$,   where powers of $\Phi$
 are neglected and the term  $\exp (r/l)/r$ in $\Phi$
 must he suppressed because of asymptotical flatness.

\bigskip
 
That means, loosely
 speaking, the right hand side of  eq.  (\ref{11.6})  will be neglected. 
Connected with this one may doubt the relevance of the term 
$cr^{-1}e^{-r/l}$  in  eq.  (\ref{11.1})  for
 $m \ne  0$ and $r \to \infty$,  because it is small compared with the 
neglected terms.
 Further, a finite $\Phi(0)$  requires $c = m$  in  eq.  (\ref{11.1}). 
Then $\Phi$  is just the Bopp-Podolsky 
 potential stemming from  4th order electrodynamics,
 and it were  Pechlaner and  Sexl  \cite{pechlaner}
 who proposed this form as representing a gravitational potential.
Now we have 
\be
\Phi(r) = m \left( \exp (-r/l) - 1 \right)/r = -m/l + rm/(2l^2) + \dots \, , 
\ee
but also this finite potential  gives rise to a singularity: For one of 
the curvature  invariants we get $R_{ij}R^{ij} \approx m^2/r^4$ as $r \to  0$,
 that means, the linearization makes no sense for this region. 
From this one can already expect that in a neighbourhood 
of Minkowski spacetime no regular solutions exist.

\bigskip

Second, for $r \to  0$, in  \cite{fiedler1983} it is shown 
 that there exists a one-parameter family 
of solutions being singularity-free and analytical in 
a neighbourhood of $r = 0$.  The parameter will be called $\varepsilon$
 and can be defined as follows: neglecting
 the terms with $r$ in the system eqs.  (\ref{11.8}),  (\ref{11.9}),   (\ref{11.10})
 one obtains a linear system with constant coefficients possessing 
 just a one-parameter family of solutions being regular at $r = 0$; it reads
\be\label{11.12}
 \eta_1 = \varepsilon k^2r \, , \quad
\eta_2 = -5\varepsilon k r \,  ,   \quad   \eta_3  = 4 \varepsilon k^2r \,  .   
\ee
The factors $k^n$  are chosen such that $\varepsilon$
 becomes dimensionless.  Now one 
can take  eq.  (\ref{11.12})  as the first term of a power series 
 $\eta_i = \Sigma_n \, a_i^{(n)} r^n$ , 
and inserting this ansatz  into that system, 
   one iteratively obtains the coefficients $a_i^{(n)}$.
For $n$ even, $a_i^{(n)}$  vanishes:  The 
$r^3$-terms are $k^3(\varepsilon/14 + 10\varepsilon^2/21)$,
 $ -k^2(\varepsilon/2 - 10\varepsilon^2/3)$  and 
$k^3(3\varepsilon/7 + 20\varepsilon^2/7)$ respectively. 
Furthermore,  the $r^{2n-1}$-term is always a suitable power of $k$
 times a polynomial in $\varepsilon$ of the order $\le n$.

\bigskip
 
Up to the $r^2$-terms the corresponding metric   (\ref{11.7}) reads
\be\label{11.13}
     ds^2 = (1 + 5k\varepsilon r^2/3) dt^2
 - (1 + 5k\varepsilon r^2/3) dr^2 - r^2d \Omega^2 \, .
\ee

\bigskip
 
Third, we look for a linearization which holds uniformly for $0 \le  r < \infty$. 
A glance at  eqs.  (\ref{11.11})   and    (\ref{11.7}) shows that $\eta_i = 0$
 gives Minkowski spacetime, and therefore we neglect
 terms containing products of  $\eta_i $  in the system.
 In other words, instead of the linearization used before we additionally
 retain the term $r^2 \eta_1$.  Then we proceed as follows: from eqs.   (\ref{11.8})
and    (\ref{11.10}) we obtain
\be\label{11.14}
\eta_3 = \eta_1' + 3 \eta_1      \qquad {\rm  and}
\ee
\be\label{11.15}
\eta_2 =\frac{- \eta_1   ' - 5 \eta_1   + (e^{2x} - 4) \eta_1
+ l^3 e^{3x} \eta_1  (\eta_1 + \eta_1' /6)
}{2k+ l e^x (\eta_1'   + 5 \eta_1   + 6  \eta_1)/3} \, .
\ee
Inserting eqs.   (\ref{11.14}),  (\ref{11.15})
 into  eq.  (\ref{11.9}) one obtains a third order equation for $\eta_1$ 
only whose linearization reads 
\be\label{11.16}
0 = \eta_1''' + 5\eta_1'' + \eta_1'(2 - e^{2x}) - \eta_1(8+ 4e^{2x}) \,  .
\ee
In  Schmidt and  M\"uller  \cite{schmuell} the same 
vacuum  field equations were discussed  for axially symmetric Bianchi type I models.
 They possess a four-parameter group of isometries, 
too,  and the essential field equation is also a 
third order equation for one function. 
The difference is that in the present case spherical
 symmetry implies an explicit coordinate dependence.

\bigskip

The solution of  eq.  (\ref{11.16})  which is bounded for $x \to  - \infty$  
reads 
\be\label{11.17}
\eta_1 = \gamma \cdot \left[ (3e^{-4x} + e^{-2x} )
 \sinh \, e^x  - 3 e^{-3x} \cosh \, e^x \right] \, .
\ee
A second
 solution is    $ \eta_1 = - 12 mr^{-4} $
leading to the Schwarzschild solution,  
this solution solves both the full and the linearized equations
 in accordance with the fact that it makes zero both sides 
of  eq.  (\ref{11.6}), and the third one can be obtained from them up
 to quadrature by usual methods.  A comparison of  eq.  (\ref{11.17})
 with  eq.  (\ref{11.12}) gives $\gamma  = 15\varepsilon l^{-3}$.

\bigskip
 
Now we insert this $\eta_1$  into eqs.    (\ref{11.15}),   (\ref{11.11})
  and neglect   again powers of $\eta_1$; then the metric reads
\bea
\beta =   5 \varepsilon\left[l r^{-1} \sinh (r l^{-1})
 - \cosh (rl^{-1}) \right] \, ,               \nonumber            \\
\lambda = \frac{5 \varepsilon}{2 l} \cdot
\int_0^r \left[   (l^2 z^{-2} - 1) \sinh (zl^{-1}) - l z^{-1} \cosh (z l^{-1})
   \right]  dz \, . \label{11.18}
\eea
This linearization has, in  contrary to  eqs.  (\ref{11.1})   and   (\ref{11.12}), 
 the following property: to each $r_0> 0$ and $\Delta > 0$ there
 exists an $\delta > 0$ such that for $- \delta < \varepsilon < \delta$ 
the relative error of the linearized solution  (\ref{11.18}) 
does not exceed $\Delta$ uniformly on the interval $0 \le  r \le  r_0$.

\subsection{Non-standard analysis}\label{s114}

In this section we will prove that in a neighbourhood of Minkowski 
spacetime  there do not exist any solutions.  ``Neighbourhood of Minkowski 
spacetime''
 is in general a concept requiring additional explanations because of the large number of 
 different topologies discussed in literature.  To make the above statement sufficiently strong 
we apply a quite weak topology here: given a $\delta > 0$
 then all spacetimes   being diffeomorphic to Minkowski spacetime  and fulfilling
$ \vert R_{ij} R^{ij} \vert < \delta k^2$ 
form a  neighbourhood about Minkowski spacetime.
 In  eq.  (\ref{11.13})  we have at $r =0$,  independently of higher order terms in $r$,  
\be\label{11.19}
R_{ij} R^{ij}
     = 100 k^2 \varepsilon^2/3 \qquad {\rm  and} \qquad  R_{00} =  5k\varepsilon \, .
\ee
Therefore it holds: the one-parameter family of solutions
 being regular at $r = 0$ is invariantly characterized
 by the real parameter $\varepsilon$, and a necessary condition for it to be within 
a neighbourhood of Minkowski spacetime  is that $\varepsilon$
 lies in a neighbourhood of $0$.  

\bigskip

Now we suppose $\varepsilon$
 to be an infinitesimal number,  i.e., a positive number which is smaller 
than any positive real number. The mathematical theory dealing with such infinitesimals 
is called non-standard analysis, cf. Robinson  \cite{robinson}. 
The clue is that one can handle non-standard 
numbers like real numbers. Further we need the so-called
 Permanence principle, which is also called the Robinson lemma: 
Let $A(\varepsilon)$  be an internal statement holding for all 
infinitesimals $\varepsilon$. Then there exists a positive standard real 
$\delta > 0$ such that $A(\varepsilon)$  holds for all $\varepsilon$
 with   $0 < \varepsilon < \delta$.
 The presumption ``internal"    says, roughly speaking, that in the formulation 
 of $A (.)$  the words ``standard''   and ``infinitesimal" do not appear. 

\bigskip

This permanence principle shall be applied as follows:  $A(\varepsilon)$ 
is the statement: ``Take  eq.  (\ref{11.12})  as initial condition for eqs. 
   (\ref{11.8}),    (\ref{11.9}),   (\ref{11.10})
 and calculate the corresponding metric 
   (\ref{11.7}),   (\ref{11.11}). Then there exists 
an $r_0 \le l$  such that $R_{ij} R^{ij} \ge  k^2$ at
$r =   r_0$.''  Remark:  At this point it is not essential whether $r_0$  is 
a standard or an infinitely large non-standard number.

\bigskip
 
 Proof  of     $A(\varepsilon)$  for infinitesimals
   $\varepsilon$    with   $\varepsilon \ne 0$: the difference 
of   eq.  (\ref{11.18})   in relation to the exact solution is of the order $\varepsilon^2$, 
i.e. the relative error is infinitesimally small. 
For increasing $r$, $\,   \vert \beta \vert $  becomes arbitrary large, i.e. $(1 + \beta)^{-1}$ 
 becomes small and $R_{ij}R^{ij}$   increases to arbitrarily large values, cf.  
eq.  (\ref{11.18}).   Now take $r_0$ such that with metric 
    (\ref{11.7}),   (\ref{11.18})    $R_{ij}R^{ij} \ge  2k^2 $
 holds. Then, for the exact solution, 
$R_{ij}R^{ij} \ge  k^2 $    holds at $r = r_0$, because their 
difference was shown to be infinitesimally small.

\bigskip

Now the permanence principle tells us  that 
there exists a positive standard real $\delta> 0$
 such that for all $\varepsilon$ with $0 < \vert \varepsilon \vert  < \delta$   
the corresponding exact solution has a  point $ r_0$  such that at $r=  r_0  $
 the inequality  $   R_{ij}R^{ij} \ge  k^2 $ 
 holds. But ``$R_{ij}R^{ij} <  k^2 $''  is another necessary 
condition for a solution to lie within a neighbourhood of  Minkowski spacetime.
 Remark:  Supposed this $r_0$ is an infinitely large non-standard number, then by 
continuity arguments also a standard finite number with the same property exists.

\bigskip

The statement is proved, 
but we have learned nothing about the actual value of the number $\delta$. 
Here,   numerical calculations may help. 
They were performed as follows: the power series for the
 functions $\eta_i$
were calculated up to the $r^6$-term, then these functions taken at $x = -4 $
  i.e. $r = 0.018 \ l$   were used as initial conditions for  a Runge--Kutta 
integration.  We got the following result: firstly, for $\varepsilon  = \pm 10^{-5}$
 and $r \le 10 \ l$, the relative difference between the
 linearized solution  eq.  (\ref{11.17})  and the numerical
 one is less than 2 per thousand. 
Secondly, for $0 < \vert \varepsilon \vert \le 1$  the behaviour 
``$\beta \to     - \infty \cdot {\rm sgn} \varepsilon$" is
 confirmed. That means, the statement made above keeps valid at
least for $\delta = 1$.  Remark:  For large values 
$\varepsilon$ the power series for the $\eta_i$   converge very slowly, 
and therefore  other methods  would be necessary to decide about asymptotical flatness.

\subsection{Discussion  -- Einstein's particle programme}\label{s115}

Fourth-order gravitational field equations could be taken as a field theoretical model 
of ordinary matter, the energy--momentum tensor of which is defined by
\be\label{11.20} 
\kappa T^\ast_{ij}= R_{ij} - \frac{1}{2} g_{ij} R   \, .
\ee
For our case one obtains at $r=  0  $ that  $\ T^\ast_{ij}$
  represents  an ideal fluid with the equation of state $p^\ast =\mu^\ast/3 $
  and, cf.  eq.  (\ref{11.19}),  
\be\label{11.21} 
\mu^\ast (0) =     5 k \varepsilon  \kappa^{-1} =
5 \varepsilon /8\pi Gl^2 \, . 
\ee
Inserting $\vert \varepsilon \vert \ge 1$
 and $l \le 1.3\cdot 10^{-13}$~cm \  into  eq.  (\ref{11.21}), we obtain
\be\label{11.22} 
\vert \mu^\ast(0) \vert \ge 1.5 \cdot 10^{53} \ {\rm g \ cm}^{-3} \, . 
\ee
Therefore it holds: if there  exists a non-trivial static spherically symmetric
 asymptotically flat  singularity-free solution of  eq.  (\ref{11.5})  at all, then the
 corresponding particle would be a very massive one: 
its phenomenological energy density exceeds that of a neutron 
star by at least 40 orders of magnitude.

\bigskip

The resulting statement can be understood as follows: 
For a small curvature the 4th order corrections to Einstein's equations are 
small, too, and  the situation should not be very different from that one we 
know from Einstein's theory. Now we want to 
refer to a problem concerning Schwarzschild coordinates: the transition 
from a general static spherically symmetric 
 line element to Schwarzschild coordinates is possible 
only in the case that the function ``invariant surface of the sphere $r =$ const. 
in dependence on its invariant radius" has not any stationary point. 
Here two standpoints are possible: either 
one takes this as a natural condition for a reasonable particle
 model or one allows coordinate singularities in  eq.  (\ref{11.7}) 
like ``$\beta \ge -1$, and $\beta = -1$  at single points"
 which require a special care. 
 For  $\beta < -1$ one would obtain a cosmological model of 
Kantowski-Sachs type. Eqs.   (\ref{11.2}) till  (\ref{11.11})
 remain unchanged for this case. 
 The discussion made above is not influenced by this.

\bigskip
 
The statement on the existence of solutions can be strengthened 
as follows: Fiedler and  Schimming  \cite{fiedler1983} 
 proved that the solutions are  analytical  in a neighbourhood of $r = 0$.
 Further, the differential equation is an analytical one and, 
therefore, in the subspace of singularity-free solutions they remain so in the limit 
$r \to \infty$. Then, there exists only a finite or countably infinite set 
of values $\varepsilon_n$  such that the corresponding solution becomes 
asymptotically flat; the question, whether this
 set is empty or not, shall be subject of further investigation.  
Furthermore, the  $\varepsilon_n$  have no   finite accumulation point. 
That means, there exists at most a discrete spectrum of solutions.

\bigskip

With  respect to this
 fact, we remark the following: as one knows, Einstein's theory is a covariant one. 
But besides this symmetry, it is  homothetically invariant, too. That means, if $ds^2$  is
 changed to   $e^{2 \chi}ds^2$  with constant $\chi$, then the tensor 
$R_{ij} - \frac{1}{2} g_{ij} R$  remains unchanged,   whereas the
 scalars $R$ and $C$  will be divided by  $e^{2 \chi}$
 and $e^{4 \chi}$  respectively. From this it follows: with one solution 
of Einstein's  vacuum equation one obtains by  homothetic  invariance 
just a one-parameter class of solutions. On the other hand, the sum $R + l^2 C$ 
has not such a symmetry and, therefore, one should 
not expect that a one-parameter family of  solutions globally exists, and  
this is just in the scope of the particle programme where a definite particle's
  mass is wanted.

\vfill 

\subsection{Non-trivial solutions of the Bach equation exist}\label{s116}
 
Following \cite{sch84b}, we show that solutions of the Bach equation exist
 which are not conformal Einstein spaces. 
In connection with fourth-order gravitational field
 equations, cf.  e.g.  Weyl 1921 \cite{weyl19}, \cite{bty}
 where the breaking of conformal invariance  was discussed, the original Bach equation,
\be\label{10.1}  
B_{ij} = 0 \, ,             
\ee
enjoys current interest, see e.g. [64], [71] and  [91].
Eq.  (\ref{10.1}) stems from a Lagrangian 
\be
L = \frac{1}{2} \,  \sqrt{-g} \,  C_{ijkl} \,  C^{ijkl}   \,  ,
\ee
and variation gives, cf. Bach 1921 \cite{bach},
\be
 \frac{1}{\sqrt{-g}}\,   \delta L / \delta g^{ij}     
  = B_{ij} = 2 \,  C^{a \ \  b}_{\ ij \ ;ba}
 +          C^{a \ \  b}_{\ ij } \,  R_{ba}        \,  .
\ee
An Einstein space, defined by 
\be\label{10.2}
     R_{ij} = \lambda \, g_{ij} \,  , 
\ee
is always a solution of the  Bach
 equation   (\ref{10.1}). But  eq. (\ref{10.1}) is conformally invariant, and therefore, 
each metric, which is conformally related to an Einstein space, 
fulfils  eq. (\ref{10.1}), too. We call such solutions  trivial ones.

\bigskip

Now the question arises whether non-trivial solutions of the  Bach
 equation  eq. (\ref{10.1})  do or do not exist. In this section we 
 will give an affirmative answer. As a by-product, some conditions  will be
 given under which only trivial solutions exist. 
Observe that  eq. (\ref{10.1}) is conformally invariant whereas 
 eq. (\ref{10.2})  is not. 
Therefore, a simple counting of degrees of freedom does not suffice.

\bigskip

Because the full set of solutions of   eq. (\ref{10.1}) 
 is not easy to describe, let us consider some homogeneous cosmological models.
 Of course, we have to consider anisotropic ones, because all Friedmann 
 models are trivial solutions of  eq. (\ref{10.1}). 
Here, we concentrate on the diagonal Bianchi type I models
\be\label{10.3}
ds^2 = dt^2  - a^2_i \, \left( dx^{i} \right) ^2  
\ee
with Hubble parameters $h_i = a_i^{-1} da_i/dt$, $h=\Sigma h_i$
and anisotropy parameters $m_i = h_i 
-    h/3$. The Einstein spaces of this kind are described in \cite{kramer}
 eq. (11.52), for $\lambda =  0$ it is just the Kasner
metric $ a_i =   t^{p_i}$,   $\Sigma p_i = \Sigma p_i^2 =   1$. 
All these solutions have the property that the quotient of two
anisotropy parameters,   $m_i/m_j$, which 
equals $ (3p_i - 1)/(3p_j - 1) $  for the Kasner metric,  is independent
of $t$, and this property is a conformally invariant one. 
Furthermore, it holds:   A solution of eqs.   (\ref{10.1}),
  (\ref{10.3}) is a trivial one, if and only if the quotients   $m_i/m_j$    are constants.

\bigskip

Restricting now to axially symmetric Bianchi type I models, 
i.e., metric   (\ref{10.3}) with $ h_1 = h_2$, the
identity $\Sigma m_i = 0$
 implies $m_1/m_2 = 1$, $m_3/m_1 =   m_3/m_2 =   -2$ ,
 i.e.,  each axially symmetric Bianchi
type I  solution of   eq. (\ref{10.1})  is  conformally related to an Einstein space.
 Analogously, all static 
spherically symmetric  solutions of   eq. (\ref{10.1})  are trivial ones, cf.
 \cite{fiedler1980}. 

\bigskip

Finally, the existence of a solution of eqs. 
  (\ref{10.1}),  (\ref{10.3})  with a non-constant
 $ m_1/m_2 $ will be shown. For
the sake of simplicity we use the gauge condition $h = 0$,
 which  is possible because of  the
conformal invariance of   eq. (\ref{10.1}). 
Then the 00 component and the 11 component of  eq. (\ref{10.1})  are
sufficient to determine the unknown functions $h_1$  and $h_2$;
 $h_3 = -h_1 - h_2$   follows from the gauge condition. 
Defining $r = (h^2_1 + h_1h_2 + h_2^2)^{1/2}$  and $p = h_1/r$, 
  eq. (\ref{10.1})  is equivalent to the system    
\be\label{10.4}
3d^2(pr)/dt^2 =  8pr^3 + 4c, \qquad c = {\rm const.,} \qquad   p^2 \le  4/3 \,  ,  
\ee
\be\label{10.5} 
9(dp/dt)^2 r^4 =   [2r d^2 r/dt^2 - (dr/dt)^2 - 4r^4] (4r^2 - 3p^2r^2) \,  . 
\ee
As one can see, solutions with a non-constant $p$
 exist, i.e., $m_1/m_2$  is not constant for this case.

\bigskip

Result:  Each solution of the  system
  eqs. (\ref{10.3}),  (\ref{10.4}),  (\ref{10.5})
  with $dp/dt \ne  0$  represents a non-trivial solution
of the Bach equation   (\ref{10.1}).

\newpage

\section{The Newtonian limit of non-linear  gravity}\label{Kap12}
\setcounter{equation}{0}

In this chapter, the weak-field slow-motion limit of fourth and 
higher-order gravity will be 
deduced. Here we follow \cite{sch86c} for the fourth-order case
and \cite{quandt} for the general case. More explicitly: we consider the 
Newtonian limit of the theory based on the Lagrangian 
\be 
{\cal L} = \left( R + \sum_{k=0}^p \, a_k \, R \Box^k R 
 \right) \sqrt{-g} \, .
\ee
The gravitational potential of a point mass turns out to be a 
combination of Newtonian and Yukawa terms. 
For fourth-order and sixth-order gravity, $p=0$ and  
$p = 1$ respectively,  the coefficients are calculated
explicitly. For general $p$ one gets the potential to be 
\be
\Phi  = m/r \, \left( 1 +  \sum_{i=0}^p \,   c_i  \exp (-r/l_i) \right)
\ee
with certain coefficients $c_i$ fulfilling the relation
\be
  \sum_{i=0}^p \,   c_i = 1/3 \, .
\ee  
Therefore, the potential is always unbounded near the origin, see
also [10].

\subsection{The Newtonian limit of 4th-order  gravity}
Let  us consider the gravitational theory defined by the Lagrangian
\be\label{12.1}
 L_{\rm g}=  (8\pi G)^{-1} \Bigl(
R/2 + (\a R_{ij}R^{ij} + \b R^2) l^2 \Bigr) \, . 
\ee
$G$ is Newton's constant, $l$ a coupling length and $\a$
 and $\b$  numerical parameters. $R_{ij}$  and $R$  
are the Ricci tensor and its trace. Introducing the 
matter Lagrangian $L_{\rm m}$  and varying  $  L_{\rm g}
   +  L_{\rm m}$   one obtains the field equation
\be\label{12.2}
     E_{ij}  + \a H_{ij} + \b G_{ij}  = 8\pi G \,  T_{ij}  \,   .         
\ee
For $\a = \b = 0$  this reduces to
 General Relativity Theory. The explicit expressions $H_{ij}$ and $ G_{ij}$
 can be found in sections \ref{s74x}
 and \ref{s112}, see also Stelle  \cite{stelle78}.

\bigskip

In a well-defined sense, the weak-field slow-motion limit 
of Einstein's theory is just Newton's theory, 
cf.  Dautcourt 1964  \cite{dautcourt1963}. In the following,  
we consider the analogous problem for fourth-order
 gravity eqs.  (\ref{12.1}),  (\ref{12.2}).
 In some  cases,  the Newtonian limit of the theory defined by 
 eq. (\ref{12.2})  has already been deduced  in the literature: 
 For the special case $\a = 0$ see Pechlaner and   
Sexl  \cite{pechlaner} or   Polijevktov-Nikoladze  \cite{poli}. For the 
case $\a + 2\b = 0$ see Havas  \cite{havas} or  Jankiewicz  
 \cite{jankiewicz},  and for 
$\a  + 3\b  = 0$ see Borzeszkowski, Treder and 
Yourgrau  \cite{bty}. Cf. also Anandan   \cite{anandan},  where torsion 
has been taken into account.

\bigskip

The slow-motion limit can be equivalently described 
as the limit $c \to \infty$, where $c$  is the  velocity of  light. 
In this sense we have to  take  the  limit 
$G \to  0$  while $G \cdot  c$  and $l$  remain constants. 
Then the energy-momentum tensor $T_{ij}$ 
 reduces to the rest mass density $\rho$:
\be\label{12.3}
     T_{ij}=    \delta^0_i \delta^0_j \rho  \,    ,    
\ee
$x^0=     t $ being the time coordinate. The metric  can be written as
\be\label{12.4}
     ds^2 =  (1 - 2\phi) dt^2
 - (1 + 2\psi) (dx^2 + dy^2 + dz^2)          \, .
\ee
Now eqs.   (\ref{12.3})  and   (\ref{12.4}) will be inserted into  
eq. (\ref{12.2}).  In our approach, products and time derivatives 
 of   $\phi$   and $\psi$ can be neglected,  i.e.,
\be
R=4 \D \psi -2 \D \phi \, ,    \qquad {\rm where} \qquad 
\D f =f_{,xx} + f_{,yy} + f_{,zz} \, .
\ee
Further  $R_{00} = - \D \phi$, $H_{00} = -2 \D R_{00} - \D R$
  and $G_{00} = -4 \D R$, where $l = 1$.

\bigskip

Then it holds: The
 validity of the $00$-component and of the trace of  eq. (\ref{12.2}),
\be\label{12.5}
 R_{00} - R/2 + \a H_{00} + \b G_{00} = 8 \pi  G \rho    
\ee
and
\be\label{12.6} 
- R - 4(\a + 3 \b ) \D R = 8 \pi G \rho \,  ,
\ee
imply the validity of the full  eq. (\ref{12.2}).

\bigskip

Now, let us discuss eqs. (\ref{12.5}) and  (\ref{12.6})
 in more details: Eq.   (\ref{12.5})  reads
\be\label{12.7}
- \D \phi    - R/2 +\a (2 \D \D \phi - \D R) - 4 \b \D R = 8 \pi G \rho \, .  
\ee
Subtracting one half of  eq. (\ref{12.6})  yields
\be\label{12.8}
- \D \phi   + 2 \a  \D \D \phi + (\a + 2 \b) \D R = 4 \pi G \rho \, .  
\ee
For $\a  + 2 \b  = 0$  one obtains
\be\label{12.9}
- (1 -  2 \a  \D ) \D \phi   = 4 \pi G \rho 
\ee
and then   $\psi  = \phi$  is a solution of eqs.  (\ref{12.5}),  (\ref{12.6}). 
For all other cases the equations for 
$\phi$  and $\psi $ do not decouple immediately, but, 
to get equations comparable with Poisson's equation 
we apply $\D$ to   eq. (\ref{12.6})   and continue as follows.

\bigskip

     For $\a + 3\b  = 0$  one gets from  eq. (\ref{12.8}) 
\be\label{12.10}
- (1 -  2 \a  \D ) \D \phi   = 4 \pi G ( 1 + 2 \a \D /3 )  \rho   \, .
\ee
The $\D$-operator applied to the source term 
in  eq. (\ref{12.10})   is only due to  the application  of 
$\D$ to the trace, the original equations
   (\ref{12.5}),  (\ref{12.6})  contain only $\rho$ itself.

\bigskip

For $\a = 0$ one obtains similarly the equation
\be\label{12.11}
- (1 + 12 \b  \D ) \D \phi   = 4 \pi G ( 1 + 16 \b \D )  \rho   \, .
\ee

\bigskip

For all other cases - just the cases not yet covered 
by the literature - the elimination of $\psi$ from  the 
system   (\ref{12.5}),  (\ref{12.6})  gives rise to a sixth-order equation
\be\label{12.12}
- \bigl(1 + 4(\a + 3 \b)   \D \bigr)   ( 1 - 2 \a \D)
  \D \phi   =   4 \pi G \bigl( 1 + 2(3 \a + 8 \b) \D \bigr)  \rho   \, .
\ee

\bigskip

Fourth-order gravity is motivated by quantum-gravity 
considerations and therefore, its long-range behaviour 
should be the same as in Newton's theory. 
Therefore, the signs of the parameters $\a$, $\b$ should 
be chosen to guarantee an exponentially vanishing and 
not an oscillating behaviour of the fourth-order terms:
\be\label{12.13}
\a \ge 0 \, , \qquad   \a + 3 \b \le 0 \, .
\ee
On the other hand, comparing parts of  eq. (\ref{12.12})  with  
the Proca equation it makes sense to  define the masses 
\be\label{12.14}
m_2 = \left(2 \a  \right)^{-1/2} \quad {\rm and} \quad 
m_0 = \left(  -4(\a + 3\b )    \right)^{-1/2} \, .
\ee
Then  eq. (\ref{12.13}) prevents  the masses of the spin 2 and spin 0 gravitons to 
become imaginary.

\bigskip

Now, inserting a delta source $\rho  = m \delta $  into  eq. (\ref{12.12})  one 
obtains for $\phi $ the same result as Stelle  \cite{stelle78}, 
\be\label{12.15} 
\phi = m r^{-1} \bigl( 1 + \exp (-m_0 r)/3 - 4 \exp (-m_2 r)/3 \bigr) \, .
\ee
To obtain the metric completely one has also to calculate $\psi$. It reads
\be\label{12.16}
\psi = m r^{-1} \bigl( 1 - \exp (-m_0 r)/3 - 2 \exp (-m_2 r)/3 \bigr) \, .
\ee
For finite values $m_0$ and $m_2$  these are both 
bounded functions,  also for $r \to 0$. In the limits $\a \to  0$, i.e. 
  $m_2 \to \infty$,    and
 $\a + 3\b \to 0$, i.e.  $m_0 \to \infty$,   the terms with   $m_0$ and $m_2$   
in eqs.   (\ref{12.15})  and  (\ref{12.16})
 simply vanish.  For these cases $\phi$  and $\psi $ become
 unbounded as $r \to 0$.

\bigskip

Inserting  eqs. (\ref{12.15}) and   (\ref{12.16}) into the metric  
  (\ref{12.4}),  the behaviour of 
the geodesics shall be studied. First, for an estimation 
of the sign of the gravitational force we take a test 
particle at rest and look whether it starts falling towards 
the centre or not. The result is: for $m_0 \le 2 m_2$,  
gravitation is always attractive,  and for $m_0> 2m_2$  it is attractive 
for large but repelling for small distances.
 The intermediate case  $m_0 = 2m_2$,  i.e., $3\a + 8\b =  0$, is
 already known to  be a special one from  eq. (\ref{12.12}).

\bigskip

Next,  let us study the perihelion advance 
for distorted circle-like orbits. Besides the general 
relativistic perihelion advance, which vanishes 
in the Newtonian limit,  we have an additional 
one of the following behaviour: for $r \to   0$  
and $ r \to \infty $  it vanishes and for  $r \approx 1/m_0$ 
and $r \approx 1/m_2$  it has local maxima,  i.e., resonances.

\bigskip

Finally,  it should be mentioned that the  gravitational field of an extended 
body can be obtained by integrating eqs.   (\ref{12.15}) and   (\ref{12.16}). 
For a spherically symmetric body 
the far field is also of  the type 
\be
 m r^{-1} \Bigl( 1 + a \exp (-m_0 r) + b  \exp (-m_2 r) \Bigr) \, ,
\ee
 and the  factors $a$ and $b$ carry 
information  about the mass distribution inside the body.

\subsection{Introduction to higher-order gravity}

One-loop quantum corrections
 to the Einstein equation can be described by curvature-squared terms and 
lead to fourth-order
gravitational field equations; their Newtonian limit is described by a 
potential ``Newton + one Yukawa term", cf. e.g. Stelle \cite{stelle78}
 and  Teyssandier \cite{teyss}. 
 A Yukawa potential has the form $\exp (-r/l)/r$  and was originally 
used by Yukawa \cite{yu}  to describe the meson field.

\bigskip

Higher-loop quantum corrections to the Einstein equation are expected to 
contain terms of the type $R \Box^k R$  in the Lagrangian, which leads to a 
gravitational field equation of order $2k + 4$, cf. \cite{gss}.
Some preliminary results to this type of equations are already due 
to Buchdahl \cite{buch1}. For $k=1$, the cosmological consequences of 
the corresponding  sixth-order field equations 
are  discussed by Berkin and  Maeda \cite{berkin}, 
 and by  Gottl\"ober,  M\"uller and  Schmidt \cite{gms}.

\bigskip

In the present chapter we deduce the 
Newtonian limit following from this higher order field equation. The Newtonian 
limit of General Relativity Theory is the usual Newtonian theory, cf.  
e.g. Dautcourt 1964 \cite{dautcourt1963} or Stephani \cite{stephani}.
 From the general structure of the linearized higher-order field equation, 
cf. \cite{sch90d}, one can expect that for this higher-order 
gravity the far field of the point mass in the Newtonian limit 
is the Newtonian  potential plus a sum of different Yukawa terms. 
And just this form is that one discussed in connection with the fifth force, 
cf. \cite{ger}, \cite{stacey} and \cite{san}. 
Here we are interested in the details of this connection 
between higher-order gravity and the lengths and coefficients in 
the corresponding Yukawa terms.

\subsection{Lagrangian and field equation}

Let us start with the Lagrangian
\be\label{12x}
{\cal L} = \left( R + \sum_{k=0}^p \, a_k \, R \Box^k R 
 \right) \cdot \sqrt{-g} 
\, ,  \qquad a_p \ne 0 \, .
\ee
In our considerations we will assume 
that for the gravitational  constant $G$
 and for the speed of  light $c$ 
it holds $G = c = 1$. This only means a special 
choice of units. In eq. (\ref{12x}), $R$ denotes the curvature scalar, 
$\Box$ the D'Alembertian, and $g$ the determinant of the metric. 
Consequently, the coefficient $a_k$ 
 has the dimension ``length to the power $2k + 2$".

\bigskip

The starting point for the deduction of the field 
equation is the principle of minimal 
action. A necessary condition for it is the stationarity of the action: 
\be
- \, \frac{\d {\cal L}}{\d g_{ij}} = 8 \pi \, T^{ij} \,  \sqrt{-g} \, ,
\ee
 where $T^{ij}$ denotes the energy-momentum tensor. The explicit
equations for 
\be 
P^{ij} \,  \sqrt{-g}= - \,  \frac{\d {\cal L}}{\d g_{ij}}
\ee
    are given in \cite{sch90d}.  Here we only need the 
linearized field equation. It reads, cf. \cite{gss}
\be\label{12y}
P^{ij}  \equiv R^{ij} - \frac{R}{2} \,   g^{ij} + 2 \sum_{k=0}^p \, a_k  [
 g^{ij} \Box^{k+1} \, R - \Box^k  R^{\, ; \, ij} ] = 8 \pi T^{ij}\, , 
\ee
and for the trace it holds:
\be
g_{ij} \cdot P^{ij} =  -  \frac{n-1}{2}  R + 2n   \sum_{k=0}^p \, a_k [
 g^{ij} \Box^{k+1} \, R ]  = 8 \pi T \, .
\ee
$n$ is the number of spatial dimensions; 
the most important application is of cause $n = 3$.
 From now on we put $n = 3$.

\subsection[Newtonian limit in higher-order gravity]{The Newtonian 
limit in higher-order gravity  }

The Newtonian limit is the weak-field static  
approximation. So we use the linearized 
field equation and insert a static metric and an energy-momentum tensor
\be
T_{ij} = \d^0_i \ \, \d^0_j \, \rho \, , \qquad \rho  \ge  0
\ee
into eq. (\ref{12y}).

\bigskip

Without proof we mention that the metric  can be brought into spatially 
conformally flat form,  and so we may  use
\bea 
g_{ij} = \eta_{ij} + f_{ij } \, ,
  \nonumber \\
\eta_{ij} = {\rm  diag} (1, \,  -1, \,  -1, \,  -1)
 \qquad {\rm and}
 \nonumber \\
f_{ij} = {\rm  diag} (-2\Phi, \, -2\Psi , \,  -2\Psi , \,  -2\Psi ) \, .
\eea
 Then the  metric equals
\be\label{12t}
ds^2 = (1 - 2\Phi) dt^2 - (1 + 2\Psi ) (dx^2 + dy^2 + dz^2) \, ,  
\ee
where $\Phi$ and $\Psi$  depend on $x$, $y$ and $z$.
 Linearization means that the metric $g_{ij}$
 has only a small difference to  $\eta_{ij}$; 
quadratic expressions in  $f_{ij}$ and its 
partial derivatives are neglected.  
We especially consider the case of a point mass. In this case it holds: 
$\Phi  = \Phi(r)$, $ \Psi  = \Psi (r)$,  with 
\be
r = \sqrt{x^2+y^2 +z^2} \, , 
\ee
 because of spherical  symmetry and $\rho = m \,  \d$.  
Using these properties, we deduce the field equation
 and discuss the existence  of solutions of the above mentioned type.

\bigskip

At first we make some helpful general considerations: The functions 
$\Phi$ and $\Psi $  are determined
 by eq. (\ref{12y}) for $i = j = 0$ and the trace of eq. (\ref{12y}). If these 
two equations hold, then  all other component-equations are automatically 
satisfied. For the 00-equation we need $R_{00}$:
\be
R_{00} = - \Delta \Phi \, .
\ee
Here, the Laplacian is given  is as usual by 
\be 
\Delta  =  \frac{\pa^2}{\pa x^2} +  \frac{\pa^2}{\pa y^2}
 +  \frac{\pa^2}{\pa z^2} \, . 
\ee
For the inverse metric we get
\be
g^{ij} = {\rm diag} \left( 1/
 (1 - 2\Phi) , \,   - 1/ (1 + 2\Psi ) , \,   - 1/ (1 + 2\Psi ) , \,   - 1/ (1 + 2\Psi )
 \right) 
\ee
and $ 1/  (1 - 2\Phi)  = 1 + 2\Phi  + h(\Phi )$, where $h(\Phi )$ is 
quadratic in $\Phi$  and vanishes  after linearization. So we get 
\be
g^{ij} = \eta^{ij} - f^{ij} \, .
\ee
 In our coordinate system, $f^{ij}$   equals $f_{ij}$  for all $i, j$. 
For the curvature scalar we get
\be
     R = 2(2 \Delta \Psi - \Delta \Phi )     \, .
\ee
Moreover, we need expressions of the type
 $\Box^k \, R$. $\Box R$  is  defined by $ \Box R  = R_{\, ; \, ij} \, g^{ij}$,   
where ``$;$"  denotes the covariant derivative. 
Remarks: Because of linearization we may replace
 the covariant derivative with  the partial one. So we get
\be
 \Box^k \,  R = (~1)^k \, 2( - \Delta^{k+1} \Phi  + 2 \Delta^{k+1} \Psi )    
\ee
and after some calculus
\be\label{12z}
          -8 \pi \rho  = \Delta  \Phi  + \Delta \Psi \,  .    
\ee
We use eq. (\ref{12z}) to eliminate $\Psi$ from the system. 
So we get   an  equation relating $\Phi$   and $\rho  = m \d$.
\be\label{12p}
     -4\pi \left( \rho + 8  \sum_{k=0}^p \, a_k
 (-1)^k  \Delta^{k+1} \,  \rho \right) 
 = \Delta  \Phi  + 6  \sum_{k=0}^p \, a_k  (-1)^k  \Delta^{k+2} \Phi \, .
\ee
In spherical coordinates it holds
\be 
\Delta   \Phi = \frac{2}{r} \Phi_{\, , \, r }  +\Phi_{\, , \, r r } \, ,  
\ee
because $\Phi$  depends on the radial coordinate $r$ only.

\bigskip

We apply the following lemma: In the sense of distributions it holds
\be
\Delta \left( \frac{1}{r} e^{-r/l} \right) = \frac{1}{rl^2}e^{-r/l} - 4 \pi \d \, . 
\ee
Now we are ready to solve the whole problem. We assume
\be
\Phi  = \frac{m}{r} \, \left( 1 +  \sum_{i=0}^q \,
   c_i  \exp (-r/l_i) \right) \, , \quad l_i > 0 \, .
\ee
Without loss of generality we may assume $l_i \ne l_j$
 for $i \ne j$. Then eq. (\ref{12p})  together with that  lemma  yield
\bea
8\pi   \sum_{k=0}^p \, a_k   (-1)^k  \Delta^{k+1} \,  \d =   \sum_{i=0}^q \, 
\left(\frac{c_i}{t_i} + 6 \sum_{k=0}^p \, a_k  (-1)^k   \frac{c_i}{t_i^{k+2}}
\right)  \frac{1}{rl^2}e^{-r/l_i} \nonumber \\  - 4 \pi   \sum_{i=0}^q 
\left(c_i  + 6 \sum_{k=0}^p  a_k   (-1)^k  
\frac{c_i}{t_i^{k+1}} \right) \d   \nonumber \\
+ 24 \,  \pi \sum_{k=0}^p  \sum_{j=k}^p \sum_{i=0}^p
 c_i a_j (-1)^{j+1} \frac{1}{t_i^{j-k}}  \Delta^{k+1} \d 
\eea
where $t_i = l_i^2$ \, ; therefore also $t_i \ne t_j$ for $i\ne  j$.
This equation is equivalent to the system 
\bea
\sum_{i=0}^q c_i = 1/3 \, , \label{12r}
\\
\sum_{i=0}^q \frac{c_i}{t_i^s} = 0\, ,  \qquad s = 1, \dots p \label{12s}
\\
t_i^{p+1} + 6 \sum_{k=0}^p a_k (-1)^k t_i^{p-k }
= 0\, ,  \qquad i=0, \dots q \, . \label{12q}
\eea
From eq. (\ref{12q}) we see that the values 
$t_i$ represent $q + 1$ different solutions of one polynomial. This 
polynominal   has the  degree $p + 1$. Therefore $q \le  p$.

\bigskip

Now we use eqs. (\ref{12r})  and (\ref{12s}) . They can be written in matrix form as
\be
 \left( \begin{array}{c}  1 \dots  1\\  1/ t_0  \dots  1/t_q
 \\ \dots \\  1/ t_0^p  \dots  1/t_q^p  \end{array} \right)  \cdot 
 \left( \begin{array}{c} c_0 \\ c_1 \\ \dots \\ c_q \end{array} 
 \right)  = \left( \begin{array}{c} 1/3 \\ 0 \\ \dots \\ 0 \end{array}  \right)
\ee
Here, the first $q + 1$ rows form a regular matrix, the Vandermonde matrix. 
Therefore, we get
\be
1/t_i^j =  \sum_{k=0}^q    \l_{jk} \, / \,  t_i^k \qquad  j = q + 1, \dots    p
\ee
with certain coefficients $    \l_{jk}$
 i.e., the remaining  rows depend on the first $q + 1$ ones. If $    \l_{j0} \ne 0$
 then the system has no solution. So   $    \l_{j0}= 0$  for all $ q + 1 \le  j \le  p$. 
But for $q < p$ we would get 
\be
1/t_i^q = \sum_{k=1}^q    \l_{q+1\, k} \, / \,  t_i^{k-1} 
\ee
and this is a contradiction to the above stated regularity. 
Therefore $p$ equals $q$. The polynomial in (14) may be written as
\be
6 \cdot  \left( \begin{array}{c} 1 \  1/t_0 \dots  1/t_0^p\\ 
 \dots \\  1 \  1/ t_p  \dots  1/t_p^p  \end{array} \right)  \cdot 
 \left( \begin{array}{c} a_0 \\  \dots \\ (-1)^p a_p \end{array} 
 \right)  = \left( \begin{array}{c}
-t_0 \\ \dots  \\  - \t_p \end{array}  \right)
\ee
This matrix is again a Vandermonde one, i.e., there exists always a unique 
solution $(a_0, \dots  a_p)$, which  are the coefficients of the quantum 
corrections to the Einstein equation,  such that the  Newtonian limit of 
the corresponding gravitational field equation is a sum of Newtonian and 
Yukawa potential with prescribed lengths $l_i$. A more explicit form of the 
solution is given in  section \ref{s126}.

\subsection{Discussion of the weak-field limit}

Let us give some special examples 
of the deduced formulas of the Newtonian 
limit of the theory described by the Lagrangian (\ref{12x}). 
If all the $a_i$ vanish we get of course the usual Newton theory
\be
\Phi = \frac{m}{r} \, , \quad      \Delta \Phi   = -4 \pi \d \, .
\ee
$\Phi $   and $\Psi $ refer to   the metric  according to eq. (\ref{12t}). For
$ p = 0$ we  get for $a_0 <0$
\be
\Phi =  \frac{m}{r} \left[  1 + \frac{1}{3} \, e^{-r/\sqrt{-6 a_0}} \right]
\ee
cf. \cite{stelle78}  and  
\be
\Psi =  \frac{m}{r} \left[  1 - \frac{1}{3} \, e^{-r/\sqrt{-6 a_0}} \right]
\ee
cf. \cite{sch86c}.   For $ a_0 > 0$  no Newtonian limit exists.

\bigskip

For $p = 1$, i.e., the theory following from sixth-order gravity
\be
{\cal L} = \left( R + a_0 R^2 +  a_1  R \Box R 
 \right) \sqrt{-g} \, , 
\ee
we get, see \cite{quandt}
\be
\Phi = \frac{m}{r} \left[  1 + c_0  e^{-r/l_0} +  c_1  e^{-r/l_1} \right]
\ee
and
\be
\Psi =     \frac{m}{r} \left[  1 - c_0 e^{-r/l_0} - c_1 e^{-r/l_1}  \right]
\ee
where
\be
c_{0,1} = \frac{1}{6} \mp \frac{a_0}{2 \sqrt{9a_0^2 + 6 a_1}}
\ee
and
\be
l_{0,1}= \sqrt{- 3 a_0 \pm \sqrt{9 a_0^2 + 6a_1  }} \, .
\ee
This result is similar in structure but has different coefficients as 
in fourth-order gravity with  included square of the Weyl tensor in the Lagrangian.

\bigskip

The Newtonian limit for the degenerated case $l_0 = l_1$
 can be obtained by a limiting procedure as follows:
 As we already know $a_0 <0$,
 we can choose the length unit such that $a_0 = - 1/3$. 
The limiting case $9 a_0^2 + 6a_1 \to 0$ 
may be expressed by $a_1 = - 1/6 + \e^2$.  
After linearization in $\e$ we get:
\be
l_i = 1 \pm \sqrt{3/2} \, \e \, c_i   = 1/6 \pm 1/(6 \sqrt 6 \e)
\ee
and applying the limit $\e \to 0$ to the corresponding fields $\Phi $ and $\Psi$ we get 
\bea
\Phi  = m/r  \{1 + (1/3 + r/6) e^{-r} \}
\nonumber \\
 \Psi =    m/r  \{1 - (1/3 + r/6) e^{-r} \} \, . 
\eea
For the general case $p > 1$, the potential is a complicated expression, 
but some properties are explicitly known, these hold also for $p = 0, 1$.
 One gets
\be
\Phi  = m/r \, \left( 1 +  \sum_{i=0}^p \,   c_i  \exp (-r/l_i) \right)
\ee
and
\be
\Psi  = m/r \, \left( 1 -  \sum_{i=0}^p \,   c_i  \exp (-r/l_i) \right)
\ee
where $\sum c_i  = 1/3$; $\sum$ means  $\sum_{i=0}^p$  and $l_i$  and $c_i$
 are, up to permutation of indices,  uniquely determined by the Lagrangian.

\bigskip

There exist some inequalities between the coefficients $a_i$, which must be fulfilled in 
order to get a physically acceptable Newtonian limit. By this phrase we mean 
that besides the above conditions, additionally the fields $\Phi$  and
 $\Psi$  vanish for $r \to \infty$  and that the derivatives $d\Phi /dr$ and $d\Psi /dr$
 behave like $O(1/r^2)$. These inequalities express essentially the fact that
the $l_i$  are real, positive, and different from each other. 
The last of these three conditions can be weakened by allowing the
 $c_i$  to be polynomials in $r$  instead of being constants, cf. the example 
with $p = 1$ calculated above.

\bigskip

The equality $\sum c_i = 1/3$  means that the gravitational potential is 
unbounded and behaves, up to a factor 4/3,  like the Newtonian 
potential for $r \approx  0$. The equation $\Phi  + \Psi = 2m/r$
 enables us to rewrite the metric as
\be
ds^2 = (1 - 2 \theta) \left[ (1-    2m/r) dt^2
- (1  + 2m/r)    (dx^2 + dy^2 + dz^2) \right] \, ,
\ee
which is the conformally transformed linearized
 Schwarzschild metric with the conformal factor $1 - 2\theta$, where
\be
\theta  = \frac{m}{r} \sum c_i e^{-r/l_i}
\ee
can be expressed as functional of  the curvature scalar, this is the linearized
 version of  the conformal transformation theorem, cf. \cite{sch90d}.
 For an arbitrary matter configuration the gravitational 
potential can be obtained by the usual integration procedure.

\subsection{A homogeneous sphere}\label{s126}
For general $p$ and characteristic lengths $l_i$ fulfilling 
$0 < l_0 < l_1 < \dots < l_p$  we write the Lagrangian as
\bea
{\cal L} =R - \frac{R}{6} \left[
(l_0^2 + \dots + l_p^2) R + (l_0^2 l_1^2 + l_0^2l_2^2
 \dots + l_{p-1}^2 l_p^2) \Box  R + \right.
\nonumber \\
\left. 
(l_0^2 l_1^2l_2^2 + \dots +  l_{p-2}^2 l_{p-1}^2
 l_p^2) \Box^2 R + \dots 
+ l_0^2 \cdot l_1^2 \cdot \dots \cdot l_p^2 \Box^p R \right] 
\eea
the coefficients in front of $\Box^i R$ in this formula read
\be
\sum_{0 \le j_0 < j_1 < \dots < j_i \le p}
 \quad  \prod_{m=0}^i \, l^2_{j_m} \, .
\ee
Using this form of the Lagrangian, the gravitational potential of a 
point mass reads
\bea
\Phi =     \frac{m}{r} \left[   1 + \frac{1}{3}
 \sum_{i=0}^p (-1)^{i+1}   \prod_{j\ne i}
   \vert \frac{l_j^2}{l_i^2} -1 \vert ^{-1} \, 
  e^{-r/l_i} \right] \, ,  \\ \Psi =     \frac{m}{r} \left[
 1 - \frac{1}{3}  \sum_{i=0}^p (-1)^{i+1}  \prod_{j\ne i}
   \vert \frac{l_j^2}{l_i^2} -1 \vert ^{-1} \,   e^{-r/l_i} \right] \, . 
\eea
For a homogeneous sphere of radius $r_0$  and mass $m$  we get
\be 
\Phi =     \frac{m}{r} \left[  1 + \frac{1}{r_0^3}  \sum_{i=0}^p
  e^{-r/l_i} \, l_i^2 \, \tilde c_i  \bigl( 
r_0 \cosh (r_0/l_i) - l_i \sinh (r_0/l_i)  \bigr)  \right] \, , 
\ee
where
\be
\tilde c_i =   (-1)^{i+1}  \prod_{j\ne i}
   \vert \frac{l_j^2}{l_i^2} -1 \vert ^{-1} \, .
\ee

\newpage

\section[Cosmic strings and gravitational waves]{Cosmic
strings and gravitational waves}
\setcounter{equation}{0}
\setcounter{page}{175}

 We consider strings with the Nambu action as extremal surfaces in a given spacetime, 
thus, we ignore their back reaction. Especially, we look for strings sharing one symmetry 
with the underlying spacetime. If this is a non-null symmetry the problem of determining 
the motion of the string can be dimensionally reduced. Following \cite{sem}, we
 get exact solutions for the 
following cases: straight and circle-like strings in a Friedmann background, straight 
strings in an anisotropic Kasner background, different types of strings in the metric 
of a gravitational wave. The solutions will be discussed.

\subsection{Introduction to cosmic strings}

To give detailed arguments for 
considering strings means carrying coals to Newcastle. So we only list the main points:
A string is, generally speaking, an object possessing a two-dimensional world surface, 
 which is also called a world sheet,
 in contrast to a point particle possessing a one-dimensional world line, 
cf. the review article by Vilenkin \cite{vil}. In details we have

\medskip

1.   One considers strings with the Nambu action in a $D$-dimensional 
flat spacetime. The theory can be consistently quantized for $D = 26$ only: 
otherwise the light cone quantization leads to a breakdown of 
Lorentz covariance, cf. Green, Schwarz and Witten \cite{gsw}.
But we consider a classical non-quantized theory only and 
require $D = 4$ henceforth.

\medskip

2.   Cosmic strings are one-dimensional topological defects in gauge field theories. 
They have a large mass and can be seeds for larger objects by accretion 
processes, cf. Zeldovich \cite{z}.

\medskip

3.   One looks for solutions of the Einstein equation with 
distribution-valued energy-momentum tensor whose support
is a two-dimensional submanifold of indefinite signature. The equation 
of state is $p_z = - \mu$ and the solution is called cosmic string. 
These are good candidates for the origin of structure, 
especially for the formation of  galaxies,  in the early universe.

\medskip

4.   If one looks for strings according to 2. or 3. in a given spacetime, i.e., 
with negligible back reaction of the string onto  spacetime geometry, one 
arrives at the Nambu action, too, see Nielsen  and Olesen \cite{no} for 2. 
and Geroch  and Traschen \cite{gt} for 3. In other words, cosmic strings are 
extremal, maximal or minimal  in dependence of the boundary conditions, 
surfaces of indefinite signature in a given spacetime. This approach is justified
 if  the diameter of the string is small compared with the curvature
 radius of the underlying  manifold and if for the string tension $\a'\ll  1/G$  
holds; we use units with  $c = 1$. Normally, one 
thinks in orders of magnitude $G \a'=   10^{-6\pm2}$,
 see Brandenberger \cite{br}. In other  words, the string is supposed to possess 
a mass per unit length of $10^{22\pm2}$ g/cm, if 
the phase transition is supposed  to be at the GUT-scale, see  Frolov and Serebriany \cite{fs}.

\medskip

In \cite{fsh},  a stationary  string in the Kerr-Newman metric has 
been considered. We use the method  developed there and apply it
 to other cases. In this chapter  we shall adopt the 4. approach and try  to 
give   some  geometric 
insights into the motion of a string. To this end we give a sample of closed-form 
solutions for a special class of string solutions: strings which  share a spacelike 
or timelike isometry with the underlying background metric.

\subsection{The main formulae}

The string is a two-dimensional extremal surface  of  indefinite signature. Let 
us take coordinates $(\t, \s) = (y^A)$, $A = 0, 1$ within the string and coordinates 
 $(x^i)$, $i=  0, \,  1,\,  2,\,  3$  for the spacetime $V_4$ with the 
metric $g_{ij}$. The string is given 
by specifying  the four functions $x^i(y^A)$. The signature for the metric
  $g_{ij}$  is $(+---)$.  The tangents to the string are
\be
     e^i_A = \pa x^i/\pa y^A
\ee
and induce the metric
\be
     h_{AB} =  e^i_Ae^j_B \,  g_{ij}
\ee
at the  string world sheet. The signature  of the string is required to be $(+-)$, thus
\be
     h \equiv {\rm  det} \,  h_{AB} <0 \, . 
\ee
The action to be varied is
\be
I = \frac{1}{2 \a ' } \int \int \sqrt{-h} \,  d\s d \t \, .
\ee
Instead of writing down the full equations we specialize to 
the case we are interested here: we require that an isometry 
for both the underlying spacetime and the string exists.  
Let $k_i$  be a non-null hypersurface-orthogonal Killing vector field, i.e.,
\be
k_i k^i \ne 0, \quad k_{[i} k_{j;k]} =0, \quad k_{(i;j)} =0 \, .
\ee
Then there exist coordinates $x^i$ \    $(x^0 = t)$ such
\be
ds^2 = g_{00} dt^2 + g_{\a\b} dx^{\a} dx^{\b} \, .
\ee
$\a, \, \b = 1, \, 2, \,  3$,
 the $g_{ij}$   do not depend on $t$,  $k_i = \pa/\pa t$. 
The sign of $g_{00}$ is  determined by the 
condition $g_{00} \, k_i k^i > 0$: for timelike $k_i$
 we have $g_{00}> 0$  and for spacelike $k_i$
 we have $g_{00}< 0$. The requirement that 
$k_i$  is also an isometry of the string gives us the possibility to specify the functions 
$x^i(y^A)$  to be $t = \t$, $x^\a$ depends  on $\s$ only. Then we get 
\be 
e^i_0 = (1, \, 0, \, 0, \, 0)\, , \quad
e^i_1 = (0, \, dx^\a/d\s) 
\ee
and  $h_{01}=0$, $ h_{00}  = g_{00}$,
\bea
h_{11}= g_{\a\b}\  dx^{\a} /d\s \ dx^{\b} /d\s \, , \nonumber \\
I = \frac{1}{2 \a '  } \int  \int 
\sqrt{ - g_{00}\  g_{\a \b} \  dx^{\a} / d \s \ dx^{\b} / d\s }
 \ d\s d \t \, . \label{13a}
\eea
The integrand does not depend on $\t$, so we 
can omit the $\t$ integration. Therefore, 
extremizing eq. (\ref{13a}) is the same as solving
the geodesic equation for the auxiliary metric $f_{\a\b}$
 of a $V_3$ defined by
 \be
f_{\a\b} = - g_{00} g_{\a\b} \, .
\ee
Remarks:  1. If $k_i$ is not hypersurface-orthogonal then we get 
contains terms with $g_{0 \a}$,  and one has to add 
 $g_{0 \a} g_{0 \b}$ to the r.h.s. of the definition of $f_{\a\b}$.  

\bigskip

2. The dimensional reduction is not fully trivial: the geodesic equation for 
metric $f_{\a\b}$  corresponds to compare the strings in eq. (\ref{13a})
 with other strings sharing the same isometry induced by $k_i$, whereas the 
Nambu action has to be compared 
with all other strings, too. But  writing down all  full
 equations or counting the degrees of freedom one can see that at our 
circumstances no difference appears.

\bigskip

3. For a timelike $k_i$, $f_{\a\b}$  is positive definite, 
and the condition $h < 0$ is 
automatically fulfilled. On the other hand, for a spacelike $k_i$, 
$f_{\a\b}$  is  of signature $(+ - -)$,   and the validity of   $h < 0$  
 requires the vector $dx^\a/d\s$
 to be timelike, i.e.,  the root in eq. (\ref{13a})  has to be real.

\subsection{The string in a Friedmann model}

Now we specify the underlying $V_4$  to be a spatially flat Friedmann model 
\be\label{13b}
ds^2 =dt^2 - a^2(t)(dx^2 +dy^2 +dz^2) \,  .
\ee

\subsubsection{The open string}

First we use the spacelike Killing vector $\pa/\pa x$ of 
 metric (\ref{13b}). We have in mind  an 
infinitely long straight string moving through the expanding universe. We apply 
the formalism deduced in the previous section 
and get the following result: we write metric (\ref{13b})  in the form  
\be
ds^2= - a^2 dx^2 + dt^2 - a^2(dy^2 + dz^2  )   \,   . 
\ee
The metric $f_{\a\b}$ then reads
\be
ds^2_{(3)} = a^2 dt^2 - a^4(dy^2 + dz^2) \, .
\ee
Without loss of generality the string is situated at $z = 0$  and moves into 
the $y$-direction according to 
\bea
ds^2_{(3)} = a^2 dt^2 - a^4dy^2     \nonumber \\
 \ddot t +  \frac{1}{a} \frac{da}{dt} \dot t^2 + 2
\frac{da}{dt} a \dot y^2 = 0 \quad {\rm with} \quad \cdot = d/d\lambda  \nonumber \\
 a^4 \dot y = M  \nonumber \\
a^2 \dot t^2 - a^4 \dot y^2 =1  \nonumber \\
y(t) = M \int \frac{dt}{a(t) \sqrt{a^4(t) + M^2}} \label{13c}
\eea
where $M$ is an integration constant. The natural distance of the string 
from the origin is
\be
s(t) = a(t) \cdot  y(t) \, .
\ee

\bigskip

 1.   Example. Let $a(t) = t^n$, $ n \ge 0$
 then for $t \gg 1$, cf. Stein-Schabes  and Burd \cite{sb},
\be 
     y(t)   \approx M t^{1-3n} \, , \quad s(t)
   \approx M t^{1-2n}  \quad   {\rm   for} \quad  n \ne 1/3
\ee
and $y(t) \approx M \ln t$ for $n= 1/3$.
 Interpretation: $   n =0$,  i.e., the absence of gravity, implies
 a linear motion as it must be the case. We have  
\be
\lim_{t \to \infty} y(t) = \infty
\ee
if and only if $n \le 1/3$,  i.e., for $n > 1/3$  the string comes to rest with respect 
to the cosmic background after a finite time. A more stringent condition to be 
discussed is that the string comes to rest in a natural frame of a suitably 
chosen reference galaxy. This means  
\be
\lim_{t \to \infty} s(t) < \infty
\ee
 and is fulfilled 
for $n \ge 1/2$. Therefore, the most interesting cases $n = 1/2$, the radiation
 model  and $n = 2/3$, the Einstein-de Sitter dust model,  have the property that 
straight open strings come to rest after a sufficiently long time independently 
of the initial conditions.

\bigskip

 2.   Example.  Let $a(t) = e^{Ht}$, the inflationary  model, $H> 0$. 
Then $y(t) =e^{-3Ht}$, $
 s(t) =e^{-2Ht}$ . We have the same result as in the first example 
with $n \gg 1$.

\bigskip

 3.   Example.  Let
\be
a(t) =  - t^{2/3}(1 + t^{-2} \cos (mt)) \, . 
\ee
This background metric is from damped oscillations 
of a massive scalar field or, equivalently, from fourth-order 
gravity   $L = R + m^{-2} \, R^2$. We get with eq. (\ref{13c})
\be
      y(t) \approx   -1/t - ct^{-4} \sin mt \, , \quad 
    c = {\rm  const. \quad  for} \quad   t \to \infty \, , 
\ee
which is only a minor modification of the result of the first example 
with $n = 2/3$, therefore, one should not expect a kind of resonance effect between 
the open string and a massive scalar field.

\subsubsection{The closed string}

We insert $dy^2 + dz^2 = dr^2 + r^2 d\Phi^2$  into metric  (\ref{13b})
 and use the spacelike Killing vector $\pa/\pa\Phi$: its trajectories are circles, 
so we have in mind 
a closed string with radius $r$  moving and  oscillating  in the 
expanding universe. The corresponding geodesic equation leads to
\bea
\frac{d}{d\lambda}
 (-a^4r^2\dot r) - a^2 r \dot t^2 + a^4 r (\dot x^2 + \dot r^2)=0 \, ,
\\
\frac{d}{d\lambda} (a^4 r^2 \dot x) =0 \,   ,
\\
\dot t^2 - a^2 (\dot x^2 + \dot r^2)= \frac{1}{a^2r^2}\, , 
\eea
where the dot denotes $d/d\lambda$, $\lambda$
 is the natural parameter along the 
geodesic. We are mainly interested in solutions not moving into 
the $x$-direction, to understand the oscillating behaviour. Inserting $x = 0$ into 
 this system of 3 equations
 and using the fact that always $\dot  t \ne  0$
 holds, this system can be brought  into the form
\be
r \cdot r''a^3- 2 r r'a^4a'-r'^2a^3+3rr'a^2a'+a=0\, , 
\ee
where the dash denotes $d/dt$.

\bigskip

4. Example. Let $a = 1$, i.e., we have the flat Minkowski spacetime. 
 Then this equation reduces to
\be\label{13d}
rr''=  r'^2 - 1\, .
\ee
  The solution reads       
\be
           r(t) = r_0 \cos ((t - t_0)/r_0) \, , \quad
r'=     -\sin((t- t_0)/r_0) \, .
\ee
This is the oscillating solution for closed strings. At points $ t$ where
 $r = 0$, we have $ \vert  r' \vert = 1$. These points are the often discussed cusp points 
of the string, where the interior string metric becomes singular and the 
motion approximates the velocity of light, cf. Thompson \cite{to}.

\bigskip

Let us compare eq. (\ref{13d}) with the analogous equation for a positive definite 
back-ground metric. It reads
\be
rr''=  r'^2 + 1\, 
\ee
and has solutions with \  cosh \   instead of \  cos. This is  the usual minimal 
surface taken up e.g. by a soap-bubble spanned between two circles.

\bigskip

 5.   Example.  Let $a(t) = t^n$, then we have to solve  
\be
rr''t^{2n}   - 2nrr'^3t^{4n-1} - r'^2t^{2n} + 3nrr't^{2n-1} + 1 = 0 \, .
\ee
This equation governs the radial motion of 
a circle-like closed string in a Friedmann background. The solutions can be 
hardly obtained by analytic methods.

\subsection{The string in an anisotropic Kasner background}

Now we take as background metric
\be
ds^2 =    dt^2 - a^2(t) dx^2 - b^2(t) dy^2 - c^2(t) dz^2
\ee
  and $\pa/\pa x$ as Killing vector. Astonishingly, the anisotropy has only 
a minor influence on the motion of the string, so we get mainly the same 
formulae  as in the isotropic case: the geodesic equations are
\bea
\frac{d}{d\lambda} (a^2 b^2 \dot y) =0 
 \quad {\rm hence} \quad a^2 b^2 \dot y = M_1   \\  
\frac{d}{d\lambda} (a^2 c^2 \dot z)  = 0 \quad {\rm hence} \quad  a^2 c^2
 \dot z  = M_2       \\ 
  a^2 \dot t^2 - a^2  c^2 \dot z^2   = 1
\eea
and can be  integrated to yield for $M_1 M_2 \ne  0$
\be
     y(r) = \int \ \left( \frac{a^2b^2}{M_1^2} + \frac{M_2^2}{M_1^2}\cdot
  \frac{b^2}{c^2} +1 \right)^{-1/2} \   \frac{dt}{b(t)} \, . 
\ee
The equation for $z(t)$ can be obtained from this equation 
 by interchanging  $b \leftrightarrow c$ and $M_1 \leftrightarrow M_2$. 

\bigskip

For the Kasner  metric we have   $a = t^p$, $b=t^q$, $c=t^r$, 
$p + q + r = p^2 + q^2 + r^2 = 1$.  We get
\be
     y(r) = \int \frac{dt}{ t^q \sqrt{t^{2(p+q)} + t^{2(q-r)}+1 } }\, . 
\ee
The behaviour for $t \to \infty$  in dependence
 of the values $p$, $q$, $r$ can be seen  from this equation.

\subsection{The string in a gravitational wave}

As background metric we use the plane-wave ansatz
\be
ds^2
=    2dudv + p^2(u)dy^2 + q^2(u)dz^2 \, .
\ee
This  metric represents  a solution of Einstein's vacuum equation if
\be
q \,  d^2p/du^2 +p \, d^2q/du^2 =   0 \, .
\ee
Let us take $\pa /\pa y$  as Killing vector. Then the geodesic equations for 
the auxiliary metric are 
\bea
\frac{d}{d\lambda} (p^2 \dot u) =0 
 \quad {\rm hence} \quad p^2  \dot u = M_1 \ne 0 \, ,    \\
\frac{d}{d\lambda} (p^2 q^2 \dot z)  = 0 \quad
 {\rm hence} \quad p^2 q^2 \dot z  = M_2    \, , \\
 -2 p^2 \dot u \dot v - p^2 q^2 \dot z^2   = 1 \, .
\eea 
If we take $u$ as new independent variable, which is possible because 
of $\dot u \ne  0$,  we get the solutions
\bea
     z(u) = \frac{M_2}{M_1} \int \frac{du'}{q^2(u')} \ ,   
\nonumber \\
 v(u) = \frac{- M^2_2}{2M^2_1} \int
\left( \frac{1}{q^2(u')} - p^2(u') \right) du'  \ .   
\eea
As an example let us take
\be
p(u) = \sin u \,  , \qquad  q(u) = \sinh u
\ee
which produces a vacuum solution.  With this choice of $p$ and $q$  we get
\be
z(u) = M \coth u\,  , \qquad 
v(u) = -M^2(4 \coth u + \sin 2u - 2u)/8 \, .
\ee 
Another Killing vector of the plane wave   is $\pa/\pa v$,
 but it is a null Killing vector and 
so the reduction used above does not work. But there exists a further 
non-null Killing vector of this plane wave: It reads
\be 
k_i   = (0, y, H, 0), \qquad {\rm where} \qquad
 H = - \int p^{-2} du\,  . 
\ee
By a coordinate transformation
\bea 
     u=t \qquad  y=w \cdot e^{-G} \qquad    G(u)= \int \frac{du}{p^2 H}
 \nonumber \\
     z = k \qquad     v =  x + \frac{1}{2H} e^{-2G} (w^2-1)
\eea 
we get the form 
\be 
ds^2 = e^{ -2G}p^2 dw^2  +  2 dx dt + 
 \frac{e^{ -2G}}{p^2H^2}      dt^2 + q^2dk^2 
\ee
and have to solve the equations
\bea  
\frac{d}{d\lambda} \left( p^2 q^2 e^{-2G} \dot k 
\right)=0 \, , \qquad  \frac{d}{d\lambda} \left( p^2 e^{-2G} \dot t 
\right)=0 \, , \nonumber \\
-p^2 q^2 e^{-2G} \dot k^2 - \frac{1}{H^2} e^{-4G}
\dot t^2 - 2p^2 e^{-2G} \dot x \dot t   = 1\, . 
\eea
For the special case $\dot k = 0$ one has finally
\be 
x(t) = - \frac{1}{2} \int e^{-2G} \left(
 \frac{p^2}{M_2^2} + \frac{1}{H^2 p^2} \right) dt  \, .
\ee

\bigskip

Let us suppose a spatially flat Friedmann model eq. (\ref{13b}) with scale 
factor $ a(t) = t^n$, and $n = 2/3$, representing the Einstein-de Sitter dust model,
 or  $n = 1/2$, representing the radiation model. 
There we consider an open string which is only a little 
bit curved such that  the approximation of a straight string is applicable. 
At time $t = t_0 > 0$ we can prescribe place and initial velocity 
$v_0$ of the  string arbitrarily and 
get for $t \to \infty$  the behaviour 
\be
s(t) \approx {\rm  const}  + M(v_0) t^{1 -2n}
\ee
  where $s$ denotes the natural distance from the origin. That means, for 
the cases $n = 2/3$ and $= 1/2$ we are interested in, the string comes to rest 
for large values $t$ at a finite distance from the origin.

\bigskip

We compare this result with the analogous motion of a point-particle in 
the same background: in the same approximation we get 
\be
s(t) \approx {\rm const}  + M(v_0) t^{1-n} \, , 
\ee
 i.e.,  $\vert s \vert \to \infty$  as $t \to \infty$, a totally other type of motion. 
It should be mentioned that in the absence of gravity, i.e., $n = 0$, 
both motions are of the same type, but otherwise not.

\bigskip

The remaining calculations above indicate that the interaction of 
the motion of the string with scalar field oscillations, 
with anisotropy, and with gravitational waves  is quite weak, 
we did not find any type of resonance effects.

\newpage

{\large 
\renewcommand{\baselinestretch}{0.965} 

{\normalsize 

\section{Bibliography}

} } }


\begin{thebibliography}{111}
%
\setcounter{page}{187}

\bibitem{x1}
Altshuler,  B.: 1990, Class. Quant. Grav. {\bf 7}, 189. {\it 123}

\bibitem{x2}
Amendola, L.: 1993, Phys. Lett.  B  {\bf  301}, 175. 1999,  Phys. Rev. D {\bf  60},  043501. 
 { \it 129}

\bibitem{x3}
Amendola, L.,   Battaglia Mayer, A., Capozziello, S.,  
Gottl\"ober,  S.,  M\"uller,  V.,  Occhionero, F., Schmidt,
H.-J.: 1993,  Class. Quant. Grav. {\bf 10}, L43.  { \it 126}

\bibitem{x4}
Amendola, L.,  Capozziello, S.,   Litterio, M.,  Occhionero, F.: 1992, 
Phys. Rev.  D  {\bf  45}, 417. { \it 126}

\bibitem{anandan}
Anandan,  J.: 1983, in: Hu,  N., ed., {\it Proc. 3. Marcel 
Grossmann Meeting B}, Amsterdam NHPC, 941. { \it 160}

\bibitem{anderson}
Anderson, P.: 1986, Phys. Lett. B {\bf 169}, 31. { \it 111}

\bibitem{arno}
Arnowitt, R., Deser, S., Misner, C.: 1962,  in: Witten, E., ed.,
 {\it Gravitation, an introduction to current research}, Wiley New York. { \it 123}

\bibitem{bach}
Bach, R.: 1921, Math. Zeitschr. {\bf  9}, 110. { \it 145, 147, 156 }

\bibitem{bachmann}
Bachmann, M.,  Schmidt, H.-J.: 2000, 
 Phys. Rev. D {\bf 62},  043515; gr-qc/9912068.  { \it 83}

\bibitem{y1}
 Barabash, O.,  Shtanov, Y.: 1999, Phys. Rev. D {\bf 60}, 064008; astro-ph/9904144. 
 { \it 159}

\bibitem{y2}
Barraco, D., Hamity, V.: 2000,   Phys. Rev. D {\bf  62},  044027. 
1999, Gen. Relat. Grav. {\bf  31},  213. { \it 112}

\bibitem{x5,x8}
Barrow, J.: 1986, Phys. Lett. B {\bf  180}, 335.
 1987, Phys. Lett. B {\bf  187}, 12. { \it 123, 125, 126, 136}


\bibitem{x6}
Barrow, J.,  G\"otz, G.:  1989,  Class. Quant. Grav. {\bf  6}, 1253. 
 { \it 122, 123}

\bibitem{barrowmatzner}
Barrow, J., Matzner, R.: 1980, Phys. Rev. D {\bf  21}, 336. { \it 59, 74}

\bibitem{x7}
Barrow, J.,  Saich, P.: 1990,  Phys. Lett. B {\bf   249}, 406. { \it 126}


\bibitem{barrowzia}
Barrow, J., Sirousse-Zia, H.: 1989, Phys. Rev. D {\bf 39}, 2187.
Erratum 1990, D {\bf  41}, 1362. { \it 61, 125, 127, 139}

\bibitem{barrowtipler}
Barrow, J., Tipler, F.: 1988, {\it The anthropic cosmological principle}, Cambridge 
University Press. { \it 75}

\bibitem{x9}
Battaglia Mayer, A.,  Schmidt, H.-J.: 1993,  Class. Quant. Grav. {\bf 10}, 2441.
 { \it 128}

\bibitem{belinskyc}
Belinsky, V., Grishchuk, L., Khalatnikov, I., Zeldovich, Y.:
 1985, Phys. Lett. B {\bf  155}, 232. { \it 89}

\bibitem{belinskyb}
Belinsky, V., Grishchuk, L., Zeldovich, Y., Khalatnikov, I.: 
1985, Zh. Eksp. Teor. Fiz. {\bf 89}, 346. { \it 71, 74}


\bibitem{belinsky}
Belinsky, V., Khalatnikov, I.: 1987, Zh. Eksp. Teor. Fiz. {\bf 93}, 784.
 { \it 59, 74}

\bibitem{x10}
Belinsky, V.,  Lifshitz, E., Khalatnikov, I.: 1972,  Sov. Phys. JETP {\bf 35}, 
838. { \it 59, 125} \footnote{The authors are 
in alphabetical order of the original Russian language version.}

\bibitem{bergmann}
Bergmann, O.:  1981, Phys. Lett. A {\bf 82}, 383. { \it 9}

\bibitem{x13,x14}
Berkin, A.: 1990, Phys. Rev.  D {\bf  42}, 1016.  1991, Phys.  Rev.  D {\bf  44}, 1020.
 { \it 127, 129}


\bibitem{berkin}
Berkin, A., Maeda, K.: 1990, Phys. Lett. B {\bf  245}, 348.
1991,  Phys. Rev. D  {\bf  44}, 1691.{ \it 129, 164}

\bibitem{y3}
Bi\v c\'ak, J: 2000, Lect. Notes  Phys. {\bf 540}, 1-126; gr-qc/0004016.
 {\it   19}

\bibitem{x15}
Bi\v c\'ak, J.,   Podolsky, J.:  1992,  in: Abstracts Conf. Gen. Relat. 13, 
Cordoba, 12. 1995, Phys. Rev. D  {\bf  52}, 887.  {\it   127}

\bibitem{y4}
Bi\v c\'ak, J.,  Pravda, V.: 1998, Class. Quant. Grav. {\bf 15},  1539.
 {\it   44}

\bibitem{bicknell}
Bicknell, G.: 1974, J. Phys. A {\bf  7}, 341. 1061.  {\it   61, 87, 103, 139}

\bibitem{bleyerliebscher}
Bleyer, U.,  Liebscher, D.-E.,  Schmidt, H.-J.,  Zhuk, A.: 1990, Wiss. Zeitschr. d. 
Univ. Jena, Naturwiss. Reihe {\bf  39},  20. {\it   85}

\bibitem{bleyerschmidt}
Bleyer, U., Schmidt, H.-J.: 1990, Int. J. Mod. Phys. A {\bf  5} (1990)  4671.
 {\it   121}

\bibitem{y5}
 Bojowald, M., Kastrup, H.,  Schramm, F.,  Strobl, T.: 2000,  Phys. Rev. D 
{\bf  62} (2000) 044026.  {\it   111}


\bibitem{x18}
Borde, A.: 1994,  Phys. Rev. {\bf D 50}, 3692. {\it   124}

\bibitem{x17}
Borde, A., Vilenkin, A.: 1994,  Phys. Rev. Lett. {\bf 72}, 3305. {\it   124}

\bibitem{hhvb}
Borzeszkowski, H. v.: 1981, Ann. Phys. (Leipz.) {\bf  38}, 239. {\it   146}

\bibitem{x19}
Borzeszkowski, H. v., Treder, H.-J.: 1994, Found. Phys. {\bf 24}, 949.
 {\it   127}

\bibitem{bty}
Borzeszkowski, H. v.,  
Treder, H.-J.,  Yourgrau, W.: 1978, Ann. Phys. (Leipz.) {\bf 35},  471.
 {\it   146, 156, 160}

\bibitem{boulware}
Boulware, D.: 1973,  Phys. Rev. D {\bf 8}, 2363. {\it   45}

\bibitem{br}
Brandenberger, R.: 1987, Int. J. Mod. Phys. A {\bf 2}, 77. {\it   176}

\bibitem{x20}
Brauer, U.,  Rendall, A.,  Reula, O.: 1994,  Class. Quant. Grav. {\bf 11}, 2283.
 {\it   123}

\bibitem{x21}
Breizman, B., Gurovich, V.,  Sokolov, V.: 1970, Zh. Eksp.  Teor. Fiz. {\bf 59}, 288.
 {\it   125}

\bibitem{buch1}
Buchdahl, H.: 1951, Acta Math. {\bf 85}, 63. {\it 128, 134, 164 }

\bibitem{buchdahl}
Buchdahl, H.: 1962, Nuovo Cim. {\bf 23}, 141. {\it   92}

\bibitem{x22}
Burd, A.: 1993,  Class. Quant. Grav. {\bf 10}, 1495. {\it   123}

\bibitem{burd}
Burd, A., Barrow, J.: 1988, Nucl. Phys. B {\bf  308}, 929. {\it   122}

\bibitem{calzetta}
Calzetta, E.: 1989, Class. Quant. Grav. {\bf 6},  L227. {\it 76, 140 }



\bibitem{x23}
Calzetta, E.,  Sakellariadou, M.: 1992,  Phys. Rev. D {\bf  45}, 2802. 
 {\it   127}

\bibitem{y6}
Campanelli, M.,  Lousto, C.: 1996, Phys. Rev. D {\bf 54}, 3854. {\it   112}

\bibitem{y7}
Capozziello, S., Lambiase, G.: 2000,   Gen. Relat. Grav. {\bf  32}, 295. 673. 
 {\it   112}

\bibitem{carfora}
Carfora, M., Marzuoli, A.: 1984, Phys. Rev. Lett. {\bf 53}, 2445. {\it   106}

\bibitem{carmeli}
Carmeli,  M., Charach, C. : 1980, Phys. Lett. A {\bf 75}, 333. {\it   9}

\bibitem{x24}
Chiba, T., Maeda, K.: 1994,  Phys. Rev.  D  {\bf  50}, 4903. {\it   124}

\bibitem{cohen}
Cohen, M.,  Cohen, J.: 1971, in: Kuper, C., ed.,  {\it Relativity  and 
Gravitation}, Gordon and Breach New York,  99.  {\it   51}

\bibitem{x26}
Coley, A., Tavakol, R.: 1992,  Gen. Relat. Grav.  {\bf 24}, 835.  {\it   125}

\bibitem{collins}
Collins,  C., Szafron, D.: 1981,  J. Math. Phys. {\bf 22}, 543. {\it   9}

\bibitem{x27}
Contreras, C.,  Herrera, R.,  del Campo, S.: 1995,  Phys. Rev. D {\bf  52}, 4349. 
 {\it   127}

\bibitem{x28}
Cotsakis, S.,  Demaret, J.,  De Rop, Y., Querella, L.: 1993, Phys. Rev. D {\bf  48}, 4595. 
 {\it   125}

\bibitem{x29}
Cotsakis, S., Flessas,  G.:  1993, Phys. Lett. B {\bf  319}, 69. {\it   127}

\bibitem{y8}
Coule, D.: 2000,  Phys. Rev. D {\bf  62}, 124010; gr-qc/0007037. 
 1995, Class. Quant. Grav. {\bf 12},  455; gr-qc/9408026. {\it   83}

\bibitem{dautcourt1963}
Dautcourt, G. : 1963,  Math. Nachr. {\bf 27}, 277.
 1964,  Acta Phys. Polon. {\bf  25}, 637. {\it   45, 54, 164}

\bibitem{y9}
Demaret, J., Querella, L., Scheen, C.: 1999, Class. Quant. Grav. {\bf 16}, 749. 
Demaret, J.,  Querella, L.: 1995, Class. Quant. Grav. {\bf 12},  3085.  {\it   112}

\bibitem{x30}
Demianski, M.: 1984, Nature {\bf 307}, 140.  {\it 123 }

\bibitem{denisov}
Denisov, V. I.: 1972,  J. Exp. Teor.  Phys. {\bf  62}, 1990. {\it 45 }

\bibitem{y10}
 Dzhunushaliev, V.,  Schmidt, H.-J.: 2000,  J. Math. Phys. {\bf  41},  
3007;   gr-qc/9908049.    {\it   156}


\bibitem{ehlers}
Ehlers, J.,  Kundt, W.:  1962,  in: Witten, E., ed.,
 {\it Gravitation, an introduction to current research}, Wiley New York, 49.
 {\it   36}

\bibitem{einstein1}
Einstein, A.: 1914, Sitzungsber. Akad. d. Wiss. Berlin, 1030.
 1916, Sitzungsber. Akad. d. Wiss. Berlin, 688.
 1918, Sitzungsber. Akad. d. Wiss. Berlin, 154.
 1921, Sitzungsber. Akad. d. Wiss. Berlin,  261. {\it   35, 145}

\bibitem{einstein1939}
Einstein, A.: 1939,  Ann. Math. {\bf 40},  922. {\it   46}

\bibitem{einsteindesitter}
Einstein, A.,  De Sitter, W.: 1932, Proc. Nat. Acad. Sci. {\bf 18}, 213.
 {\it   19, 24}

\bibitem{einsteinpauli}
Einstein, A., Pauli, W.: 1943, Ann. Math. {\bf 44}, 131. {\it   146}

\bibitem{ellis}
Ellis, G.: 1984, Ann. Rev. Astron. Astrophys. {\bf  22}, 157.
Ellis,  G.:  1967,  J. Math. Phys. {\bf 8}, 1171. {\it 13, 104 }

\bibitem{y11}
 Esteban, E., Kazanas, D.: 2001,   Gen. Relat. Grav. {\bf  33}, 1281.   {\it 156 } 

\bibitem{y12}
 Fabris, J., Pelinson, A., Shapiro, I.: 2001, Nucl. Phys. B {\bf  597}, 539; hep-th/0009197.
 {\it   39}

\bibitem{y13}
Fabris, J.,  Reuter, S.: 2000,  Gen. Relat. Grav. {\bf  32}, 1345.  {\it   83}

\bibitem{y14}
Faraoni, V., Gunzig, E.,  Nardone, P.: 
1999, Fund. Cosmic Physics  {\bf  20}, 121; gr-qc/9811047. {\it   112}

\bibitem{x31}
Feldman,  H.: 1990, Phys. Lett.  B  {\bf  249}, 200. {\it   127}

\bibitem{ferraris}
Ferraris, M.: 1986, in: Fabbri, R., Modugno, M., eds., {\it VI Conv. 
Naz. Rel. Gen. e Fisica della Grav.}, Pitagora Edit. Bologna,  127.
 {\it   110}

\bibitem{fiedler1980}
Fiedler, B.,  Schimming, R.: 1980, Rep. Math. Phys. {\bf 17}, 15. {\it   39}

\bibitem{fiedler1983}
Fiedler, B., Schimming, R.: 1983, Astron. Nachr. {\bf 304}, 221.
 {\it 146, 148, 149, 155 }

\bibitem{frehland}
Frehland, E.: 1972,  Ann. Phys. (Leipz.) {\bf 28},  91. {\it   45}

\bibitem{fs}
Frolov, V.,  Serebriany, E.: 1987, Phys. Rev. D {\bf 35}, 3779.
 {\it   176}

\bibitem{fsh}
Frolov, V.,  Skarzhinsky, V., Zelnikov, A., 
Heinrich, O.: 1988, Preprint Potsdam-Babelsberg, PRE-ZIAP 88-14.
 {\it   176}

\bibitem{fulling}
Fulling, S.: 1973, Phys. Rev. D {\bf 7}, 2850. {\it   58}

\bibitem{fullingparker}
Fulling, S., Parker, L.: 1974, Ann. Phys. NY {\bf  87}, 176. {\it   58}

\bibitem{y15}
Gemelli, G.: 1997,  Gen. Relat. Grav. {\bf 29}, 1163.  {\it   39}

\bibitem{ger}
Gerbal, D., Sirousse-Zia, H.: 1989,   C. R. Acad. Sci. Paris {\bf  309}, 353.
 {\it   165}

\bibitem{geroch}
Geroch, R.: 1969, Commun. Math. Phys. {\bf  13}, 180. {\it 42, 64, 73, 115 }

\bibitem{gt}
Geroch, R.,  Traschen, J.: 1987, Phys. Rev. D {\bf  36}, 1017. {\it   176}

\bibitem{x32}
Gibbons, G. and  Hawking, S.: 1977,  Phys. Rev. D {\bf  15}, 2738. {\it   123}

\bibitem{gibbons}
Gibbons, G.,  Hawking, S.,  Stewart, J.: 1987,  Nucl. Phys. B {\bf 281}, 736.
 {\it   85}

\bibitem{goenner}
Goenner, H.: 1987, in:  MacCallum, M., ed., {\it Proc. 11th Int. 
Conf. Gen. Rel. Grav.}, Cambridge University Press,  262. {\it 110 }

\bibitem{gor}
Gorbatenko, M., Pushkin, A.,  Schmidt, H.-J.: 2002,   Gen. Relat. 
Grav. {\bf 34}, in press; gr-qc/0106025. {\it 156 }

\bibitem{gott1984a}
Gottl\"ober, S.: 1984, Ann. Phys. (Leipz.) {\bf  41}, 45.  1984, Astron. 
Nachr. {\bf 305}, 1.  1987, Astrophys. Space Sc. {\bf  132}, 191.{\it   59, 107, 109}


\bibitem{gottmuell}
Gottl\"ober, S., M\"uller, V.: 1987, Class. Quant. Grav. {\bf 4}, 1427.   {\it 59}

\bibitem{gms}
Gottl\"ober, S., M\"uller, V., Schmidt, H.-J.: 1991, 
Astron. Nachr. {\bf  312},  291.   {\it 129, 164}

\bibitem{x33}
Gottl\"ober, S.,  M\"uller, V.,  Schmidt, H.-J.,  
Starobinsky,  A.: 1992, Int. J. Mod. Phys.  D  {\bf  1}, 257.   {\it 126}

\bibitem{gottschmidt}
Gottl\"ober, S.,  Schmidt, H.-J.: 1984, Potsdamer Forschungen B {\bf 43}, 117.
    {\it 59}

\bibitem{gss}
Gottl\"ober, S.,  Schmidt, H.-J, 
Starobinsky, A.: 1990, Class. Quantum Grav. {\bf  7}, 893.
   {\it 129, 164, 165}

\bibitem{gsw}
Green, M., Schwarz, J., Witten, E.: 1987, {\it Superstring theory},
 Cambridge University Press.    {\it 175}

\bibitem{griff}
Griffiths, J.: 1995, 
Gen. Relat. Grav. {\bf 27}, 905.   {\it 13, 26}


\bibitem{grishchuk}
Grishchuk, L.,  Sidorov, J.: 1988,  in: Markov, M., ed., {\it Proc. 4. Sem. 
Quantum Gravity Moscow}, WSPC Singapore,  700.    {\it 85}

\bibitem{guro}
 Gurovich, V.,  Schmidt, H.-J.,  Tokareva, I.: 2001,  
 Gen. Relat. Grav. {\bf 33},  591;  gr-qc/0007046.   {\it 83}

\bibitem{hall1}
Hall, G.: 1988, Gen. Relat. Grav. {\bf 20},  671.
 1990,  J. Math. Phys. {\bf 31},  1198.  1991,  Diff. Geom. Appl.  {\bf 1},  35.   
   {\it 40}

\bibitem{halliwell}
Halliwell, J., Hawking, S.: 1985,   in: Markov, M., ed., {\it 
Proc. 3. Sem. Quantum Gravity Moscow}, WSPC Singapore,  509.   {\it 85}

\bibitem{harvey}
Harvey, A.: 1990, Class. Quant. Grav. {\bf 7},  715.   {\it 41}

\bibitem{havas}
Havas, P.: 1977,   Gen. Relat. Grav. {\bf  8},  631.   {\it 160}

\bibitem{hawkinga}
Hawking, S.: 1984, Nucl. Phys. B {\bf 239}, 257.
 1984, in: De Witt, B., Stora, R., eds., 
 {\it Relativity, Groups, and Topology II},  North Holland  PC Amsterdam.
   {\it 59}

\bibitem{hawkingellis}
Hawking, S., Ellis, G.: 1973, {\it  The large scale structure of
space-time}, Cambridge University Press.   {\it 36}

\bibitem{hawkingluttrell}
Hawking, S., Luttrell, J.: 1984, Nucl. Phys. B {\bf 247}, 250.   {\it 59}

\bibitem{x36}
Hawking, S.,  Moss, I.: 1982, Phys. Lett. B  {\bf  110}, 35.   {\it 123}

\bibitem{y16}
Hehl, F., McCrea, J.,  Mielke, E., Ne'eman, Y.: 1995, Phys. Repts. {\bf 258}, 1.
   {\it 112}

\bibitem{y17}
 Hindawi, A., Ovrut, B.,  Waldram, D.: 1996,  Phys. Rev. D {\bf 53}, 5583.
   {\it 112}

\bibitem{x37}
Hoyle, F.,  Narlikar, J.: 1963,  Proc. R. Soc. Lond. A {\bf  273}, 1.   {\it 123}

\bibitem{x38}
H\"ubner, P.,  Ehlers, J.: 1991,  Class. Quant. Grav. {\bf 8}, 333.   {\it 123}

\bibitem{x39}
Hwang, J.: 1991,  Class. Quant. Grav. {\bf 8}, L133.    {\it 126}

\bibitem{y18}
 Hwang, J.,  Noh, H.: 2001,  Phys. Lett. B {\bf  506}, 13.    {\it 112}

\bibitem{ishihara}
Ishihara, R.: 1986, Phys. Lett. B {\bf 179}, 217.   {\it 112}

\bibitem{israel}
Israel, W.: 1966, Nuovo Cim. B {\bf 44}, 1.   {\it 45, 54}

\bibitem{israelstewart}
Israel, W.,  Stewart, J.: 1980, in: Held, A., ed., {\it  General  Relativity 
and Gravitation II}, Plenum New York,  491.   {\it 46}

\bibitem{y19}
Ivashchuk, V., Melnikov, V.: 1995, Grav. Cosmol. {\bf 1}, 133; hep-th/9503223.
    {\it 111}

\bibitem{jakubiec1988}
Jakubiec, A.,  Kijowski, J.: 1988,   Phys. Rev. D {\bf  37}, 1406.
 1989,   J. Math. Phys. {\bf 30}, 1073.   {\it 110}

\bibitem{jankiewicz}
Jankiewicz,  C.: 1981,  Acta Phys. Polon. {\bf  13},  859.   {\it 160}

\bibitem{x40}
Jensen, L.,  Stein-Schabes, J.: 1987,  Phys. Rev. D  {\bf  35}, 1146.   {\it 127}

\bibitem{jordan}
Jordan, P.,  Ehlers, J.,  Kundt, W.: 1960, Abh. Akad. Wiss. Mainz,
Math./Nat. Kl. {\bf 2}, 21.   {\it 36}

\bibitem{y20}
Kao, W.: 2000,  Phys. Rev. D {\bf  61},  047501.    {\it 112}

\bibitem{kerner}
Kerner, R.: 1982, Gen. Relat. Grav. {\bf  14}, 453.   {\it 94}

\bibitem{y21}
Kiefer, C.: 2001, in: Janke, W.,  Pelster, A.,  
Schmidt, H.-J.,  Bachmann, M., 
eds., {\it Fluctuating paths and fields}, WSPC Singapore, 729.   {\it 83}


\bibitem{x41}
Kirsten, K.,  Cognola, G., Vanzo, L.: 1993,  Phys. Rev. D {\bf  48}, 2813.
    {\it 129}

\bibitem{kleinert}
 Kleinert, H.,  Schmidt, H.-J.: 2000, {\it Cosmology  with curvature--saturated 
gravitational   lagrangian $R/\sqrt{1 + l^4 R^2}$}, Preprint gr-qc/0006074.
    {\it 112}

\bibitem{k}
Kluske, S.,  Schmidt, H.-J.: 1996, Astron. Nachr. {\bf  317}, 337; gr-qc/9503021.
   {\it 121}

\bibitem{x42}
Kofman, L.,  Linde, A.,   Starobinsky, A.: 1996, Phys. Rev. Lett. {\bf 76}, 1011. 
   {\it 123}

\bibitem{x43}
Kottler, F.: 1918,  Ann. Phys. (Leipz.) {\bf 56}, 410.    {\it 128}

\bibitem{kowalski}
Kowalski, O.: 1993, 
     Nagoya Math. J. {\bf 132}, 1.   {\it 31}

\bibitem{kowalskin}
Kowalski, O.,. Nikcevic, S.: 1995, {\it  On Ricci eigenvalues of locally homogeneous
Riemannian 3--manifolds}, preprint.   {\it 32}

\bibitem{kramer}
Kramer, D., Stephani, H., Herlt, E., MacCallum, M.: 1980, {\it Exact solutions of 
 Einstein's field  equations}, Verl. d. Wiss. Berlin.     {\it 11} 

\bibitem{krasilnikov}
Krasilnikov, V. A.,  Pavlov, V. I.: 1972, Vestnik  Univ. Moskau  {\bf  27}, 235. 
   {\it 46}

\bibitem{krasinski}
Krasinski, A.:  1980, Proc. Conf. General  Relativity GR 9 Jena,  44.   {\it 9}

\bibitem{krasinskib}
Krasinski, A.: 1997, {\it Inhomogeneous Cosmological Models}, 
Cambridge University Press.   {\it 9, 19}

\bibitem{kuchar}
Kucha\v{r}, K.: 1968, Czech. J. Phys. B  {\bf  18}, 435.   {\it 45}

\bibitem{y22}
 Kung, J.: 1996, Phys. Rev. D {\bf 53}, 3017.   {\it 112}

\bibitem{lake1979}
Lake, K.: 1979,  Phys. Rev. D  {\bf 19}, 2847.
 1981, Lett. Nuovo Cim.  {\bf 30}, 186.  1993, J. Math. Phys. {\bf 34},  5900. 
   {\it 41, 42, 45, 46}

\bibitem{lanczos}
Lanczos, K.: 1924, Ann. Phys. (Leipz.)  {\bf 74}, 518.   {\it 45}

\bibitem{lee}
Lee,  C. W.:  1978, Proc. R. Soc. Lond. A {\bf 364}, 295.   {\it 10}

\bibitem{lopes}
L\'opez, C.: 1981,  Nuovo Cim. B {\bf 66}, 17.   {\it 45}

\bibitem{x44}
Lu, M.  and  Wise, M.: 1993,  Phys. Rev. D {\bf  47}, R3095.    {\it 129}

\bibitem{lukash}
Lukash, V. N., Schmidt, H.-J.:  1988, Astron. Nachr. {\bf 309}, 25.
   {\it 71, 72, 107, 109}

\bibitem{maccallum}
MacCallum,  M.:  1971, Commun. Math. Phys. {\bf 20}, 57.  
 1972, Phys. Lett. A {\bf  40}, 385.   {\it 15, 127, 139}

\bibitem{x45}
MacCallum, M.,  Stewart, J.,  Schmidt, B.: 1970, Commun. Math. Phys. {\bf 17}, 343.
   {\it 127, 139}

\bibitem{y23}
Macias, A.: 1999, Gen. Relat. Grav. {\bf  31},  653.    {\it 83}

\bibitem{madsen}
Madsen, M.,  Ellis, G.: 1988, Mon. Not. R. Astron. Soc. {\bf  234}, 67.
   {\it 76}

\bibitem{x48x47}
Maeda, K.: 1988, Phys. Rev. D {\bf  37}, 858.
 1989,  Phys. Rev. D {\bf  39}, 3159.   {\it 126}

\bibitem{maeda}
Maeda, K.,  Sato, H.: 1983,  Progr. Theor. Phys. {\bf  70}, 772.    {\it 46, 54}

\bibitem{x49}
Maeda, K.,  Stein-Schabes, J.,  Futamase, T.: 1989, Phys. Rev. D  {\bf  39}, 2848.
   {\it 126}

\bibitem{x50}
Martin, G.,  Mazzitelli, F.: 1994,  Phys. Rev. D {\bf  50}, R613.    {\it 129}

\bibitem{y24}
 Mignemi, S.: 1996, Ann. Phys. NY 245 (1996) 23.    {\it 111}

\bibitem{x51}
Mijic, M.,  Stein-Schabes, J.: 1988, Phys. Lett. B  {\bf  203}, 353.   {\it 126}

\bibitem{x52}
Mirzabekian, A.,  Vilkovisky, G.,  Zhytnikov, V.: 1996,  Phys. Lett. B {\bf  369}, 215.
   {\it 142}

\bibitem{x53}
Mondaini, R., Vilar, L.: 1993, Int. J. Mod. Phys. D {\bf  2}, 477.   {\it 124}

\bibitem{y25}
Mongan, T.: 2001, Gen. Relat. Grav. {\bf 33}, 1415.    {\it 83}

\bibitem{x54}
Moniz,  P.: 1993, Phys. Rev. D  {\bf  47}, 4315.   {\it 128}


\bibitem{morgan}
Morgan, L.,  Morgan, T.: 1970,  Phys. Rev. D {\bf 2}, 2756.   {\it 45}

\bibitem{x55}
Morris, M.: 1989,  Phys. Rev. D {\bf  39}, 1511.   {\it 127}

\bibitem{x57}
M\"uller, V.: 1986,  Ann. Phys. (Leipz.) {\bf 43}, 67.   {\it 125}

\bibitem{muell1985}
M\"uller, V., Schmidt, H.-J.: 1985, Gen. Relat. Grav. {\bf  17}, 769.
1989, Gen. Relat. Grav. {\bf  21}, 489.   {\it 28, 71, 98, 99, 105}

\bibitem{x56}
M\"uller, V.,  Schmidt, H.-J., Starobinsky, A.: 1988,  Phys. Lett. B {\bf  202}, 198. 
   {\it 125}

\bibitem{x58}
Murphy, G.: 1973,  Phys. Rev. D {\bf  8}, 4231.   {\it 123}

\bibitem{x59}
Nakao, K.,  Nakamura, T., Oohara, K.,  Maeda, K.: 1991, 
Phys. Rev. D  {\bf  43}, 1788.   {\it 127}

\bibitem{x60}
Nakao,  K.,  Maeda, K.,  Nakamura, T.,  Oohara, K.: 1993,
Phys. Rev. D  {\bf  47}, 3194 (1993).   {\it 127}


\bibitem{x61}
Nakao, K.,  Shiromizu, T.,  Maeda, K.: 1994, Class. Quant. Grav. {\bf  11}, 2059. 
   {\it 123}

\bibitem{nariai1973}
Nariai, H.: 1973, Progr. Theor. Phys. {\bf  49}, 165.
 1974, Progr. Theor. Phys. {\bf 51}, 613.   {\it 87}

\bibitem{neugebauer}
Neugebauer, G.: 1980, {\it Relativistische Thermodynamik}, Akad.-Verl. Berlin. 
   {\it 46}

\bibitem{no}
Nielsen, H.,  Olesen, P.: 1973, Nucl. Phys. B {\bf  61}, 45.   {\it 176}

\bibitem{novotny}
Novotny, J.: 1984, Coll. J. Bolyai Math. Soc. Diff. Geom. Debrecen {\bf 46}, 959.
   {\it 94}



\bibitem{x62}
Oleak, H.: 1974, Astron. Nachr. {\bf 295}, 107. 1987,  Ann. Phys. (Leipz.) 
{\bf 44}, 74.   {\it 124}

\bibitem{ono}
Ono, S., Kondo, S.: 1960, in:  Fl\"ugge, S., ed., {\it  Handbuch der Physik}
 {\bf  10},  Springer Berlin,  134.    {\it 46, 47}

\bibitem{page1984}
Page, D.: 1984, Class. Quant. Grav. {\bf  1}, 417.  1987, Phys. Rev. D {\bf  36}, 1607.
   {\it 57, 59, 61, 67, 74}

\bibitem{y26}
 Paiva, F., Reboucas, M.,  Hall, G., MacCallum, M.: 1998,  
Class. Quant. Grav. {\bf 15}, 1031.   {\it 44}


\bibitem{papa}
Papapetrou, A., Hamoui, A.: 1968,  Ann. Inst. H. Poincar\'e A  {\bf 9}, 179.
   {\it 45}

\bibitem{papatreder}
Papapetrou, A.,  Treder, H.-J.: 1959,  Math. Nachr. {\bf 20}, 53.   {\it 45}

\bibitem{y27}
 Parker, L., Raval, A.: 1999, Phys. Rev. D {\bf  60}  063512.     {\it 83}


\bibitem{pechlaner}
Pechlaner, E.,  Sexl, R.: 1966,  Commun. Math. Phys. {\bf 2}, 165.
   {\it 149, 160}

\bibitem{x64}
Peter, P., Polarski, D.,  Starobinsky,  A.: 1994, Phys. Rev. D {\bf  50}, 4827.
    {\it 123}

\bibitem{y28}
 Pimentel, L.: 2001,  Gen. Relat. Grav. {\bf  33}, 781;  gr-qc/0009046.   {\it 83}

\bibitem{poli}
Polijevktov-Nikoladze, N.: 1967,  J.  Exp. Teor.  Fiziki {\bf  52},  1360.
   {\it 160}

\bibitem{y29}
Pollney, D.,  Skea, J.,  D'Inverno, R.: 2000, Class. Quant. Grav. {\bf 17}, 2885.
   {\it 44}

\bibitem{x65}
Pollock, M.: 1987,  Phys. Lett. B  {\bf  192}, 59.   {\it 125, 126}

\bibitem{x67}
Prigogine, I.,  Geheniau, J., Gunzig, E., Nardone, P.: 1988, 
Proc. Nat. Acad. Sci. {\bf 85}, 7428.  1989,  Gen. Relat. Grav. {\bf 21}, 767. 
1989, Int. J. Theor. Phys. {\bf 28}, 927.   {\it 124}

\bibitem{quandt}
Quandt, I., Schmidt, H.-J.: 1991, Astron. Nachr. {\bf 312}, 97; gr-qc/0109005.
   {\it 128, 159, 171 }

\bibitem{y30}
Rainer, M.: 1995, Grav. Cosmol. {\bf 1}, 121.
 1995, Int. J. Mod. Phys. D {\bf 4}, 397.   {\it 44}

\bibitem{rainer}
Rainer, M.,  Schmidt, H.-J.: 1995,  Gen. Relat. Grav.  {\bf 27}, 1265; gr-qc/9507013.
   {\it 14, 32, 33, 34}

\bibitem{y31}
Rainer, M., Zhuk, A.: 2000,  Gen. Relat. Grav. {\bf 32}, 79.    {\it 111}

\bibitem{robinson}
Robinson, A.: 1966, {\it Non-standard analysis}, North Holland Amsterdam.
   {\it 152}

\bibitem{x70}
Rogers, B., Isaacson, J.: 1991,  Nucl. Phys. B {\bf  364}, 381.
 1992,  Nucl. Phys. B {\bf  368}, 415.   {\it 127}

\bibitem{y32}
Romero, C., Tavakol, R., Zalaletdinov, R.: 1996, Gen. Relat. Grav. {\bf 28},  365. 
   {\it 112}

\bibitem{x71}
Saa, A.: 1996, J. Math. Phys. {\bf 37}, 2346.    {\it 122}

\bibitem{z1}
Sakharov, A.: 2000,  Gen. Relat. Grav. {\bf  32}, 365; 1967, 
Dokl. Akad. Nauk SSSR {\bf 177}, 70.    {\it 128}

\bibitem{y33}
 Sanchez, M.: 1998,  Gen. Relat. Grav. {\bf 30}, 915.   {\it 29}

\bibitem{san} 
Sanders, R. : 1984, Astron. Astrophys. {\bf  136}, L21.
1986, Astrophys. J. {\bf  154}, 135.   {\it 165}

\bibitem{y34}
Sanyal, A., Modak, B.: 2001,  Phys. Rev. D {\bf 63}, 064021; gr-qc/0107001. 
   {\it 83}

\bibitem{x72}
Schimming, R.,  Schmidt, H.-J.: 1990, NTM-Schriftenr.
Geschichte der Naturwiss., Technik, Medizin  {\bf 27}, 41.     {\it 128}

\bibitem{sch82a}
Schmidt, H.-J.: 1982,  Astron. Nachr. {\bf 303},  227; gr-qc/0105104.
   {\it 9, 15}

\bibitem{sch82b}
Schmidt, H.-J.: 1982,  Astron. Nachr. {\bf 303},  283;  gr-qc/0105105.   {\it 19}

\bibitem{sch83}
Schmidt, H.-J.: 1983, in: Bertotti, B., ed., {\it  Contributed Papers
Conf. General Relativity GR10  Padova} {\bf  1},    339.   {\it 45, 46}

\bibitem{sch84a}
Schmidt, H.-J.: 1984, Gen. Relat. Grav.  {\bf 16},  1053; gr-qc/0105106. 
   {\it 46}

\bibitem{sch84b}
Schmidt, H.-J.: 1984, Ann. Phys. (Leipz.) {\bf  41}, 435; gr-qc/0105108. 
   {\it 156}

\bibitem{sch85}
Schmidt, H.-J.: 1985, Astron. Nachr. {\bf 306},  67;   gr-qc/0105107.   {\it 145}

\bibitem{sch86a}
Schmidt, H.-J.: 1986,  Conf.  General Relativity GR 11 Stockholm,   117.   {\it 87}

\bibitem{sch86b}
Schmidt, H.-J.: 1986, {\it  Physikalische Konsequenzen von Feldgleichungen
vierter  Ordnung bei der Beschreibung des fr\"uhen Universums
und isolierter Gravitationsfelder}, Thesis B, 
Akad. d. Wiss.  Berlin; 2nd ed.: 1994,  Tectum-Verlag Marburg.   {\it 
87, 88, 89, 106}

\bibitem{sch86c}
Schmidt, H.-J.: 1986,  Astron. Nachr. {\bf 307},  339;  gr-qc/0106037.
   {\it 159, 170}

\bibitem{sch87a}
Schmidt, H.-J.: 1987,  Astron. Nachr. {\bf 308},  183; gr-qc/0106035.   {\it 87,
 103, 110}

\bibitem{sch87b}
Schmidt, H.-J.: 1987, J. Math. Phys. {\bf  28}, 1928. Addendum: 1988, 
J. Math. Phys. {\bf 29},  1264.   {\it 14, 42, 73}


\bibitem{sch88a}
Schmidt, H.-J.: 1988,  Class. Quant. Grav. {\bf 5}, 233.   {\it 110}

\bibitem{sch88b}
Schmidt, H.-J.: 1988, Astron. Nachr. {\bf 309},  307; gr-qc/0106036.   {\it 87}

\bibitem{sch88c}
Schmidt, H.-J.: 1988, Phys. Lett. B {\bf 214}, 519.   {\it 75, 121}

\bibitem{sch89}
Schmidt, H.-J.: 1989, Class. Quant. Grav. {\bf 6}, 557.  1994, 
Phys. Rev. D {\bf 50}, 5452; gr-qc/0109006. 1995, 
Phys. Rev. D {\bf  52},  6198; gr-qc/0106034.  {\it 112}

\bibitem{sch90a}
Schmidt, H.-J.: 1990,  Astron. Nachr. {\bf 311},  99; gr-qc/0108087.   {\it 58}

\bibitem{sch90b}
Schmidt, H.-J.: 1990,  Astron. Nachr. {\bf 311},  165; gr-qc/0109004.   {\it 113}

\bibitem{sch90c}
Schmidt, H.-J.: 1990, in: Janyska, J., Krupka, D., eds., {\it Proc. Conf. Brno, 
Diff.  Geom. Appl.},  WSPC Singapore, 405; gr-qc/0109001.    {\it 77}

\bibitem{sch90d}
Schmidt, H.-J.: 1990, Class. Quantum Grav. {\bf  7}, 1023.
   {\it 128, 139, 164, 165, 172}

\bibitem{x74}
Schmidt,  H.-J.: 1992, in:  Sato, H.,  Nakamura, T., eds., 
 { \it  Proc. Sixth Marcel Grossmann Meeting on General Relativity Kyoto}, 
WSPC  Singapore,  92.   {\it 129}

\bibitem{x75}
Schmidt, H.-J.: 1993,  Gen. Relat. Grav. {\bf 25}, 87.  Erratum {\bf 25},  863. 
   {\it 127}

\bibitem{schv}
Schmidt, H.-J.: 1994, Phys. Rev. D {\bf 49},  6354. 
 Erratum: 1996,  Phys. Rev. D {\bf 54},  7906; gr-qc/9404038.   {\it 27}

\bibitem{sch95}
Schmidt, H.-J.: 1995, Abstr. Conf. General Relativity GR 14 Florence; gr-qc/0109007.
   {\it 31}

\bibitem{sch96a}
Schmidt, H.-J.: 1996, in:  Janyska, J.,  Kolar, I., Slovak, J., eds., 
{\it  Proc. Conf. Brno Diff.  Geom. Appl.},  119;  gr-qc/0109008.    {\it 31}



\bibitem{sch96b}
Schmidt, H.-J.: 1996, in:  Sardanashvily, G., ed., {\it New frontiers 
in gravitation}, Hadronic Press Palm Harbor, 337; gr-qc/9404037.    {\it 31, 44}

\bibitem{sch96c}
Schmidt, H.-J.: 1996, 
 Gen. Relat. Grav. {\bf  28}, 899; gr-qc/9512006.   {\it 25}

\bibitem{schx}
Schmidt, H.-J.: 1997, Grav.  Cosmol. {\bf  3}, 185; gr-qc/9709071.    {\it 19}      

\bibitem{z2}
 Schmidt, H.-J.: 1998,  Int. J. Theor. Phys. {\bf  37}, 691; gr-qc/9512007.
   {\it 29}

\bibitem{schy}
Schmidt, H.-J.: 1998, in:  Rainer, M.,  Schmidt, H.-J., eds., {\it 
Current topics in mathematical cosmology},  WSPC
Singapore, 288; gr-qc/9808060. 1993,  Fortschr. Phys. {\bf  41}, 179.   {\it 27}

\bibitem{sch}
Schmidt, H.-J.: 1999,  Gen. Relat. Grav. {\bf 31}, 1187;  gr-qc/9905051. 
    {\it 111}



\bibitem{schz}
 Schmidt, H.-J: 2000,  Ann. Phys. (Leipz.) {\bf 9}, SI-158.
gr-qc/9905103.   {\it 44}

\bibitem{schmuell}
Schmidt, H.-J., M\"uller, V.: 1985,  Gen. Relat. Grav. {\bf  17}, 971.
   {\it 150}

\bibitem{sem}
Schmidt, H.-J., Semmelmann, U.: 1989,  Astron. Nachr. {\bf 310}, 103; 
gr-qc/0106089.    {\it 175}

\bibitem{y35}
 Senovilla, J.: 1998, Gen. Relat. Grav. {\bf 30},  701.   {\it 27}

\bibitem{schr1938}
Schr\"odinger, E.: 1938, Comment. Vatican. Acad. {\bf 2}, 321.
 1956, {\it Expanding Universes}, Cambridge University Press.   {\it 58}

\bibitem{x77}
Shibata, M.,  Nakao, K.,  Nakamura, T.,  Maeda, K.: 1994, Phys. Rev.  D {\bf  50}, 708.
   {\it 124}



\bibitem{x78}
Shinkai, H.,  Maeda, K.: 1993,  Phys. Rev. D  {\bf  48}, 3910.
  1994,  Phys. Rev. D {\bf  49}, 6367.     {\it 127}

\bibitem{x79}
Shiromizu, T.,  Nakao, K.,  Kodama, H.,  Maeda, K.: 1993,  Phys. Rev. D {\bf  47}, R3099. 
   {\it 124, 127}

\bibitem{singh}
Singh, T.,   Padmanabhan, T.: 1988, Int. J. Mod. Phys. A {\bf  3}, 1593.   {\it 75}

\bibitem{x80}
Sirousse-Zia, H.: 1982,  Gen. Relat. Grav. {\bf 14}, 751.   {\it 125}

\bibitem{spero}
Spero, A.,   Baierlein, R.: 1978, J. Math. Phys. {\bf 19}, 1324.   {\it 9}

\bibitem{x81}
Spindel,  P.: 1994, Int. J. Mod. Phys. D {\bf  3}, 273.   {\it 125, 127, 139}

\bibitem{stacey}
Stacey, F., Tuck, G., Holding, S., Maher, A., 
Morris, D.: 1981, Phys. Rev. D {\bf  23},  1683.   {\it 165}


\bibitem{staro78}
Starobinsky, A.: 1978, Pis'ma Astron. Zh. {\bf 4}, 155.   {\it 58, 74, 107}

\bibitem{x83}
Starobinsky,  A.: 1983, Sov. Phys. JETP Lett. {\bf  37}, 66.   {\it 126, 127}

\bibitem{starosch}
Starobinsky,  A., Schmidt, H.-J.: 1987, Class. Quant. Grav. {\bf  4}, 695.
   {\it 61, 103, 106, 127}

\bibitem{y36}
Starobinsky, A., Tsujikawa, S.,  Yokoyama, J: 2001, {\it  Cosmological perturbations 
from   multi-field inflation 
in generalized Einstein theories},   astro-ph/0107555; Nucl. Phys. B in press.
   {\it 112}

\bibitem{sb}
Stein-Schabes, J., Burd,  A.: 1988, Phys. Rev. D {\bf  37}, 1401.   {\it 179}

\bibitem{stelle77}
Stelle, K.: 1977, Phys. Rev. D {\bf  16}, 953.   {\it 90, 122, 128}

\bibitem{stelle78}
Stelle, K.: 1978,  Gen. Relat. Grav. {\bf  9}, 353.   {\it 146,
 149, 160, 163, 164, 170 }

\bibitem{stephani}
Stephani,  H.: 1982, {\it General Relativity}, Cambridge University Press. 
   {\it 7, 36, 164}

\bibitem{szekeres}
Szekeres, P.: 1975, Commun. Math. Phys. {\bf 41}, 55.   {\it 9, 21, 25}

\bibitem{taub}
Taub, A.: 1980, J. Math. Phys. {\bf 21}, 1423.   {\it 49}

\bibitem{teyss}
Teyssandier, P.:  1989, Class. Quant. Grav. {\bf  6}, 219.   {\it 164}

\bibitem{to} 
Thompson, J.: 1988, Phys. Rev. D {\bf  37}, 283.   {\it 182}

\bibitem{thomson}
Thomson, W. (Lord Kelvin): 1858, Proc. R. Soc. London  {\bf 9}, 255. 
   {\it 47}

\bibitem{tolman}
Tolman, R. C.: 1934, {\it Relativity Thermodynamics and Cosmology}, 
Clarendon Press Oxford.   {\it 19}

\bibitem{treder75}
Treder, H.-J.: 1975, Ann. Phys. (Leipz.) {\bf 32}, 383.  
 1977,  in: Brauer, W., ed., {\it 75 Jahre Quantentheorie},
Akad.-Verl.  Berlin, 279.  1991, Found. Phys. {\bf 21}, 283.
   {\it 129, 146}

\bibitem{turner}
Turner, M.: 1983, Phys. Rev. D {\bf  28}, 1243.   {\it 59}

\bibitem{turnerwidrow}
Turner, M., Widrow, L.: 1988, Phys. Rev. D {\bf 37}, 3428.   {\it 75}

\bibitem{urbandtke}
Urbandtke, H.: 1972, Acta Physica Austriaca {\bf 35}, 1.   {\it 45}

\bibitem{vil}
Vilenkin, A.: 1985, Phys. Rep. {\bf 121}, 263.   {\it 175}

\bibitem{x87}
Vilenkin, A.: 1992, Phys. Rev. {\bf D 46}, 2355.   {\it 124}

\bibitem{x88}
Vilkovisky, G.: 1992,  Class. Quant. Grav. {\bf  9}, 895.    {\it 128, 129}


\bibitem{voorhees}
Voorhees, B.: 1972,  Phys. Rev. D {\bf 5}, 2413.   {\it 45}

\bibitem{wain}
Wainwright, J.: 1981, J. Phys. A {\bf 14}, 1131.   {\it 9, 19}

\bibitem{x89}
Wands,  D.: 1994,  Class. Quant. Grav. {\bf 11}, 269.    {\it 129}

\bibitem{wein}
Weinberg, S.: 1979, in: Hawking, S., Israel, W., eds., {\it  General Relativity}, 
 Cambridge University Press.,  790.   {\it 105}

\bibitem{weyl19}
Weyl, H.: 1919, Ann. Phys. (Leipz.) {\bf  59}, 101.
 1921, {\it Raum, Zeit, Materie},  Springer Berlin.   {\it 145, 156}

\bibitem{x90}
Weyl, H.: 1927, {\it  Handbuch der Philosophie}, chapter
``Philosophie  der Mathematik und Naturwissenschaft'',  Oldenburg.
   {\it 122}

\bibitem{whitt}
Whitt, B.: 1984, Phys. Lett. B {\bf  145}, 176.   {\it 88}

\bibitem{wunsch}
W\"unsch, V.: 1976, Math. Nachr. {\bf 73}, 37.   {\it 147}

\bibitem{x91}
  Xu, J.,  Li, L.,  Liu, L.: 1994,  Phys. Rev. D {\bf  50}, 4886.    {\it 125}

\bibitem{x92}
Yokoyama, J. and  Maeda, K.: 1990, Phys. Rev. D {\bf  41}, 1047.   {\it 125}

\bibitem{yu}
Yukawa, H.: 1935, Proc. Math. Phys. Soc. Japan {\bf 17}, 48.   {\it 164}

\bibitem{zala}
Zalaletdinov, R.: 1996,  Gen. Relat. Grav. {\bf  28},  953.    {\it 112}

\bibitem{z}
Zeldovich, Y.: 1980, Mon. Not. R. Astron. Soc. {\bf  192}, 663.   {\it 175}

\end{thebibliography}
\end{document}